\documentclass[submission, Phys]{SciPost}
\pdfoutput=1

\usepackage[english]{babel}
\usepackage[utf8]{inputenc}
\usepackage[T1]{fontenc} 
\usepackage{amsfonts,amsmath,amssymb,amsbsy,amsthm,bm,mathtools,fixmath,mathrsfs,amscd,slashed}
\usepackage{comment,enumerate,footnote,float,subfloat,relsize,footnote,verbatim,pifont}
\usepackage{array,tabularx,tabu,multirow,framed,afterpage}
\usepackage{graphicx,graphics,epsfig,color,cancel,times,soul}
\usepackage[normalem]{ulem}

\def\ann		{{\rm ann}}
\def\BSF		{\mathsmaller{\rm BSF}}
\def\dec		{{\rm dec}}
\def\DM			{\mathsmaller{\rm DM}}

\def\med		{{\rm med}}
\def\rec		{{\rm \bf rec}}
\def\e			{{\bf e}}
\def\ebar		{{\bf \bar{e}}}
\def\p			{{\bf p}}
\def\pbar		{{\bf \bar{p}}}
\def\H			{{\bf H}}

\def\V			{{\bf V}}
\def\mDM		{m_{\DM}}
\def\me			{m_{\e}}
\def\mpd		{m_{\p}}
\def\mH			{m_{\mathsmaller{\H}}}
\def\mV			{m_{\mathsmaller{\V}}}
\def\EV         {E_{\mathsmaller{\V}}}
\def\muD		{\mu_{\mathsmaller{D}}}

\def\ED		    {{\cal E}_{\mathsmaller{D}}}
\def\np			{n_{\p}}
\def\npbar		{n_{\pbar}}
\def\ne			{n_{\e}}
\def\nebar		{n_{\ebar}}
\def\nH			{n_{\mathsmaller{\H}}}

\def\fion		{f_{\rm ion}}
\def\aD			{\alpha_{\mathsmaller{D}}}
\def\gammaD		{\gamma_{\mathsmaller{D}}}
\def\sps		{s_{\rm ps}}
\def\FD			{F_{\mathsmaller{D}}}

\def\sv			{(\sigma \vrel)}

\def\TD			{T_{\mathsmaller{D}}}
\def\TSM		{T_{\mathsmaller{\rm SM}}}

\def\vrel		{v_{\rm rel}}

\renewcommand{\(}{\left(}
\renewcommand{\)}{\right)}

\newcommand{\<}{\left\langle}
\renewcommand{\>}{\right\rangle}

\newcommand{\be}{\begin{equation}}
\newcommand{\ee}{\end{equation}}
\newcommand{\bea}{\begin{equation}\begin{aligned}}
\newcommand{\eea}{\end{aligned}\end{equation}}

\usepackage[capitalize]{cleveref}
\crefformat{pluralequation}{#2\black{Eqs.~(}#1\black{)}#3}
\Crefformat{pluralequation}{#2\black{Equations~(}#1\black{)}#3}
\crefformat{pluralfigure}{#2\black{Figs.~}#1#3}
\Crefformat{pluralfigure}{#2\black{Figures~}#1#3}
\numberwithin{equation}{section}
\begin{document}

\begin{center}
ULB-TH/20-09, LAPTH-032/20, Nikhef-2020-023
\end{center}

\begin{center}{\Large \textbf{
Indirect searches for dark matter bound state formation
\\ and level transitions
}}\end{center}

\begin{center}
Iason Baldes,\textsuperscript{1*}
Francesca Calore,\textsuperscript{2}
Kalliopi Petraki,\textsuperscript{3,4}
Vincent Poireau,\textsuperscript{5}
and Nicholas L. Rodd\textsuperscript{6,7}
\end{center}

\begin{center}
{\bf 1} Service de Physique Th\'eorique, Universit\'e Libre de Bruxelles, \\ Boulevard du Triomphe, CP225, B-1050 Brussels, Belgium
\\
{\bf 2} Universit\'e Grenoble Alpes, USMB, CNRS, LAPTh, F-74940 Annecy, France
\\
{\bf 3} Sorbonne Universit\'e, CNRS, Laboratoire de Physique Th\'eorique et Hautes Energies, \\ LPTHE, F-75252 Paris, France
\\
{\bf 4} Nikhef,  Science  Park  105,  1098  XG  Amsterdam,  The  Netherlands
\\
{\bf 5}  Universit\'e Grenoble Alpes, USMB, CNRS, LAPP, F-74940 Annecy, France
\\
{\bf 6} Berkeley  Center  for  Theoretical  Physics,  University  of  California, \\  Berkeley,  CA  94720,  USA
\\
{\bf 7} Theoretical  Physics  Group,  Lawrence  Berkeley  National  Laboratory, \\ Berkeley,  CA  94720,  USA
\\

* iason.baldes@ulb.ac.be
\end{center}

\begin{center}
August 5, 2020
\end{center}


\section*{Abstract}  
{\bf

Indirect searches for dark matter (DM) have conventionally been applied to the products of DM annihilation or decay. If DM couples to light force carriers, however, it can be captured into bound states via dissipation of energy that may yield detectable signals. We extend the indirect searches to DM bound state formation and transitions between bound levels, and constrain the emission of unstable dark photons. Our results significantly refine the predicted signal flux that could be observed in experiments. As a concrete example, we use Fermi-LAT dwarf spheroidal observations to obtain constraints in terms of the dark photon mass and energy which we use to search for the formation of stable or unstable bound states.
}

\setcounter{tocdepth}{2}
\vspace{10pt}
\noindent\rule{\textwidth}{1pt}
\tableofcontents\thispagestyle{fancy}
\noindent\rule{\textwidth}{1pt}
\vspace{10pt}

\section{Introduction}


Most of the dark matter (DM) research in the past decades has focused on DM with \emph{contact-type} interactions, {\it i.e.}~interactions mediated by particles of similar or larger mass than the DM itself, $m_{\med} \gtrsim m_{\DM}$. 
Indeed, in the prototypical WIMP (Weakly Interacting Massive Particle) scenario, DM was envisioned to couple to the weak interactions of the Standard Model (SM) and have mass $m_{\DM} \sim m_{\mathsmaller{W,Z}} \sim 100$~GeV. The current collider, direct detection, and indirect detection searches strongly constrain this scenario. Nevertheless, they still allow for WIMP DM around or beyond the TeV scale. The same conclusion essentially holds for a variety of models in which DM communicates with the SM particles via non-SM mediators. However, for WIMP DM with $m_{\DM} \gtrsim $~TeV~$\gg m_{\mathsmaller{W,Z}}$, the weak interactions manifest as \emph{long-range}~\cite{Hisano:2002fk}.

On a more fundamental and model-independent level, the unitarity of the $S$-matrix suggests that the long-range character of the interactions is a generic feature of viable thermal-relic DM models in the multi-TeV mass range and above~\cite{Baldes:2017gzw}. Indeed, unitarity sets an upper bound on the partial-wave inelastic cross sections, whose physical significance is the saturation of the probability for inelastic scattering. This in turn implies an upper bound on the mass of DM produced via thermal freeze-out~\cite{Griest:1989wd}, of the order of 100~TeV~\cite{vonHarling:2014kha,Baldes:2017gzw}. The parametric dependence of the unitarity limit on the inelastic cross sections shows that the 100~TeV regime can be approached or reached only by interactions that manifest as long-range~\cite{Baldes:2017gzw}.

Long-range interactions imply the emergence of non-perturbative effects that can affect significantly the DM phenomenology.  
A long-range force distorts the wave function of a pair of DM particles, and consequently affects all their interaction rates at low velocities. This is the well known Sommerfeld effect~\cite{Sommerfeld:1931,Sakharov:1948yq}, which has been extensively studied in the DM literature, both in WIMP and hidden-sector models (see {\it e.g.}~\cite{Hisano:2002fk,Hisano:2003ec,ArkaniHamed:2008qn,Pospelov:2008jd,Cassel:2009wt,Feng:2010zp,Hryczuk:2011tq,Abazajian:2011ak,Beneke:2014hja,Cirelli:2015bda,ElHedri:2016onc,Bringmann:2016din,Cirelli:2016rnw,Harz:2017dlj,Baldes:2017gzw,Baldes:2017gzu}).  
It has been shown to decrease the DM density in the early Universe, and enhance the expected indirect detection signals.  Another potentially more consequential implication of long-range interactions is the existence of \emph{bound states}~\cite{Pospelov:2008jd,vonHarling:2014kha,Petraki:2015hla}. 
Bound-state formation (BSF) -- which is also affected by the Sommerfeld effect -- can alter the DM phenomenology in a variety of ways.

To delineate the consequences of DM bound states, we discern two broad categories.
\begin{description}
\item [Unstable bound states.]

The formation of particle-antiparticle (positronium-like) bound states that can decay into radiation opens a new two-step DM annihilation channel. In models that feature co-annihilation between different species and/or non-Abelian forces, a variety of unstable bound states may exist. The formation and decay of unstable bound states diminish the DM density in the early Universe~\cite{vonHarling:2014kha}, thereby altering the expected DM mass and couplings and affecting all experimental signatures~\cite{vonHarling:2014kha,Kim:2016zyy,Harz:2018csl,Harz:2019rro,Biondini:2018pwp,Binder:2018znk,Binder:2019erp,Ko:2019wxq,Oncala:2019yvj,Binder:2020efn}. During the CMB (Cosmic Microwave Background) period and inside galaxies today, the bound state decay products enhance the high-energy radiative signals searched by telescopes~\cite{Pospelov:2008jd,MarchRussell:2008tu,An:2016gad,An:2016kie,Cirelli:2016rnw,Kouvaris:2016ltf,Asadi:2016ybp,Baldes:2017gzu}.

\item [Stable bound states.]

The formation of stable bound states alters -- typically screens or curtails -- the DM self-interactions inside halos~\cite{Petraki:2014uza,Wise:2014jva}, which are expected to affect the galactic structure~\cite{Spergel:1999mh,Tulin:2017ara}. Moreover, stable bound states affect the DM direct detection signatures~\cite{Cline:2012is,Kahlhoefer:2020rmg}. 
Stable bound states arise typically either due to confining forces (hadronic bound states), and/or due to weak forces in models of asymmetric DM. The latter scenario hypothesizes that the DM relic density is, analogously to ordinary matter, due to an excess of dark particles over antiparticles~\cite{Nussinov:1985xr,Petraki:2013wwa,Zurek:2013wia} that cannot be annihilated in the early Universe even if the DM annihilation cross section is very large. It follows that DM may possess significant couplings to light force mediators that in turn may render BSF rather efficient.

\end{description}

The DM capture into bound states, be they stable or unstable, invariably necessitates the dissipation of energy. The amount dissipated -- of the order of the binding energy of the bound state -- 
is significantly lower than that radiated in the typical DM annihilation and decay of unstable bound states. Pearce and Kusenko first suggested that the energy dissipated during BSF at late times may give rise to novel signals detectable via indirect searches~\cite{Pearce:2013ola}. Transitions between bound levels may also produce similar signals. Indeed, the available energy may be dissipated radiatively, most commonly via emission of a force mediator. In multi-TeV WIMP models, BSF inside halos can occur via emission of a mono-energetic photon in the multi-GeV range~\cite{Asadi:2016ybp}. In models where DM couples to non-SM forces, the emitted mediators may decay into SM particles whose cascades produce a more extended spectrum of photons and other stable SM particles~\cite{Pearce:2015zca,Cirelli:2016rnw,Baldes:2017gzu}.\footnote{We note in passing that in models and thermodynamic environments where the force mediators couple also to a plasma of relativistic particles, the DM capture into bound states may occur efficiently also by scattering on the relativistic bath, via exchange of an off-shell mediator~\cite{Kim:2016zyy,Biondini:2017ufr,Binder:2019erp,Binder:2020efn}. While BSF via scatterings~\cite{Binder:2019erp,Binder:2020efn} and other rearrangement processes~\cite{Geller:2018biy,Geller:2020tyv} can be quite efficient in the early Universe, such processes have not been shown to be efficient inside galaxies and produce DM indirect signals.}
The purpose of the present work is to initiate a systematic investigation of indirect constraints on BSF.

The indirect signals emanating from BSF and other level transitions can provide a powerful probe of asymmetric DM models, where late-time DM annihilation is highly suppressed due to the absence of antiparticles~\cite{Graesser:2011wi,Bell:2014xta,Murase:2016nwx,Baldes:2017gzu}, unless a mechanism exists that erases the asymmetry at late times~\cite{Cirelli:2011ac,Agrawal:2016uwf,Dessert:2018khu}. In contrast to annihilation, and because asymmetric DM can accommodate large DM-mediator couplings, BSF can be quite efficient. Part of the parameter space where this occurs is in fact interesting for an additional reason, that it provides a viable framework of self-interacting DM~\cite{Petraki:2014uza,Baldes:2017gzu}. Indirect signals from the formation of stable bound states in asymmetric DM models have been proposed in Refs.~\cite{Pearce:2013ola,Pearce:2015zca,Cline:2014eaa,Detmold:2014qqa,Mahbubani:2019pij}. As a concrete example that we consider below in more detail, we mention here the formation of dark atoms via emission of light dark photons that subsequently decay into SM particles via kinetic mixing with hypercharge~\cite{Pearce:2015zca}. While the parameter space of the model is broader, it has been shown that  dark atomic transitions between levels with MeV-scale splittings could inject low-energy positrons in the Milky Way at a sufficient rate to account for the observed 511~keV line~\cite{Pearce:2015zca}.

Even in the context of symmetric or self-conjugate DM, BSF signals may provide an important probe, since they may exhibit different spectral features and resonant structure than direct annihilation~\cite{Petraki:2016cnz}. Moreover, for very heavy DM whose annihilation signals fall outside the energy range of the various telescopes, the low-energy radiation could fall within the energy range probed by telescopes, and could thus be employed to constrain a wider range of DM masses.\footnote{This situation is in fact more subtle. Photons above 100 TeV can initiate electromagnetic cascades via an interaction with, for example, the CMB~\cite{Murase:2012xs,Esmaili:2015xpa,Cohen:2016uyg}, producing lower energy radiation. Similar results hold for electrons, but not for neutrinos.}

The radiative BSF cross sections can be comparable to or even significantly larger than the direct annihilation cross sections~\cite{vonHarling:2014kha,Petraki:2015hla,Petraki:2016cnz,Harz:2018csl,Harz:2019rro,Oncala:2019yvj}. In fact, the BSF cross sections in galactic environments may exceed the so-called canonical annihilation cross section, $\sigma \vrel \approx 3 \times 10^{-26} {\rm cm}^3 / {\rm s}$, by orders of magnitude due to different reasons. These include a large Sommerfeld enhancement at low velocities, and possibly the associated parametric resonances in the case of massive mediators~\cite{Petraki:2016cnz}, as well as, in the case of asymmetric DM, a larger DM-mediator coupling than that required to attain the observed DM density via freeze-out in the symmetric limit~\cite{Baldes:2017gzw}. However, the accurate estimation of the expected BSF signals, and indeed of any DM experimental signature, necessitates computing the cosmology first~\cite{Petraki:2013wwa}. If bound states exist, then they may form efficiently in the early Universe. As already mentioned, the formation and decay of unstable bound states in the early Universe decreases the DM density~\cite{vonHarling:2014kha}, and therefore alters the predicted DM parameters that determine the late-time BSF rate~\cite{Cirelli:2016rnw}. The formation of stable bound states in the early Universe changes the density of particles available to participate in the corresponding processes inside halos, and thus again affects the expected indirect signals~\cite{Pearce:2015zca}.

The structure of the paper is as follows. To flesh out the above, we begin in \cref{sec:darkQED} by introducing an atomic DM scenario with a light albeit massive dark photon that mixes kinetically with hypercharge. After summarising the cosmology of the model, we estimate the indirect signals expected from the DM capture into dark atoms via emission of dark photons. Compared to previous studies~\cite{Pearce:2015zca}, we employ improved numerical calculations of the BSF cross sections~\cite{Petraki:2016cnz}, and compute the $\gamma$-ray flux from the cascades of the charged particles produced in the dark photon decays.  
In \cref{sec:constraints}, we briefly consider the recasting of existing constraints on DM annihilation into SM particles for the purpose of constraining BSF, before deriving new constraints on BSF occurring via dark photon emission. We use Fermi-LAT observations of dwarf spheroidal (dSph) galaxies, which provide a DM rich environment with relatively lower background compared to the Galactic Centre. The constraints are cast in terms of the DM mass and the energy dissipated, such that they can be used in models with different underlying dynamics.
They are applicable to BSF, as well as excitation processes occuring via DM collisions and followed by de-excitations. 
The predictions of the atomic DM model of \cref{sec:darkQED} are confronted with the derived constraints in \cref{sec:comparison}, where we also discuss further applications. Some general remarks are then drawn in the conclusion.

\section{Atomic dark matter with a massive dark photon}
\label{sec:darkQED}

\subsection{The model}

\begin{figure}[t!]
\begin{center}
\includegraphics[width=350pt]{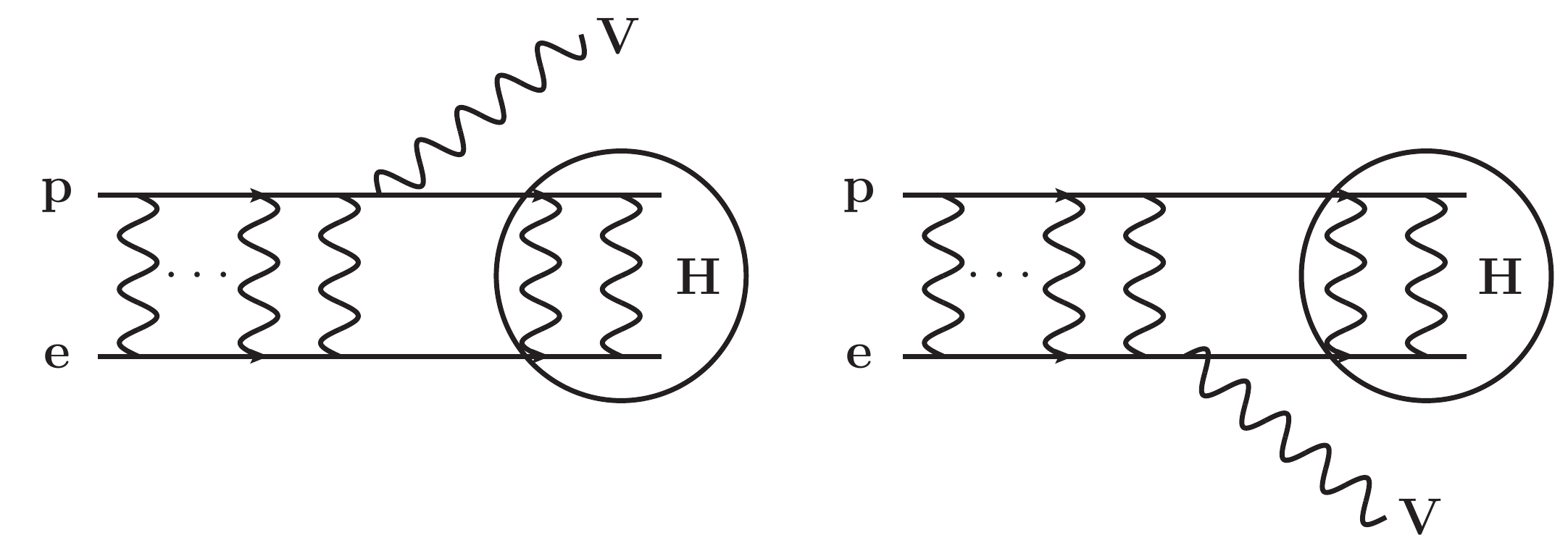}
\end{center}
\caption{\small The process targeted in this project. A dark proton and dark electron combine to form dark hydrogen through the emission of a massive mediator (dark photon). The mediator decays into SM particles through its kinetic mixing with hypercharge, which eventually yields lower energy photons, which can be searched for by Fermi. \bigskip}
\label{fig:HD}
\end{figure}

We assume that DM is charged under a dark $U(1)_{\mathsmaller{D}}$ gauge symmetry, and that it carries a particle-antiparticle asymmetry conserved at low energies due to a global dark baryonic symmetry governing the interactions of the dark sector. If $U(1)_{\mathsmaller{D}}$ is unbroken, then gauge invariance mandates that there must be at least two dark particle species with compensating asymmetries, such that the dark electric charge of the Universe vanishes. This remains true if the dark photon acquires a mass via the St\"uckelberg mechanism, 
or if $U(1)_{\mathsmaller{D}}$ is broken via a Higgs mechanism that operated in the early Universe after the dark baryogenesis took place. The latter implies that the generated dark photon mass is sufficiently small. We refer to~\cite{Petraki:2013wwa} for the detailed considerations. 

Considering the above, we will assume that the dark photon $\V$ has a small non-zero mass $\mV$, and that DM consists of two species of fundamental Dirac fermions, the dark protons $\p$ and the dark electrons $\e$, with opposite charges and masses $\mpd \geqslant \me$. Thus, the low-energy physics of the dark sector we explore in this paper is summarised by the Lagrangian:
\begin{align}
\mathcal{L} \; = 
& \; \- \frac{1}{2}\mV^2 \V_\mu \V^{\mu} - \frac{1}{4}{\FD}_{\mu\nu} \FD^{\mu\nu} - \frac{\epsilon}{2c_{w}}{\FD}_{\mu\nu} F_{Y}^{\mu\nu} 
+ \pbar (i \slashed{D} - \mpd)\p + \ebar (i \slashed{D} - \me) \e.
\end{align}
The covariant derivative is $D^\mu = \partial^\mu \pm i g_{\mathsmaller{D}} \V^\mu$  for $\p$ and $\e$ respectively. The field strength tensor is $F_{\mathsmaller{D}}^{\mu\nu} = \partial^\mu \V^\nu - \partial^\nu \V^\mu$, and $\aD \equiv g_{\mathsmaller{D}}^2/(4\pi)$ is the dark fine-structure constant.  
The dark photons may decay into SM particles through the kinetic mixing with hypercharge, controlled by the dimensionless parameter $\epsilon$. Here $c_{w} \equiv \cos{\theta_w}$ where $\theta_w$ is the Weinberg angle. Constraints on the dark photon and on DM direct detection via dark photon exchange are compiled in \cref{sec:darkphotonconstraints,sec:directdetection} respectively.

High energy completions of this scenario, including mechanisms for the generation of the dark matter-antimatter asymmetry, that could be potentially related to that of ordinary matter, can be found {\it e.g.}~in Refs.~\cite{Kaplan:2009de,Petraki:2011mv,vonHarling:2012yn,Baldes:2014gca,Choquette:2015mca}, and the DM freeze-out has been previously studied in Refs.~\cite{Baldes:2017gzw,Baldes:2017gzu}. 
Here we shall only use that the dark proton-antiproton and dark electron-positron asymmetries are equal, and that the dark antiparticles were efficiently annihilated in the early Universe with an equal amount of dark particles, thereby leaving a Universe that contains globally (nearly) equal densities of dark protons and dark electrons, $\np \cong \ne \ggg \npbar, \nebar$. The exact number of residual antiparticles depends on the effective annihilation cross section in the early Universe, here controlled by the coupling $\aD$; in order for the DM density to be set largely by the primordial asymmetry, it is sufficient that $\aD$ is somewhat higher than that for symmetric thermal relic DM of the same mass~\cite{Graesser:2011wi,Baldes:2017gzw,Baldes:2017gzu}. 

The symmetric DM realisation of this scenario, containing only one dark species, has been studied in Refs.~\cite{Cirelli:2016rnw,Cirelli:2018iax}, with particular emphasis on the indirect constraints due to late-time DM annihilations. Remarkably, annihilation constraints arising from the small but non-zero residual density of dark antiparticles, exist also in the asymmetric regime for late-time asymmetries $\npbar / \np \gtrsim 10^{-3}$, due to the large Sommerfeld enhancement of the annihilation cross section that compensates in part for the suppression of the annihilation rate due to the small residual density of antiparticles~\cite{Baldes:2017gzw,Baldes:2017gzu}. For larger asymmetries, \emph{i.e.}~larger values of $\aD$, the annihilation rate falls below the sensitivity of the current observations. Nevertheless, a larger $\aD$ implies that the formation of stable bound states may be possible for a larger range of $\mpd, \me$, giving rise to radiative signals~\cite{Pearce:2015zca} that we shall now explore.

If the dark photons are sufficiently light, then the dark protons and the dark electrons can form dark hydrogen atoms. The capture into atomic bound states may occur via emission of a dark photon, 
\begin{equation}
\p+\e \to \H + \V,
\end{equation}
provided it is kinematically allowed, and as illustrated in \cref{fig:HD}. The dark atom formation may occur in the early Universe (dark recombination), as well as at late times, during the CMB period or inside galaxies today.
In the following, we specify the relevant BSF cross section, the ionized fraction of DM that may participate in this process today, the branching fractions of the dark photons into SM particles, and the $\gamma$-ray spectrum resulting from the cascades of the latter.

\subsection{Formation of dark atoms}

For convenience, we define the following parameters
\begin{align}
\zeta &\equiv \aD / \vrel , \label{eq:zeta_def} \\
\xi &\equiv  \aD \muD / \mV , \label{eq:xi_def} \\
\muD &\equiv m_{\e}m_{\p} /(m_{\e}+m_{\p}) . \label{eq:mu_def}
\end{align}
The first is important in determining the strength of the Sommerfeld enhancement and the overlap of the scattering-state and bound-state wave functions. The second is the ratio of the dark atom Bohr radius to the range of the dark-photon-mediated interaction, and parametrises how long range this interaction manifests. The last parameter is the $\p-\e$ reduced mass.

\begin{figure}[t]
\begin{center}
\includegraphics[width=185pt]{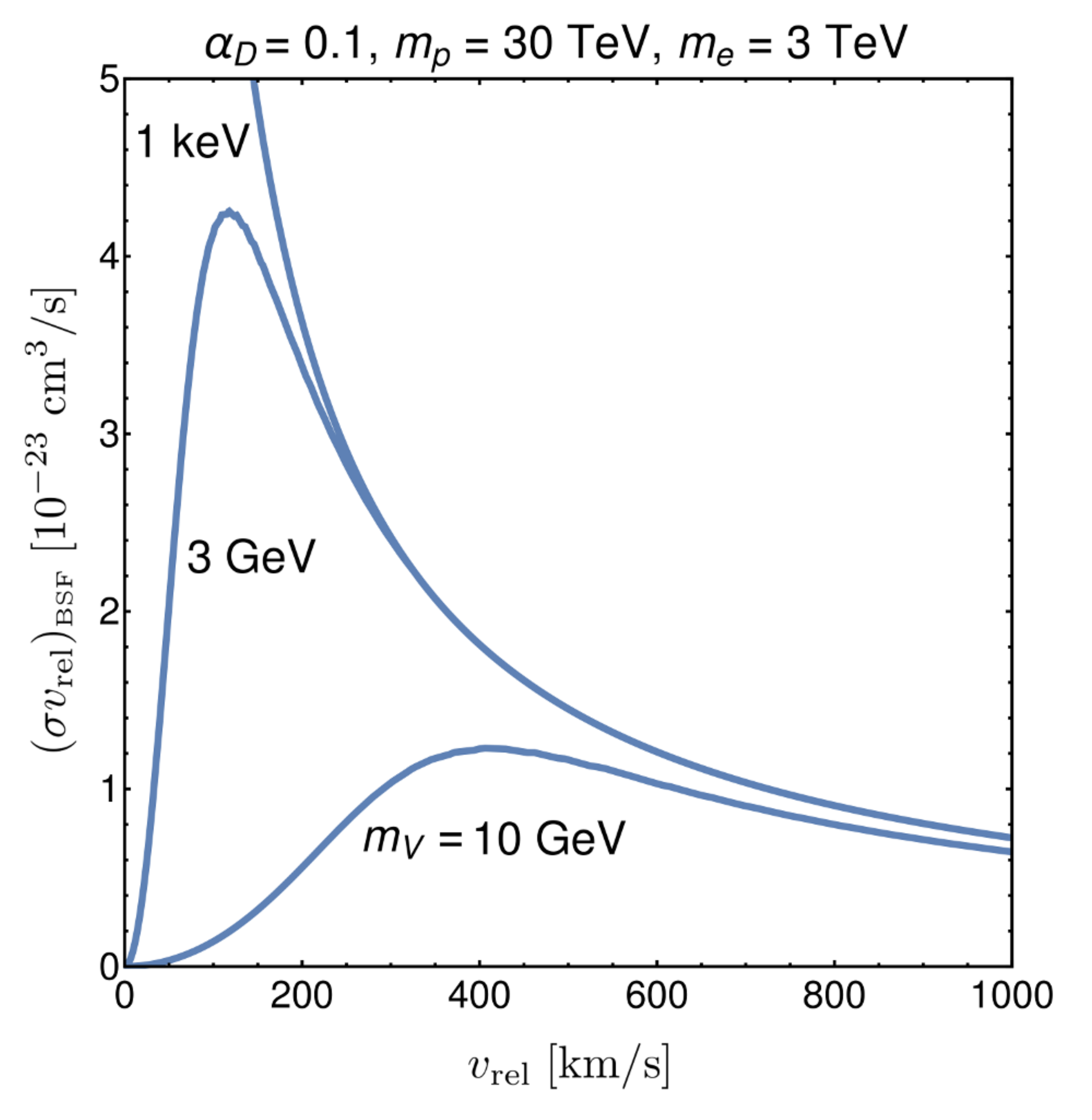}
\end{center}
\caption{\small The bound state formation cross section for different choices of dark photon mass. The scaling of the cross section changes from approximately $\sv_{\BSF} \propto \vrel^{2}$ to $\sv_{\BSF} \propto 1/\vrel$ at $\vrel \sim \mV/\mu_D$. 
\bigskip}
\label{fig:BSFsigma}
\end{figure}

Bound levels of $\p\e$ pairs exist if $\xi > 0.84$~\cite{Petraki:2016cnz}. They may form radiatively, via emission of a dark photon. We will consider capture into the ground-state only, which is the dominant BSF process and most exothermic transition, with the energy available to be dissipated being~\cite{Petraki:2015hla}
\begin{equation}
\omega \simeq \ED +\muD \vrel^2/2 
= (\muD/2) [\gammaD^2(\xi) \aD^2 +\vrel^2],
\label{eq:omega}
\end{equation}
where $\ED \equiv \gammaD^2(\xi) \times \muD \aD^2/2$ is the absolute value of the binding energy. The factor $\gammaD(\xi) \leqslant 1$ parametrises the departure from the Coulomb value. The cross section for capture into the ground state is~\cite{Petraki:2016cnz}
\begin{align}
\sv_{\BSF} 
= \frac{\pi \aD^2}{4\muD^2} 
\frac{\sqrt{\sps} (3-\sps)}{2} 
\times S_{\BSF} (\zeta, \xi) , 
\label{eq:sigmav_BSF}
\end{align}
where
\begin{align}
\sps \equiv 1- \mV^2/ \omega^2 ,
\label{eq:s_ps}
\end{align}
parametrises the phase-space suppression due to the massive dark photon. Both $\gammaD(\xi)$ and the function $S_\BSF (\zeta, \xi)$ in \cref{eq:sigmav_BSF} are computed numerically according to Ref.~\cite{Petraki:2016cnz}. In the Coulomb limit $\xi \to \infty$, the latter takes the analytical form~\cite{Petraki:2014uza,Petraki:2015hla,Petraki:2016cnz}
\begin{align}
S_{\BSF} (\zeta) \simeq 
\(\frac{2\pi \zeta}{1-e^{-2\pi \zeta}}\)
\frac{2^{10}}{3}  \frac{\zeta^4}{(1 + \zeta^2)^2} \ e^{-4 \zeta {\rm arccot}(\zeta)} .
\label{eq:SBSF_Coul}
\end{align}
In fact, the Coulomb approximation is satisfactory for $\muD \vrel \gtrsim \mV$, or equivalently $\xi \gtrsim \zeta$~\cite{Petraki:2016cnz}. We note that the capture into the ground state is a $p$-wave process. While in the Coulomb regime and for $\aD \gtrsim \vrel$, $\sv_{\BSF}$ exhibits the characteristic Sommerfeld scaling $\propto 1/\vrel$ as seen from \cref{eq:SBSF_Coul}, the finite $\mV$ implies that at velocities $\vrel \lesssim \mV/\muD$, the BSF cross section recovers the perturbative scaling $\sv_{\BSF} \propto \vrel^{2}$~\cite{Petraki:2016cnz}. An example of the cross section including the effects of the finite mediator mass is shown in \cref{fig:BSFsigma}.
Because the BSF cross section is suppressed at $\vrel \gg \aD$, as seen from \cref{eq:SBSF_Coul}, the phase-space suppression \eqref{eq:s_ps} implies that $\xi \gg 1$ whenever BSF is kinematically allowed and significant, thus to a good approximation $\gammaD(\xi) \simeq 1$.

\begin{figure}[h!]
\begin{center}
\includegraphics[width=185pt]{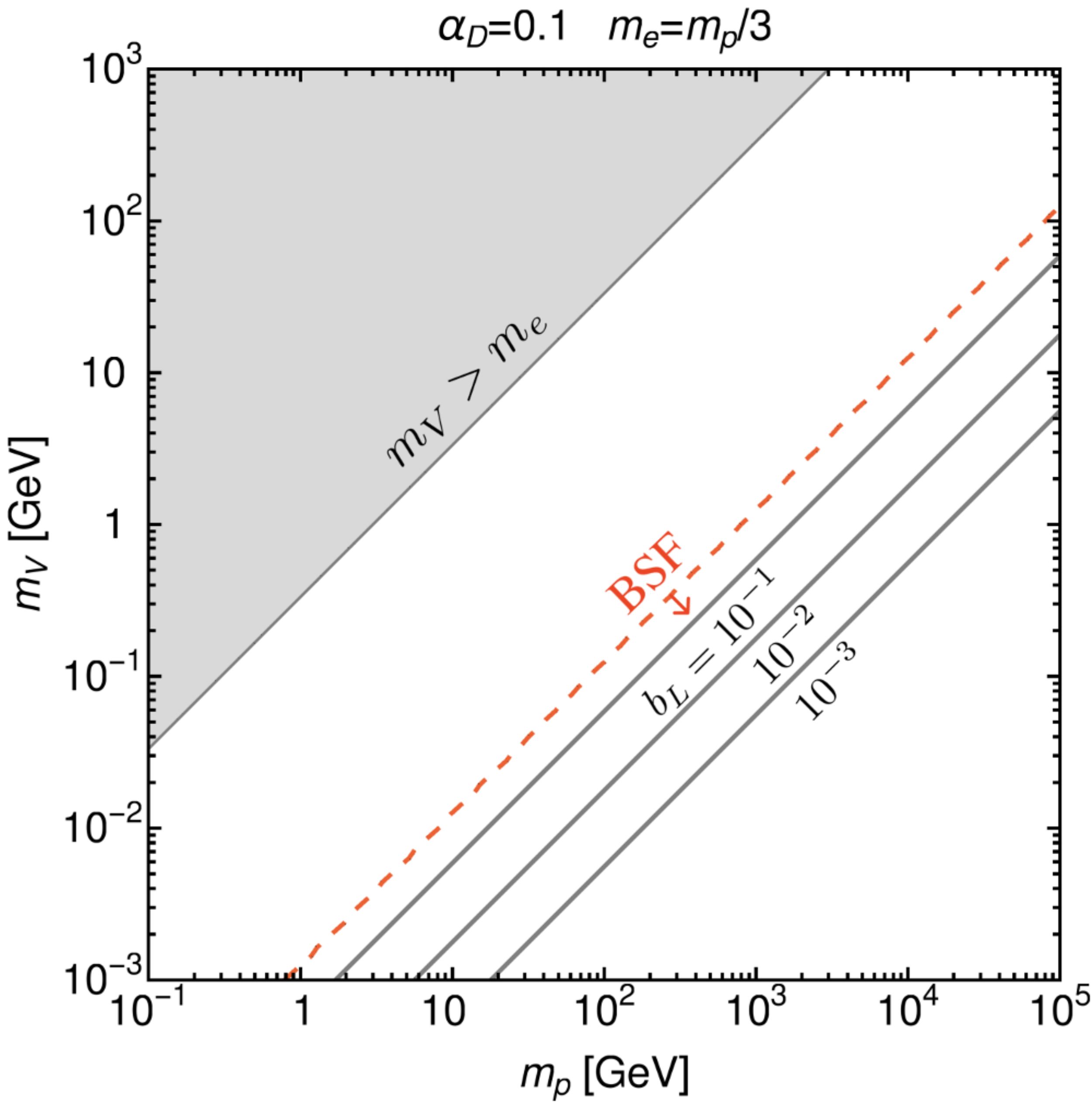}
\includegraphics[width=185pt]{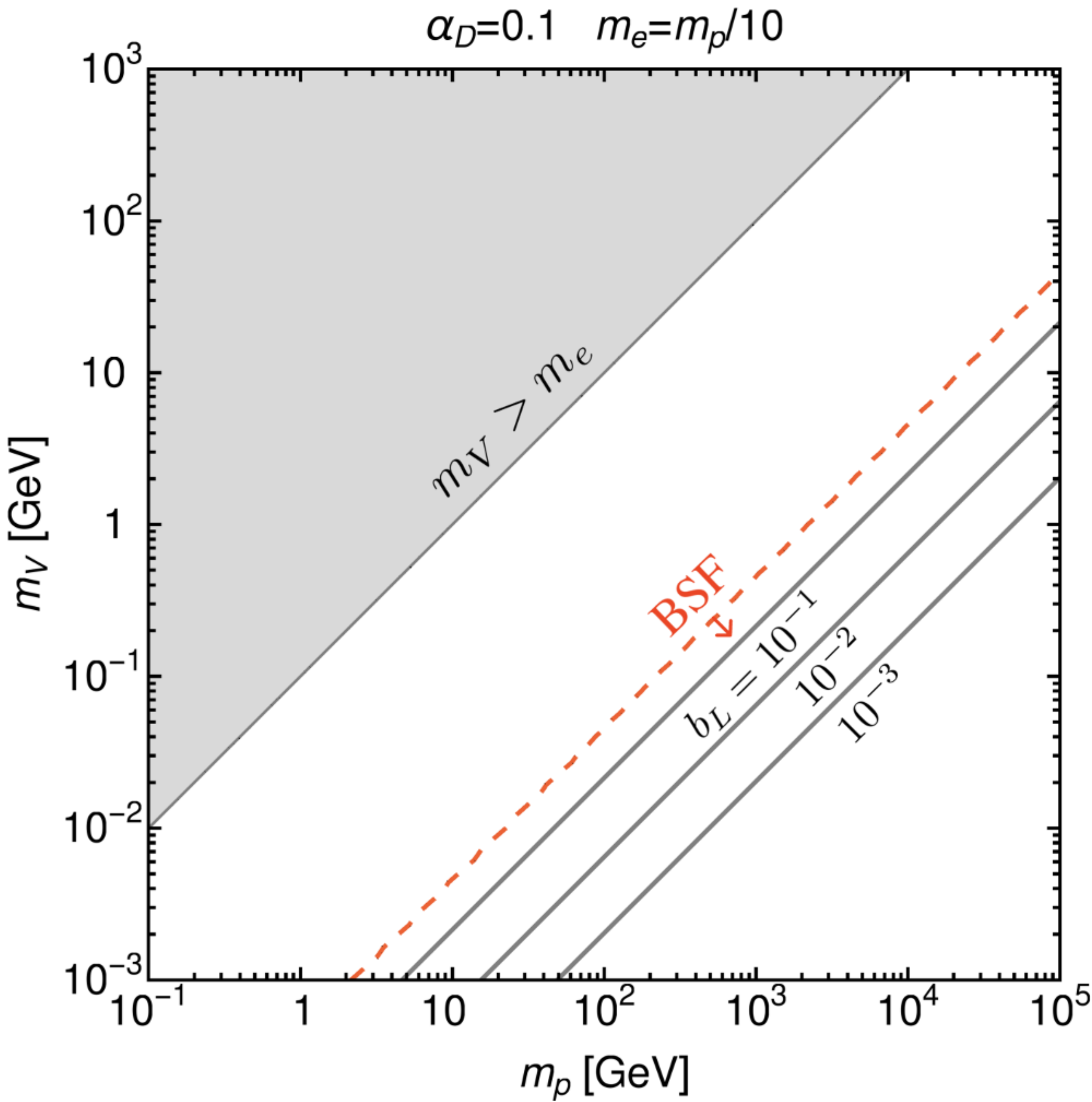}
\end{center}
\caption{\small The branching ratio of the BSF into the longitudinal polarisation state of the dark photon, $b_L$, for two sets of parameters. Bound state formation is kinematically possible below the red dashed line.}
\label{fig:longbr}
\end{figure}

\medskip

The factor $3-\sps$ in \cref{eq:sigmav_BSF} accounts for the contribution of the transverse and longitudinal dark photon polarisations to BSF. The corresponding branching fractions are
\begin{align}
b_{\mathsmaller{T}}	= \frac{2}{3 - \sps} ,
\qquad
b_{\mathsmaller{L}}	= \frac{1 - \sps}{3 - \sps} .
\label{eq:Polarisations_BranchingRatios}
\end{align}
At threshold, the dark photon is produced at rest and the ratio of $b_{\mathsmaller{T}} \colon b_{\mathsmaller{L}}$ reads $\frac{2}{3} \colon \frac{1}{3}$. In the limit of $\ED + \muD \vrel^2/2 \gg \mV$ we instead recover  $1 \colon 0$, as anticipated since the longitudinal polarisation is unphysical in the limit $\mV \to 0$.
The significance of this is that the angular distribution of dark photon decay products and the final-state ordinary photons depend on the dark-photon polarisation. Hence, once we boost from the dark photon rest frame into the observer frame, the energy spectrum of the final-state photons will also depend on the polarisation. 
The branching ratios \eqref{eq:Polarisations_BranchingRatios} are illustrated in \cref{fig:longbr} for some favourable choices of parameters for returning a significant indirect detection signal.

\subsection{Residual ionisation}

\begin{figure}[t]
\begin{center}
\includegraphics[width=190pt]{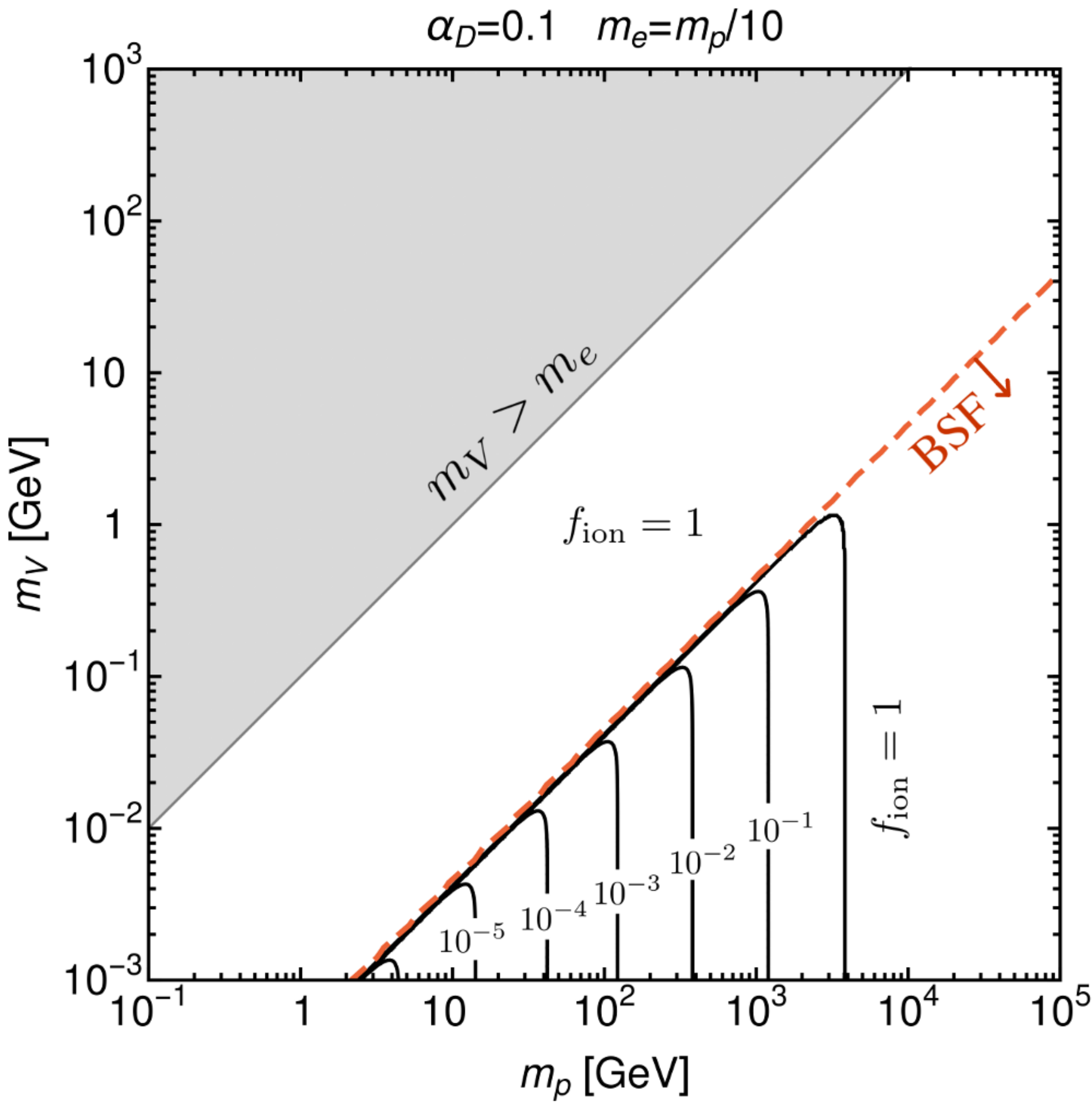}
\end{center}
\caption{\small Example of the ionization fraction. Bound state formation is possible below the red dashed contour. \bigskip}
\label{fig:fion}
\end{figure}

The formation of dark atoms may also occur in the early Universe, thereby reducing the density of dark ions today. To compute the BSF rate inside galaxies, we must therefore consider the ionized fraction of DM. This is defined as
\begin{equation}
\fion \equiv \frac{ \np }{\nH + \np},
\end{equation}
where $\np$ is the number density of unbound dark protons and $\nH$ is the number density of dark hydrogen. After dark recombination in the early Universe, the residual ionized fraction can be estimated  under the assumption of Saha equilibrium and freeze-out as 
\begin{equation}
\fion \approx \mathrm{min}
\left[1, 
10^{-10} \  \frac{\tau_{\rec}}{\aD^{4}} 
\left(\frac{\mH\muD}{\mathrm{GeV}^{2}} \right) 
\frac{2}{\sqrt{\sps} (3-\sps)} 
\right],
\label{eq:ionfrac}
\end{equation}	
where we have included the phase space factor of the BSF cross section \eqref{eq:sigmav_BSF}. The factor $\tau_{\rec}=\mathrm{min}[1,\TD/\TSM]_{\rec}$ takes into account the potentially different temperatures of the dark sector and the SM plasma during dark recombination ({\it cf.}~\cref{sec:TD}), which occurs at $\TD \sim 0.007\ED$~\cite{CyrRacine:2012fz}. 
A more detailed computation of the dark recombination that takes into account multi-level transitions has been performed in Ref.~\cite{CyrRacine:2012fz}, according to which the approximation of \cref{eq:ionfrac} is satisfactory in the regimes where $\fion = 1$ and $\fion \ll 1$. Moreover, due to the sensitive dependence of \cref{eq:ionfrac} on $\aD$, the intermediate region occupies only a small area of $\aD$ parameter space. 
It is also possible that dark atom-atom or atom-ion collisions inside halos partially reionise DM. If this is significant ({\it cf.}~\cref{foot:reion}), then the use of \cref{eq:ionfrac} underestimates the indirect signals due to BSF, thus leading to conservative constraints. 
Considering the above, and given also the various other sources of error --- notably the $J$-factors --- that will enter our analysis below, we shall proceed using \cref{eq:ionfrac} throughout. An example of $\fion$ is shown in \cref{fig:fion}.

\subsection{Interactions between the species inside halos}

If $\p-\e$ interactions are sufficiently strong, then the dark electrons may receive a kick and escape the DM halo. This would in turn suppress the expected BSF rate and indirect signals. One may wonder if the build-up of a net charge in an area of the halo is consistent with the long-range Yukawa interaction. However, for the mediator masses considered here, $\mV>$~MeV, the range of the interaction is at most of the order of picometer. Hence ejected electrons will not be drawn back into the region by the dark gauge force.

\begin{figure}[t]
\begin{center}
\includegraphics[width=185pt]{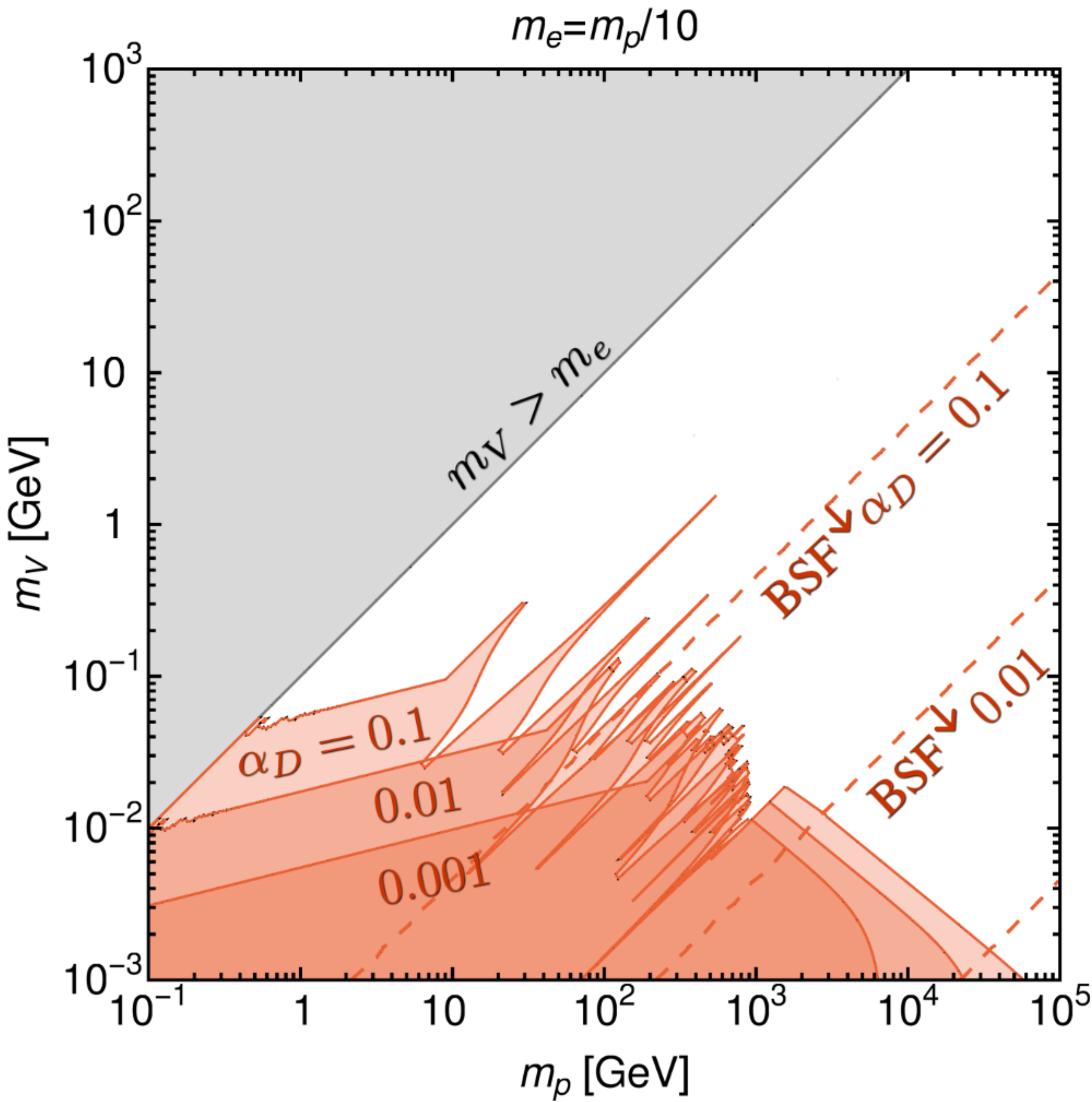}
\includegraphics[width=185pt]{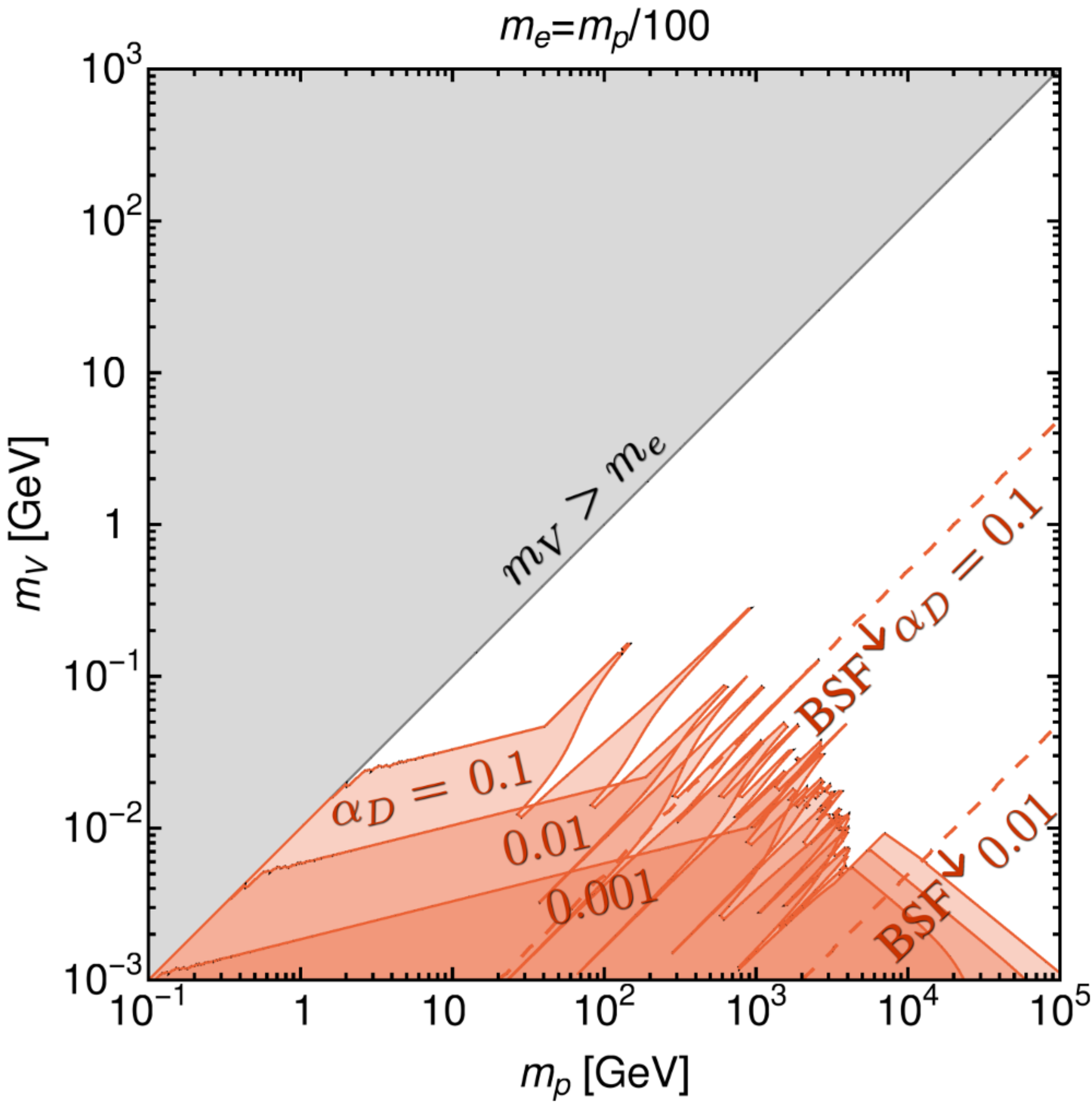}
\end{center}
\caption{\small Parameter space in which dark electrons thermalise with the dark protons, for two examples of dark electron masses. We take $\rho_{\DM} \approx 10 \; \mathrm{GeV}/\mathrm{cm}^3$ and $\vrel \approx 20 \; \mathrm{km}/\mathrm{s}$.  Efficient thermalisation requires $\mV \lesssim \mathcal{O}(0.01)$ GeV, which is anyway highly constrained from other measurements, {\it cf.} \cref{fig:darkphoton}.  \bigskip}
\label{fig:edthermal}
\end{figure}

The interaction between the DM species can be found from the formulas given in~\cite{Tulin:2013teo}, which take into account the possible long-range interaction due to the light mediator, by making the replacement $m_{X}/2 \to \muD$. Using these cross sections, we estimate the $\p-\e$ scattering rate in the areas of the halo of interest. For fully ionised DM, the scattering rate of a dark electron on dark protons is given by
\begin{align}
\label{eq:SIDM}
\Gamma_{\rm scat} 
&= (\sigma_{\rm elast} \vrel) \, n_{\p} 
=  (\sigma_{\rm elast} \vrel) \frac{\rho_{\DM} }{\mpd+\me}.
\end{align}
We take the typical DM density of a dSph at the region of interest to be $\rho_{\DM} \approx 10 \; \mathrm{GeV}/\mathrm{cm}^3$~\cite{Martinez:2013els},\footnote{For an NFW profile $\sim$ 90\% of the annihilation comes from $r < r_s$, \emph{i.e.}~from within the scale radius. The typical dSph has $r\sim 0.2$ kpc, giving a density of about $2\times10^8  \; \mathrm{M_\odot}/\mathrm{kpc}^3 \sim 8 \; \mathrm{GeV}/\mathrm{cm}^3$.}
and the typical relative velocity $\vrel \sim 20 \; \mathrm{km}/\mathrm{s}$~\cite{Burkert:2015vla}.
Assuming a lifetime of 10 billion years, in \cref{fig:edthermal} we show parameter regions where the electrons undergo on average one or more scatterings and could therefore thermalise. We see that for $\mV \gtrsim \mathcal{O}(0.01)$ GeV, the thermalisation is inefficient and we expect that the density of dark electrons in the halo is essentially the same as that of the dark protons. This rough bound on $\mV$ is not affected much by considering a larger $\rho_{\DM}$ in \cref{eq:SIDM} because the elastic cross section drops very rapidly with increasing $\mV$.

Note the estimate can be refined by including $\fion$ in \cref{eq:SIDM}, which would reduce $\Gamma_{\rm scat}$, although we should then also take into account the $\H-\e$ scatterings, whose cross section is however more suppressed due to screening. Considering our later results in \cref{sec:constraints}, and the pre-existing constraints on dark photons summarised in \cref{sec:darkphotonconstraints} --- which allow mostly for $\mV \gtrsim \mathcal{O}(0.1)$~GeV --- it becomes clear that inclusion of such effects will not change the parameter space of interest in the present study. We therefore do not include such complications here.\footnote{\label{foot:reion}  This estimation suggests also that in the same parameter space, the DM ionisation fraction can be estimated using the primordial value \eqref{eq:ionfrac}. Due to screening, atom-atom and atom-ion collisions are characterised in general by lower cross sections than ion-ion collisions. It follows that even if most of DM is predicted to be in the form of atoms after dark recombination in the early Universe, collisions in the dSph galaxies cannot reionise DM significantly. Note however that reionisation via atom-atom or atom-ion collisions may be efficient in the $\mV \sim$~few~MeV region and/or in different environments, such as the Milky Way. This could be relevant for explaining the 511~keV line~\cite{Pearce:2015zca}.}

\subsection{Dark photon decay}

\begin{figure}[t]
\begin{center}
\includegraphics[width=185pt]{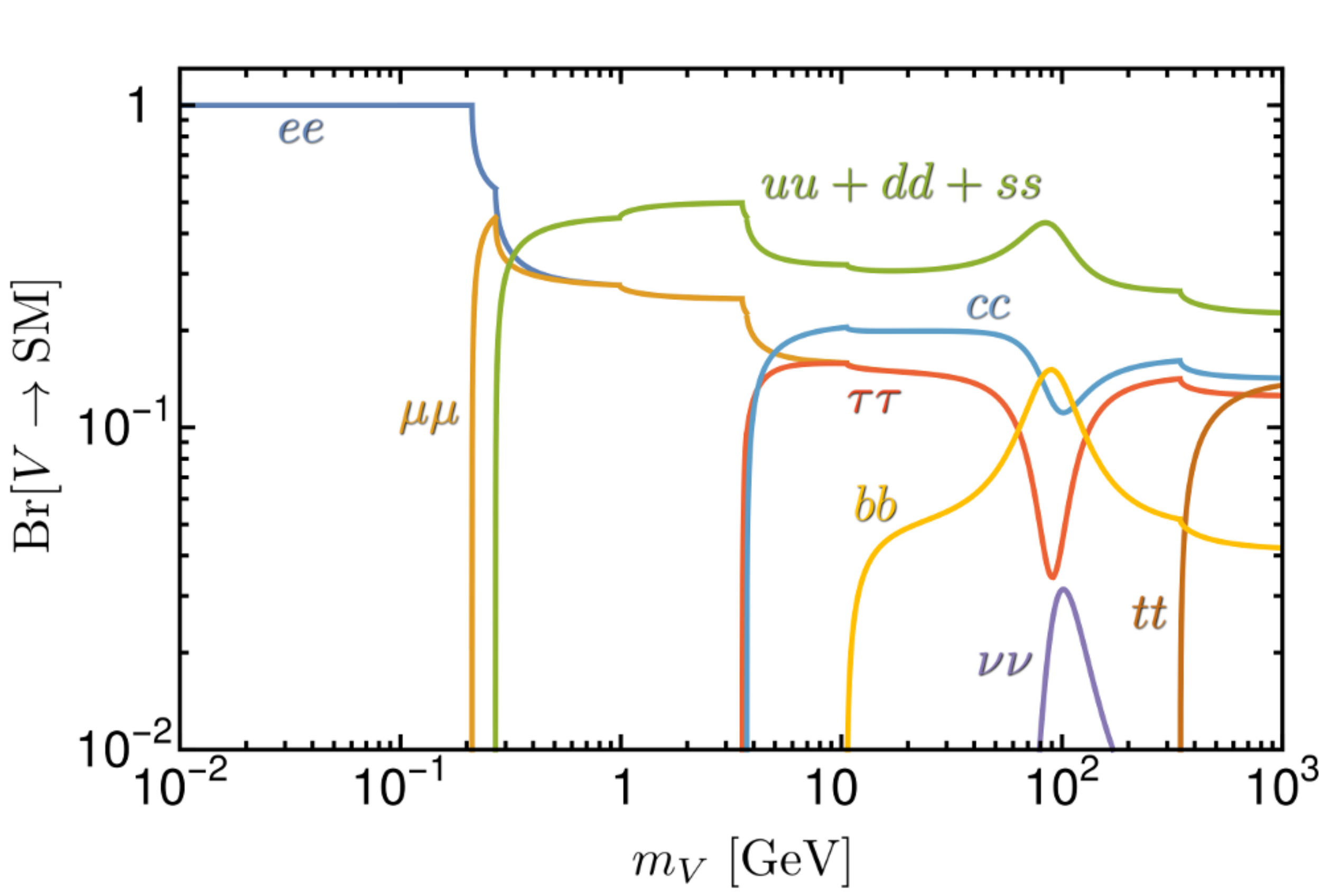}
\includegraphics[width=185pt]{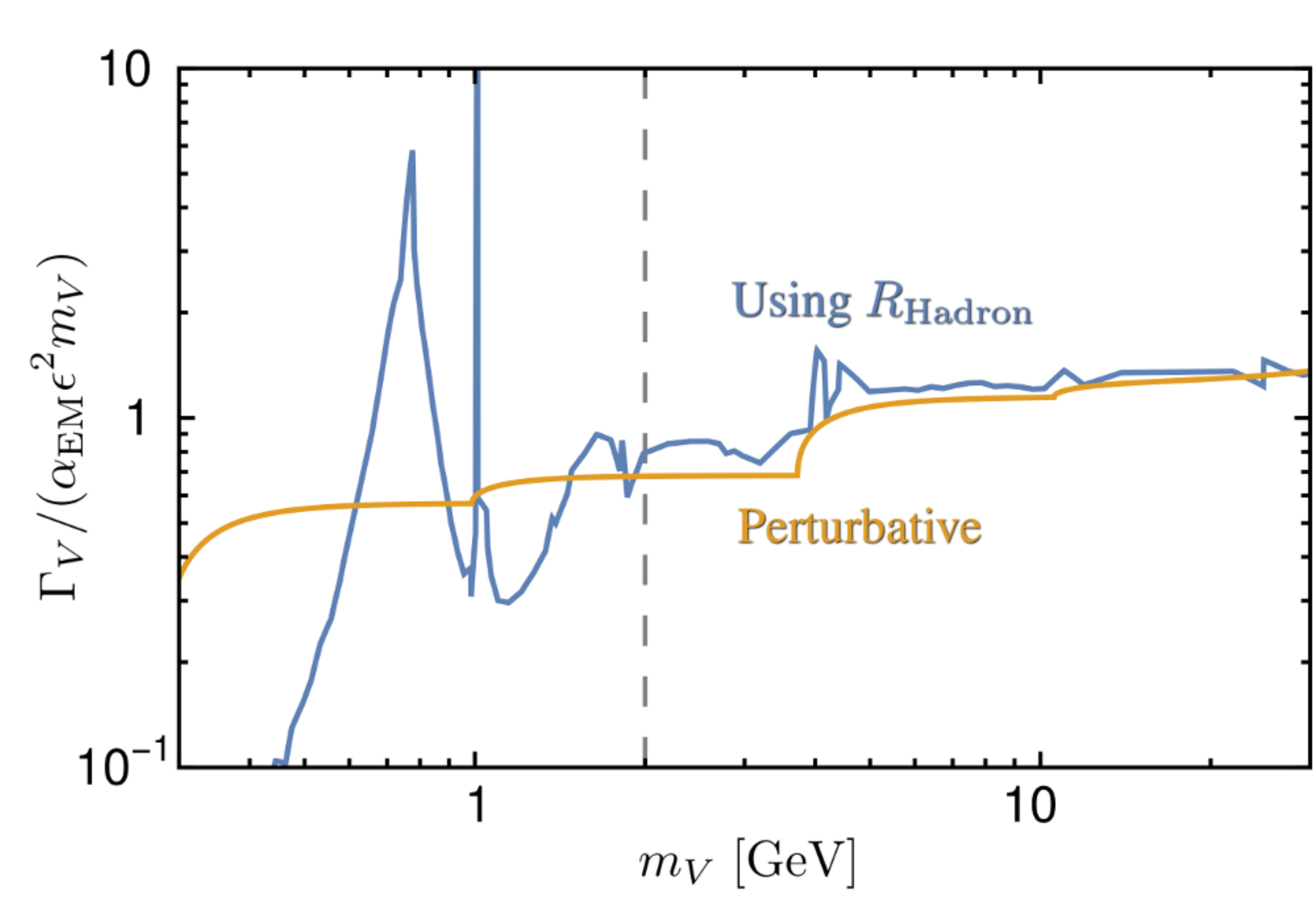}
\end{center}
\caption{\small Left: dark photon branching fraction into various primary final states. A simple perturbative calculation is used with phase space factors for the quarks set to the lightest respective meson mass. The neutrinos are summed over all three flavours. Right: comparison of the experimentally determined hadronic width and the tree-level perturbative result. Above $\mV = 2$~GeV the agreement is suitably close for our purposes. \bigskip}
\label{fig:DPBR}
\end{figure}

To obtain the expected signal we need the dark photon branching fractions.
A standard perturbative, tree-level, calculation allows one to find the partial widths into the individual decay channels using the couplings of the dark photon to the SM fermions
\begin{equation}
\mathcal{L} \supset c_{L}V_{\mu} \overline{f_{L}}\gamma^{\mu}f_{L}+c_{R}V_{\mu} \overline{f_{R}}\gamma^{\mu}f_{R}.
\end{equation}
The couplings to the chiral components of the fields are given by~\cite{Curtin:2014cca}
\begin{align}
c_{L(R)} =  \frac{g}{c_{w}} 
\left(-s_{\alpha} [ c_{w}^{2}T_{3f} - s_{w}^{2}Y_f ] + \eta c_{\alpha} s_{w} Y_f \right), 
\label{eq:Vcoupling1}
\end{align}
where $T_{3f}$ ($Y_f$) is the eigenvalue of the weak isospin (weak hypercharge) of the chiral field $f_{L(R)}$ (with normalisation such that the electric charge is $Q_{f} = T_{3f} + Y_f$), $g$ is the $SU(2)_{L}$ gauge coupling, and we make use of the definition
\begin{equation}
\label{eq:eta}
\eta \equiv \frac{\epsilon}{c_{w} \sqrt{1- \epsilon^{2} / c_{w}^{2}}}.
\end{equation}
Here $c_{\alpha} \equiv \cos{\alpha}$, $s_{\alpha} \equiv \sin{\alpha}$ and $\alpha$ is a mixing angle which brings the massive neutral gauge bosons into diagonal form. Its full expression can be found in~\cite{Curtin:2014cca}, but in the limit of small mixing it is well approximated by
\begin{equation}
\label{eq:appalpha}
\alpha \simeq - \frac{ \epsilon t_{w}  M_{Z}^{2} }{ M_{Z}^{2} - \mV^{2} },
\end{equation}
where $t_{w} \equiv \tan{\theta_w}$. Having the coupling \eqref{eq:Vcoupling1} it is straight forward to calculate the decay rate into fermions,
\begin{align}
\Gamma(V \to f \bar{f}) 
= &\frac{1}{24 \pi \mV} \sqrt{1-\frac{4m_f^2}{\mV^2}} 
\left[ (c_L^2+c_R^2) \mV^2 - (c_L^2+c_R^2-6c_Lc_R) m_f^2 \right]\,,
\end{align} 
and hence find the branching ratios of the dark photon (\cref{fig:DPBR}, left panel). Note we have replaced the quark masses in the charm and beauty phase space suppression factors with the lightest respective meson masses.

One complication, however, is the existence of hadronic resonances between the pion threshold and $\sim~5$~GeV. To gain some insight of the errors introduced, we extract the total hadronic width using the experimentally determined $R_{\rm Hadron}$ factor, which is a measurement of the off-shell photon branching into muon pairs compared to hadrons~\cite{Tanabashi:2018oca}. Note that, unlike for the CMB constraints in~\cite{Cirelli:2016rnw,Baldes:2017gzu}, we cannot not use $R_{\rm Hadron}$ directly, as we require the detailed final state photon spectrum for our calculation of the BSF limits. A comparison of the experimentally determined hadronic width to the perturbative calculation is shown in the right panel of \cref{fig:DPBR}. Given other uncertainties, such as the $J$-factors, which will enter into our limits, we deem the error introduced is acceptable for $\mV \gtrsim 2$~GeV.\footnote{As our analysis was well underway, a more careful treatment of the hadronic resonances was completed with implementation in \texttt{Herwig}~\cite{Plehn:2019jeo}, specifically for the case of light dark photons. As we are using \texttt{Pythia} to find the final state photon spectrum, we leave the incorporation of these details relevant for $\mV \lesssim 2$~GeV for future work (see also~\cite{Coogan:2019qpu}).}

\subsection{Visible photons from the decay}

After having found the dark photon couplings and decay rate into SM final states we now need to find the resulting $\gamma$-ray spectrum. This is done in two steps. Firstly, for a given polarisation of $\V$, we determine the angular distribution of decay products in the vector rest frame. We then outline how to boost this spectrum into the observer frame, where now the angular distributions of the decay products in the $\V$ rest frame converts to an energy distribution for those same final states. We now address these two issues in turn.

\begin{figure}[t]
\begin{center}
\includegraphics[width=185pt]{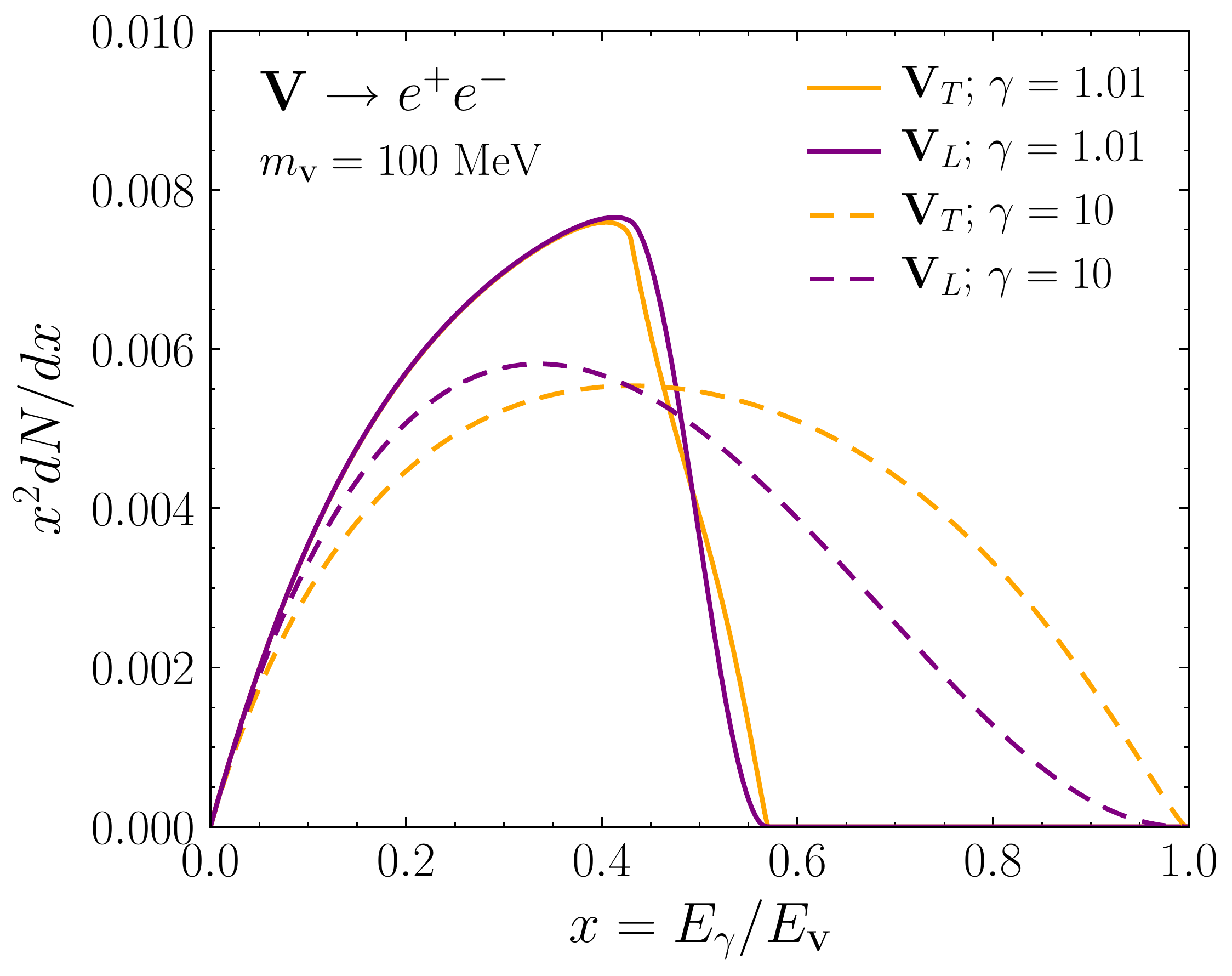}
\includegraphics[width=185pt]{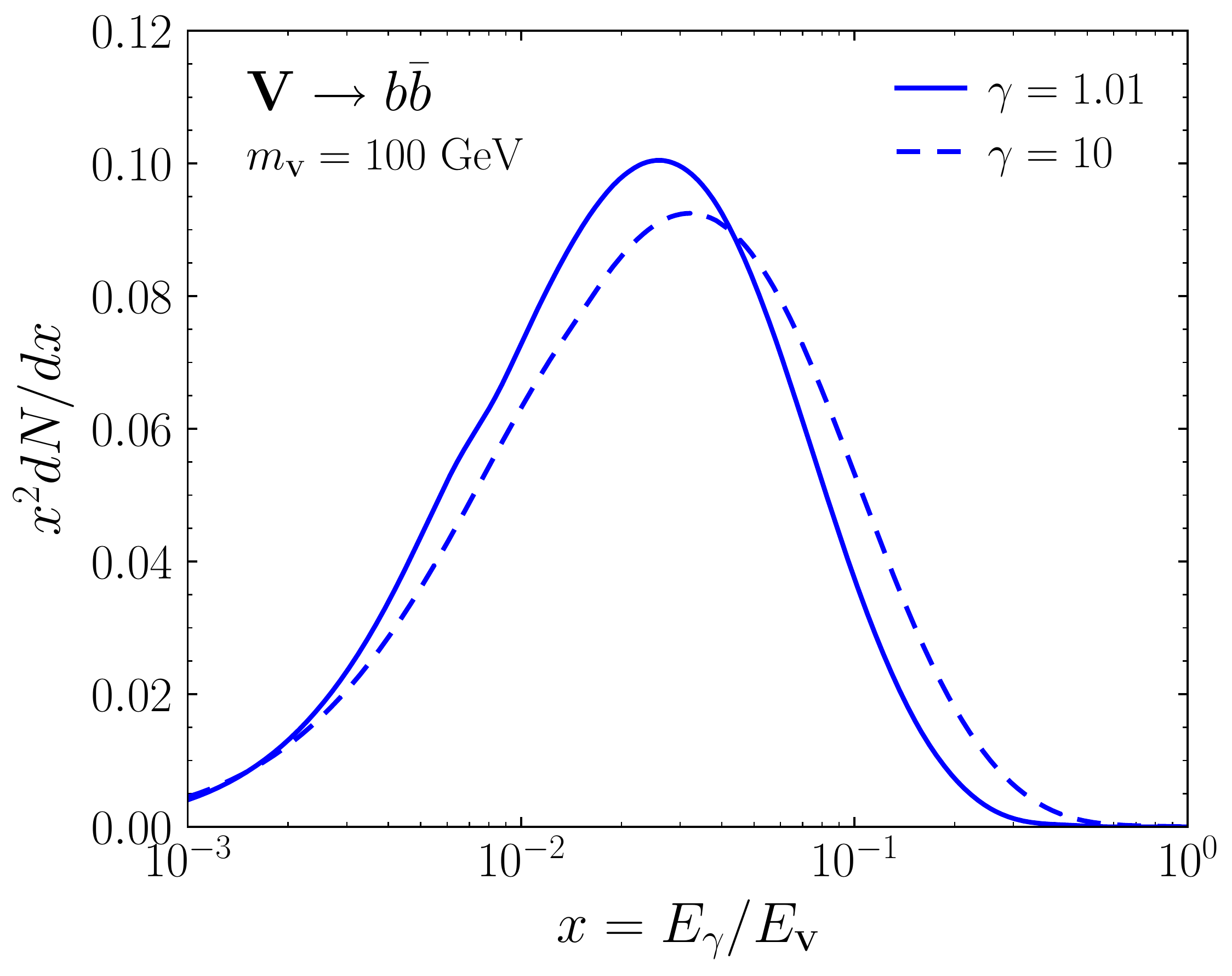}
\end{center}
\caption{\small Example observer frame photon spectra for the case of a vector decaying to $e^+ e^-$ (left) and $b \bar{b}$ (right). For the electron final state, we take $\mV = 100$ MeV, and show the spectrum for transverse and longitudinally polarised $\V$, which in this case can have a significant impact on the spectrum. For the coloured final state, we take $\mV = 100$~GeV, and now do not distinguish between polarisations (as described in the text there is not an appreciable difference between these for hadronic final states). In both cases we show results for two dark photon boosts, $\gamma = \EV/\mV$. Note that for $\gamma = 1$, $x = E_{\gamma}/\EV \leq 0.5$, and therefore in the left plot a clear transition to that regime is observed for a small boost. \bigskip}
\label{fig:photonspectrum}
\end{figure}

Example outputs from the procedure described below is shown in Fig.~\ref{fig:photonspectrum}. There, we depict observer frame photon spectra for decays into electrons and $b$-quarks, for several parameter choices. Note the full spectrum is determined by weighting all the relevant final states by the branching fractions given in Fig.~\ref{fig:DPBR}.

\subsubsection{Angular distribution of $\V$ decays}

Consider the angular dependence of decays of $\V \to f \bar{f}$ in the $\V$ rest frame. We define our coordinates such that $\hat{z}$ represents the axis along which the vector is boosted in the observer frame. We are then interested in determining the distribution of decay products with respect to this axis, and accordingly define $\theta \in [0,\pi]$ to be the angle between the fermion and the boost axis in the $\hat{x}-\hat{z}$ plane. Taking the two circular transverse polarisations to have the explicit form $\epsilon^{\mu}_{\pm} = (0,1,\pm i,0)$, we can determine the angular dependence as 
\begin{align}
& p_{\pm}(\cos{\theta}) \equiv \frac{1}{\Gamma} \frac{d\Gamma}{d\cos \theta} (\V_\pm \to f \bar{f}) \nonumber \\ 
& \qquad = \frac{3}{8} \frac{(c_L^2+c_R^2) (2-\beta^2 \sin^2\theta) \mp 2 \beta (c_R^2-c_L^2) \cos \theta - 4 (c_L-c_R)^2 (m_f/\mV)^2}{(c_L^2+c_R^2) - (c_L^2+c_R^2-6c_Lc_R) (m_f/\mV)^2}\,, \label{eq:ppm}
\end{align}
where we have defined the fermion boost 
\begin{equation}
\beta = \sqrt{1-\frac{4m_f^2}{\mV^2}}\,.
\end{equation}
For the longitudinal polarisation, $\epsilon^{\mu}_0 = (0,0,0,1)$ in the rest frame, the equivalent expression is given by
\begin{align}
& p_{0}(\cos{\theta}) \equiv \frac{1}{\Gamma} \frac{d\Gamma}{d\cos \theta} (\V_0 \to f \bar{f}) \nonumber
\\&\qquad= \frac{3}{4} \frac{(c_L^2+c_R^2) \left( 1 - \beta^2 \cos^2 \theta \right) - 2 (c_L-c_R)^2 (m_f/\mV)^2}{(c_L^2+c_R^2) - (c_L^2+c_R^2-6c_Lc_R) (m_f/\mV)^2}\,.
\end{align}
In detail it is clear that the angular distribution of the fermions varies between the polarisations. When we boost to the observer frame, discussed next, this will translate into different energy distributions.

\subsubsection{Boost of the photon spectrum}

From the above, we can determine the fermion energies in the observer frame for each of the vector polarisations. However this is not the experimental quantity of interest. Instead, we aim to determine the distribution of {\it photons} that result from the initial hard decay $\V \to f \bar{f}$. These two will coincide in the limit that the photons are produced collinearly with the fermions. Given the collinear enhancement of photon emission off a charged fermion, for certain final states this is a good approximation. For the moment let us simply assume this is true and determine the modification to the spectrum, returning to the question of when this should apply next.

We define the spectrum of photon energies, $E_0$, in the $\V$ rest frame as
\begin{equation}
\frac{dN}{dE_0}(E_0)\,.
\label{eq:restframespec}
\end{equation}
Assuming the photon is collinear with the fermions, then in the observer frame where the vector has an energy $\EV \simeq \ED$ (as the initial kinetic energy is negligible) the photon energy, $E$, is now
\begin{equation}
E  = E_0 \frac{\EV}{\mV} 
\left( 1 + \cos{\theta} \sqrt{1 - \frac{\mV^2}{\EV^2}} \right)\,.
\label{eq:boostedphotonenergy}
\end{equation}
Importantly, we see that this energy is determined not only by the distribution of rest frame energies in~\cref{eq:restframespec}, but also by the angle with respect to the rest frame, which is drawn from a distribution that depends on the polarisation of $\V$, as determined above. In detail, and as determined in \cref{sec:boostdetails}, the spectrum in the boosted frame depends on the angular distribution $p(\cos \theta)$, and takes the form
\begin{align}
\frac{dN}{dx}
= &\frac{2}{\sqrt{1 - \epsilon_B}} \int_{x_0^{\rm min}}^{x_0^{\rm max}} \frac{dx_0}{x_0}\, p\left(\frac{2 x/x_0 - 1}{\sqrt{1 - \epsilon_B}} \right)\, \frac{dN}{dx_0}(x_0)\,,
\label{eq:boostedspec}
\end{align}
where the terminals of integration are
\begin{equation}
x_0^{\rm min} =  \frac{2x}{\epsilon_B} (1-\sqrt{1 - \epsilon_B})\,, \qquad
x_0^{\rm max} =  {\rm min} \left[ 1,\, \frac{2x}{\epsilon_B} (1+\sqrt{1 - \epsilon_B}) \right]\,.
\end{equation}
These expressions are written in terms of dimensionless quantities, in particular a boost parameter $\epsilon_{B} = (\mV/\EV)^2$, and energy fractions $x_0 = 2 E_0/\mV$ and $x = E/\EV$. Note the absence of a factor of $2$ in $x$ arises, as after boosting in principle the photon can carry the full energy of the vector, whereas in the rest frame $E_0 \leq \mV/2$.

\subsubsection{Photon spectra in the $\V$ rest frame}

Equation~\eqref{eq:boostedspec} provides the photon spectrum in the observer frame, assuming the photons in the rest frame are collinear with the initial fermions. In this case, it is clear that the vector polarisation enters centrally through $p(\cos \theta)$ (note that $p[\cos \theta] = 1/2$ corresponds to the unpolarised decays). Further, note that this result does not assume $\EV \gg \mV \gg m_f$, as in parts of the parameter space that will not be true.

To determine the full spectra for a given set of model parameters, we will need to use this result for the appropriate set of fermions weighted by the branching fractions given in \cref{fig:DPBR}. Working below $\mV = 100$ GeV, we can neglect decay to $t \bar{t}$, however we will still need to consider six final states: $e$, $\mu$, $\tau$, $q=(u+d+s)/3$, $c$, and $b$. In practice we will approximate $q \approx d$, as the spectra for each of the light quarks is similar. We determine the rest frame spectra for $V \to ee\gamma$ and $V \to \mu \mu\gamma$ analytically, without assuming $\mV \gg m_{l}$ (see ~\cref{sec:analyticfsr}). For muons there is also a contribution from the radiative decay $\mu \to e \bar{\nu}_e \nu_{\mu} \gamma$ which is also included, following~\cite{Mardon:2009rc}. For the hadronic final state, including the $\tau$ and quarks, we use \texttt{Pythia} to generate the spectra.\footnote{To generate spectra below 10 GeV in \texttt{Pythia} we use the procedure in which the two beams are set separately as in~\cite{Leane:2018kjk}.}

With the rest frame spectra in hand, we can now revisit the question of how good an assumption it is to treat the photons as collinear with the initial fermions, as assumed in the derivation of \cref{eq:boostedspec}. In particular, all of the final states above (except for the radiative decay of the muon) can be simulated in \texttt{Pythia}, and then boosted for each final state photon to determine the observer frame distribution. In order to simulate the distribution of initial fermion angles according to the various vector polarisations, we weight the events according to the distributions $p_{\pm,0}(\cos \theta)$ determined above.

The results of this procedure are then compared against the output of \cref{eq:boostedspec}. We find very good agreement for leptonic final states, $e$, $\mu$, and $\tau$, which is unsurprising as we find the photons in this case to be predominantly collinear with the leptons. For hadronic final states, the correlation is less defined, and accordingly the collinear approximation breaks down. Nevertheless, we find that the distribution in this case is well approximated by the assumption of an unpolarised decay, $p(\cos \theta)=1/2$ or equivalently treating the vector as a scalar.

\section{Constraints from Fermi-LAT $\gamma$-ray data}
\label{sec:constraints}

We now seek to derive observational constraints on BSF. We first briefly consider how existing constraints on DM annihilation into SM particles can be recast to apply to processes that occur with emission of low-energy radiation. We then employ Fermi-LAT data to derive new constraints on level transitions occurring via emission of dark photons.

\subsection{Recasting constraints on DM annihilation for BSF and level transitions}

Existing constraints on DM annihilation assume that the emitted radiation has energy $E \approx \mDM$. Let $\langle\sigma_\ann \vrel \rangle_{\max}^{xx} @ M$ be the maximum observationally allowed cross section for annihilation of DM with mass $M$ into the channel $\bar{X}+ X \to xx$, where $X,\bar{X}$ denote the DM particles and $xx$ the products of the DM annihilation. 
If $XX$, $\bar{X}\bar{X}$ or $X\bar{X}$ bound states form 
via emission of an $x$ particle of energy ${\cal E}$, then the corresponding constraint is found via the rescaling~\cite{Mahbubani:2019pij,Mahbubani:2020knq}
\begin{equation}
\langle\sigma_\BSF \vrel\rangle_{\max}^{x} =
\left[\langle \sigma_\ann \vrel \rangle_{\max}^{xx}~@~\{M = {\cal E}\} \right]
\times 2 \left(\frac{\mDM}{\cal E} \right)^2 \,,
\label{eq:Rescaling}
\end{equation}
where $\mDM$ is the DM mass of interest. The factor $(\mDM/{\cal E})^2$ accounts for the different number densities of DM with mass $\mDM$ and $M={\cal E}$. The constraint on BSF is relaxed further by a factor 2 since only one $x$ is emitted during BSF (in contrast to $xx$ emitted in annihilation).  
\Cref{eq:Rescaling} applies also to exothermic level transitions that follow collisional excitations of DM bound states. In this case, $\sigma_\BSF$ should be replaced by the cross section of the scattering process that causes the excitation, while ${\cal E}$ corresponds to the energy dissipated in the de-excitation. 
Note that in the case of multicomponent DM, \cref{eq:Rescaling} may have to be adjusted to account for the potentially different densities of the DM components participating in the BSF or collisional excitation processes.

\begin{figure}[t]
\centering
{}\hfill
\includegraphics[width=0.45\textwidth]{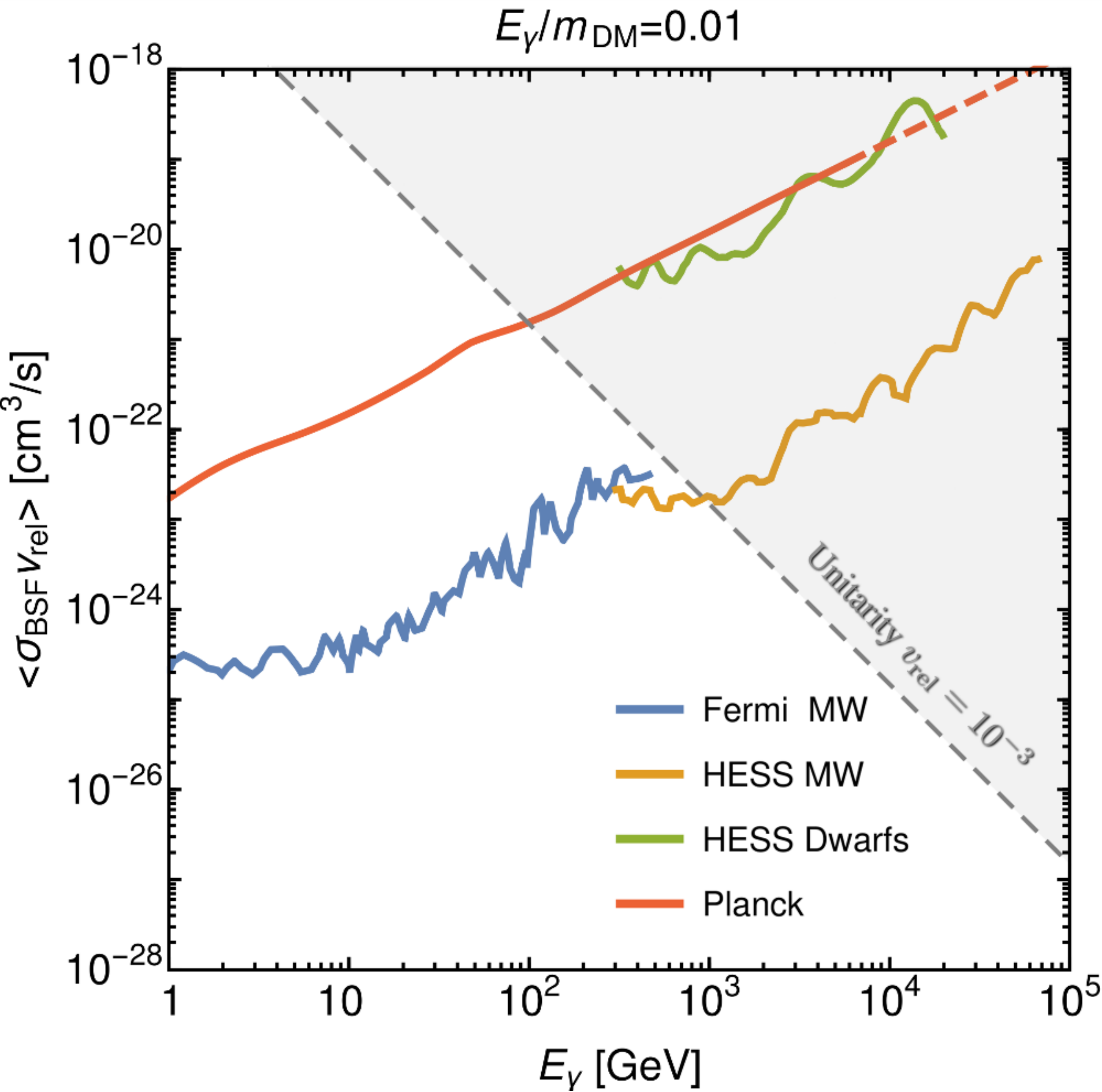} 
\hfill
\includegraphics[width=0.45\textwidth]{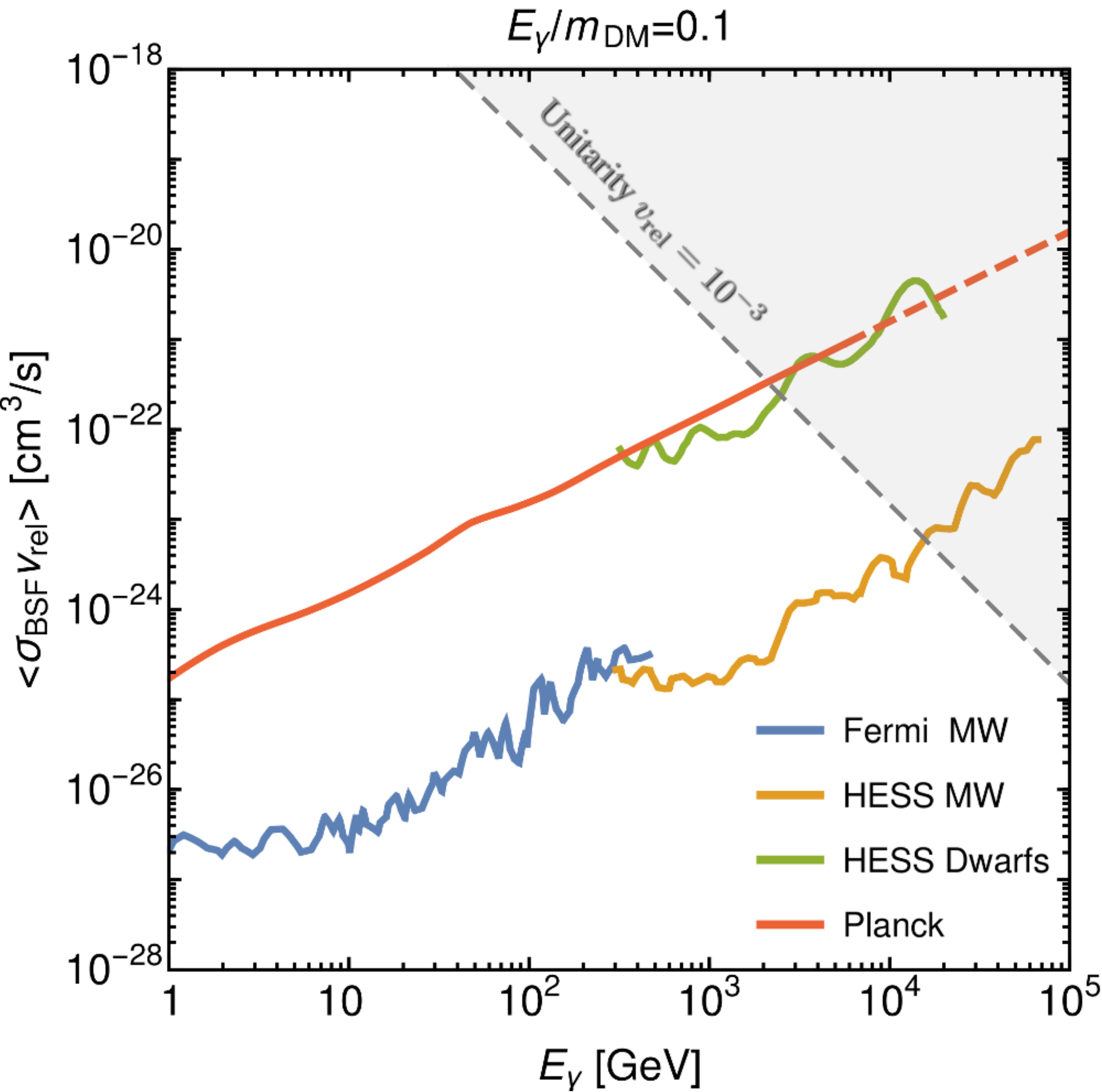}
\hfill
\caption{Recasting of indirect detection limits from Planck~\cite{Aghanim:2018eyx}, H.E.S.S.~\cite{Abdalla:2018mve,Abdallah:2018qtu} and Fermi~\cite{Ackermann:2015lka} on DM annihilation into two photons, for bound state formation via photon emission, for two examples of $E_\gamma / \mDM$ ratio. Also shown is the $s$-wave unitarity constraint for DM annihilation, $\langle \sigma \vrel \rangle < 4\pi/(\mDM^{2}\vrel)$~\cite{Griest:1989wd}, with typical relative velocity for DM in the Milky Way. The dashed line for the Planck constraint indicates where we have extrapolated the efficiency of energy deposition, $f_{\rm eff}$, beyond the tables~\cite{Slatyer:2015jla} used by Planck.}
\label{fig:Constraints_LE_fixedEnergyRatio} 
\end{figure}

An example recasting for $xx =\gamma \gamma$ is shown in \cref{fig:Constraints_LE_fixedEnergyRatio}. The observational constraints come from the Planck~\cite{Aghanim:2018eyx}, H.E.S.S.~\cite{Abdalla:2018mve,Abdallah:2018qtu}, and Fermi~\cite{Ackermann:2015lka} collaborations. It is simple to repeat this exercise for different channels. 
As seen from \cref{fig:Constraints_LE_fixedEnergyRatio}, the constraints weaken for lower ${\cal E} / \mDM$, due to the number density factor.

In many models, however, such as the one considered in \cref{sec:darkQED}, the annihilation channel is an exotic one involving non-SM mediators that subsequently decay into SM particles. Although indirect searches have been applied to DM annihilation into exotic channels (see {\it e.g.}~\cite{Cirelli:2016rnw,Baldes:2017gzu,Bringmann:2016din,Kahlhoefer:2017umn}), the resulting bounds are typically given in terms of a number of fundamental model parameters and are difficult to recast for the purposes of level transitions and BSF. Constraining such processes necessitates reanalysing the observational data and casting the results in terms of the energy dissipated in the transitions. In the following, we carry out such an analysis for transitions occurring via emission of dark photons decaying into SM particles.

\subsection{The BSF rate and photon flux}

We now return to the specifics of the model of \cref{sec:darkQED}.
Outside the parameter space where dark electrons may thermalise and get ejected from the halo, we can assume that the local dark proton and dark electron densities are equal, $\np = \ne$. Then, the total DM mass density is
\begin{align}
\rho_{\DM} 
&= \np\mpd+\ne\me+\nH\mH 
= \np \left( \mpd+\me+\frac{[1-\fion]}{\fion}\mH \right),
\end{align}
where $\fion$ is the ionisation fraction \eqref{eq:ionfrac}. 
The BSF rate per unit volume is
\begin{equation}
\frac{d^{2}N_{\BSF}}{dVdt} 
= n_{\p}\ne \<\sigma \vrel\>_{\BSF}  
= \frac{ \fion^{2} \rho_{\DM}^{2} \<\sigma \vrel\>_{\BSF}}
{\left( \fion\mpd+\fion\me+[1-\fion]\mH \right)^{2}},
\end{equation}
where $\<\sigma \vrel\>_{\BSF}$ is the averaged BSF cross section \eqref{eq:sigmav_BSF}. 
In the case of level transitions, this factor must be appropriately adjusted. Provided that the level transitions follow collisional excitation processes, then it remains true that $d^2N/(dV dt) \propto \rho_{\DM}^2$, which ensures that the following analysis applies with the appropriate rescaling. Here we focus on BSF and shall not elaborate on the specifics of excitation and de-excitation processes.

Next we define the differential photon flux incident on the detector as
\begin{equation}
d \mathrm{\Phi}_{\gamma} \equiv \frac{ d^{2} N_{\gamma} }{dA dt},
\end{equation}
where $dA$ is an infinitesimal surface area of the detector. For a source at proper distance $r$ only $dA/(4\pi r^{2})$ of the produced photons will reach the detector. We thus have
\begin{equation}
\frac{ d^{2} \mathrm{\Phi}_{\gamma} }{  dV dE } 
= \frac{ \fion^{2}  \<\sigma \vrel\>_{\BSF}  }
{ 4\pi \left( \fion\mpd+\fion\me+[1-\fion]\mH \right)^{2} } 
\frac{dN_{\gamma} }{ dE } 
\frac{ \rho_{\DM}^{2} }{ r^{2} },
\label{eq:diffluxPerVolume}
\end{equation}
where $dN_{\gamma}/dE$ is the visible photon spectrum resulting from BSF. Going to spherical coordinates $dV = r^{2} dr d\mathrm{\Omega}$ we find
\begin{equation}
\frac{ d \mathrm{\Phi}_{\gamma} }{  dE } 
= \frac{ \fion^{2} \<\sigma \vrel\>_{\BSF}  }
{ 4\pi \left( \fion\mpd+\fion\me+[1-\fion]\mH \right)^{2} } \frac{dN_{\gamma} }{ dE } J_0,
\label{eq:difflux}
\end{equation}
where the $J_0$-factor is given by
\begin{equation}
J_0 = \int_{0}^{\infty} dr \int_{\Sigma} d\mathrm{\Omega} \, \rho_{\DM}(r,\mathrm{\Omega})^{2},
\end{equation}
and $\Sigma$ is the observed area of the sky. Note in the limit $ \me \to \mpd$ and $\fion=1$ we recover, as required, the $1/16\pi$ prefactor for annihilation of non-self-conjugate DM. 
Going from \cref{eq:diffluxPerVolume} to \eqref{eq:difflux} assumes that either the velocity distribution of the DM particles remains the same along the line of sight, or that $(\sigma \vrel)_{\BSF}$ is velocity independent. In the present case, none of these assumptions is generally true, since $(\sigma \vrel)_{\BSF}$ is velocity dependent as discussed in \cref{sec:darkQED}, and the DM velocity distribution within the halos varies along with $\rho_{\DM}$. This implies that a more refined treatment may be necessary, as we discuss next.

\subsection{The $J$-factor velocity dependence}

Let us write the BSF cross section of \cref{eq:sigmav_BSF}, as $\sv_{\BSF} \equiv (\sigma \vrel )_{0}S(\vrel)$, where $\sv_{0}$ is velocity independent. We can rewrite the differential photon flux arising from the BSF, \cref{eq:difflux}, to take into account the velocity dependence:
\begin{equation}
\frac{d \mathrm{\Phi}_{\gamma}}{ d E} 
= \left[ \frac{ \fion^{2} \sv_{0} }{ 4\pi\left( \fion\mpd+\fion\me+[1-\fion]\mH \right)^{2} } \right] 
\frac{dN}{dE_{\gamma}} J,
\label{eq:flux}
\end{equation}
where $J$ is now the effective $J$-factor, which encodes the DM density, and in which the velocity dependence of the cross section has been absorbed. The full expression is~\cite{Boddy:2017vpe}
\begin{equation}
J = \int_{0}^{\infty} dr \int_{\Sigma} d\mathrm{\Omega} \int d^{3} v_1 \int d^{3} v_2 f_{\rm ps}(r,\mathrm{\Omega},v_1) f_{\rm ps}(r,\mathrm{\Omega},v_2) S(\vrel),
\end{equation}
where $f_{\rm ps}$ is the phase-space density of the dark protons and the dark electrons; since indirect signals are expected only from the regions where $\p$ and $\e$ do not thermalise, $f_{\rm ps}$ is independent of the ion mass and thus the same for both species.
As we have seen in a previous section $dN/dE_{\gamma}$ is a function of $\mV$ and the binding energy.\footnote{Strictly speaking the binding energy plus the initial kinetic energy, but the latter is sub-dominant and can be ignored to a good approximation, as $\aD \gg \vrel$ in the parameter space of interest.} By using appropriate $J$-factors, we hope to scan over some choices of $\mV$ and the binding energy, and use Fermi-LAT data to constrain the combination of factors in the square brackets in \cref{eq:flux}. This factor can then be written in terms of the underlying parameters of the model and hence eventually constrain the scenario.

The $J$-factors have been derived for $S(\vrel) = \vrel^{-1}, \, \vrel^{0}, \, \vrel^{2}, \vrel^{4}$ in Ref.~\cite{Boddy:2019qak}, where the DM density and velocity dispersion were determined as functions of the radial coordinate $r$ through a spherical Jeans analysis. 
Nominally these four cases are termed the Sommerfeld-enhanced (SE), $s$-wave, $p$-wave, and $d$-wave $J$-factors respectively. Due to the finite mediator mass, however, $\sv_{\BSF}$ scales as $\vrel^{-1}$ for $\vrel \gtrsim \mV/\muD $, but as $\vrel^{2}$ for  $\vrel \lesssim \mV/\muD$, as discussed in \cref{sec:darkQED}. (Note though that $\sv_{\BSF}$ is Sommerfeld enhanced even in the latter velocity range.)

To fully take this into account, we would need to re-calculate the $J$-factor for each choice of $\mV/\muD $. This introduces further technical difficulties. The photon spectra depend only on $\mV$ and $\EV \simeq \ED $, which is convenient for extracting the limits on the flux, as introducing further parameters is computationally expensive. We want to avoid doing this. Furthermore the $J$-factors carry a large uncertainty. So we proceed by estimating the error incurred by using the pre-calculated $J$-factors as a simplifying approximation.

To gain some insight into this error, we can estimate the implied averaged velocity dispersion by comparing the $J$-factors for the different cases. If the DM density could be factored out of the velocity integral, the respective $J$-factors would scale as
\begin{align}
\text{SE} & \propto \sqrt{\frac{x^3}{4\pi}}  
\int_{0}^{\infty}  \vrel 
\, \mathrm{Exp}\left[-\frac{ x \vrel^2}{4}\right] d\vrel 
= \sqrt{\frac{x}{\pi}}, 
\label{eq:Jfactor_SE} \\
s\text{-wave} & \propto \sqrt{\frac{x^3}{4\pi}} 
\int_{0}^{\infty}  \vrel^{2} 
\, \mathrm{Exp}\left[-\frac{ x \vrel^2 }{4}\right] d\vrel 
= 1, 
\label{eq:Jfactor_swave} \\
p\text{-wave} & \propto \sqrt{\frac{x^3}{4\pi}} 
\int_{0}^{\infty} \vrel^{4} 
\, \mathrm{Exp}\left[-\frac{ x \vrel^2 }{4}\right] d\vrel 
= \frac{6}{x}, 
\label{eq:Jfactor_pwave}
\end{align} 
where $x \equiv 2/v_{c}^{2}$ parametrises the velocity dispersion $v_{c}$. We can extract the implied velocity dispersion, following the above assumption, by taking a ratio of $J$-factors. For example, the central values of the $J$-factor for Draco I given in~\cite{Boddy:2019qak} are
\begin{equation}
\mathrm{log}_{10}\left(J/\mathrm{GeV}^{2}\mathrm{cm}^{-5}\right) = 22.93, \; 18.84, \; 11.15,
\end{equation}
for the SE, $s$-wave, and $p$-wave cross sections respectively. Taking the ratio of these values and comparing with the corresponding  ratios of the $J$-factors of \cref{eq:Jfactor_SE,eq:Jfactor_swave,eq:Jfactor_pwave}, we find the effective velocity dispersions
\begin{align}
v_{c} & \approx 19 \; \mathrm{km/s} \qquad \qquad [\text{SE-to-}s\text{-wave}],      \label{eq:vel1} \\
v_{c} & \approx 25 \; \mathrm{km/s} \qquad \qquad [p\text{-wave-to-}s\text{-wave}].  \label{eq:vel2}
\end{align}
We next substitute the value of $v_c$ found in \cref{eq:vel1} into \cref{eq:Jfactor_pwave} and find the $J$-factor is changed by a factor of 0.62. Similarly, a factor 0.79 difference is found by substituting the $v_c$ found in \cref{eq:vel2} into \cref{eq:Jfactor_SE}. The discrepancy in the $\sv_{\BSF}$ constraints would then be a factor of 0.79 (0.62) using the $p$-wave (SE) velocity dispersion in the SE ($p$-wave) $J$-factor. Repeating the exercise for the other dSphs given in table I of~\cite{Boddy:2019qak}, we find the largest discrepancy to be a factor of 0.57 for Hydrus I (SE velocity dispersion in the $p$-wave $J$-factor).  We thus estimate the uncertainty introduced by 
neglecting the $r$-dependence of the velocity dispersion as a factor of a few. 

The reason we can do this is that if $\rho_{\DM}$ did indeed factor out of the $J$-factor, {\it i.e.} there is no velocity dependence on $\rho_{\DM}$, then the $v_{c}$ would match when using the different ratios of $J$-factors above. By using the mis-matched velocity dispersion in the (incorrectly) factorised $J$-factor, we therefore obtain an estimate on the size of the effect of the $r$-dependence of the velocity dispersion 
on the $J$-factor. To be somewhat conservative, we will derive constraints using both $s$- and $p$-wave $J$-factors below, which will provide further insight into the error incurred, and from which the weaker limit can be chosen.

\subsection{Limits from Fermi-LAT observations of dSphs}
\label{sec:fermilimit}
We use Fermi-LAT observations towards dSphs to set constraints on the expected photon flux, and, ultimately, on $(\sigma \vrel)_0$ in \cref{eq:flux}.

We use about 10 years of Fermi-LAT data, collected from 500 MeV up to 500 GeV. 
The data set and analysis pipeline strictly follows the procedure presented in Ref.~\cite{Alvarez:2020cmw}. In particular, we adopt data-driven $s$-wave $J$-factors obtained through a new dynamical analysis of dSphs which does not impose any prior knowledge (nor parameterisation) about the dSph DM density profile. A similar data-driven approach is applied for the determination of the background probability distribution function at the dSph position (we refer the interested reader to methodological details presented in~\cite{Alvarez:2020cmw}).
To set constraints on the model under study, we use a standard profile-likelihood method  by fully profiling over $J$-factor and background uncertainties. To improve the statistics (and sensitivity), we stack together the four most constraining dSphs (Draco, Sculptor, Ursa Minor, and Leo II), as explained in~\cite{Alvarez:2020cmw}. We conveniently normalise the signal using the combination $\fion^{2}/(\fion\mpd + \fion\me+[1-\fion]\mH)^{2} = 1/(100 \; \mathrm{GeV})^{2}$, and we therefore set a 95\% C.L. upper limit on $(\sigma \vrel)_0$. This can easily be rescaled when comparing the limit to the prediction at a given point in model parameter space. 

\begin{figure}[t]
\begin{center}
\includegraphics[width=195pt]{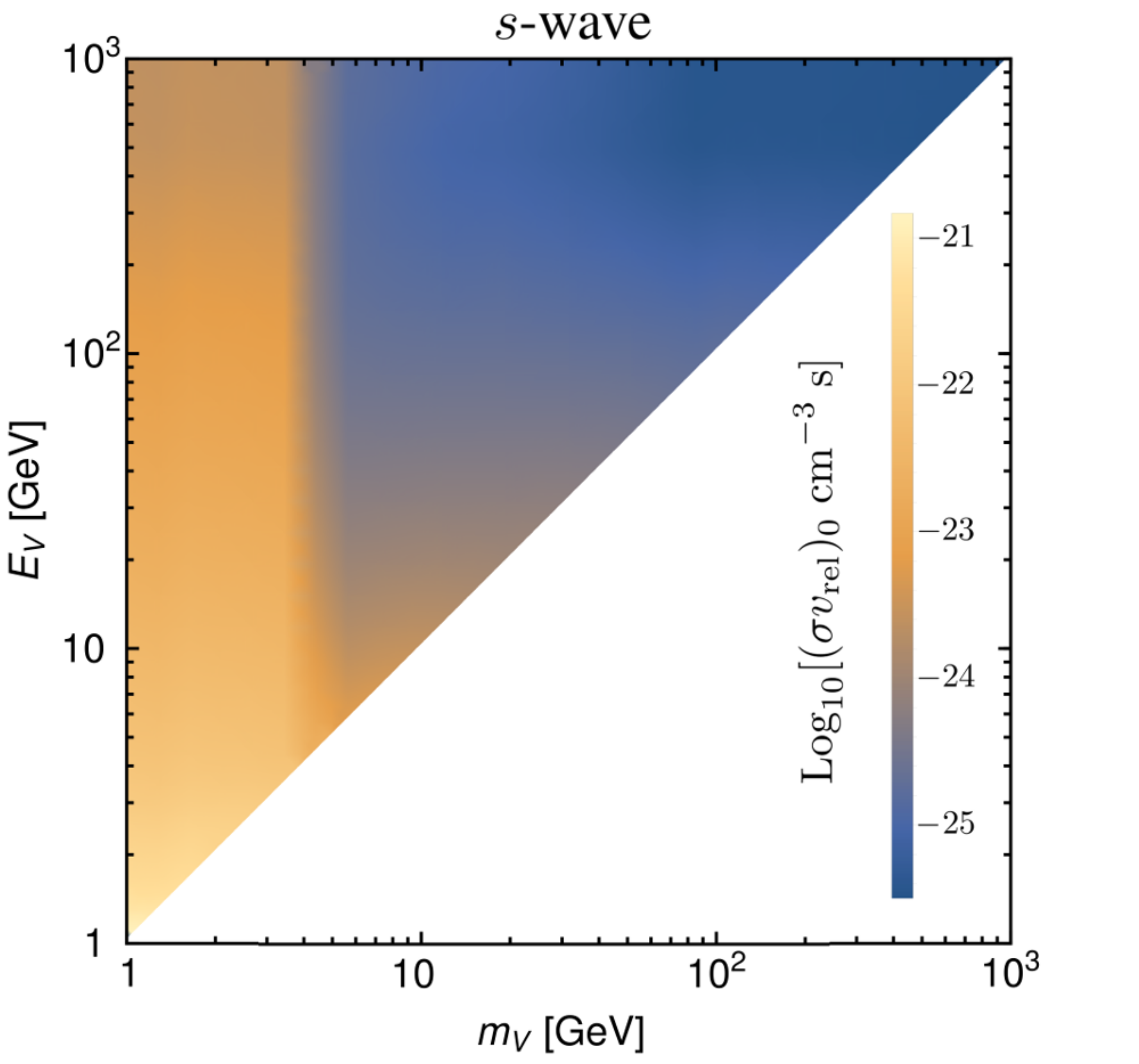}
\includegraphics[width=195pt]{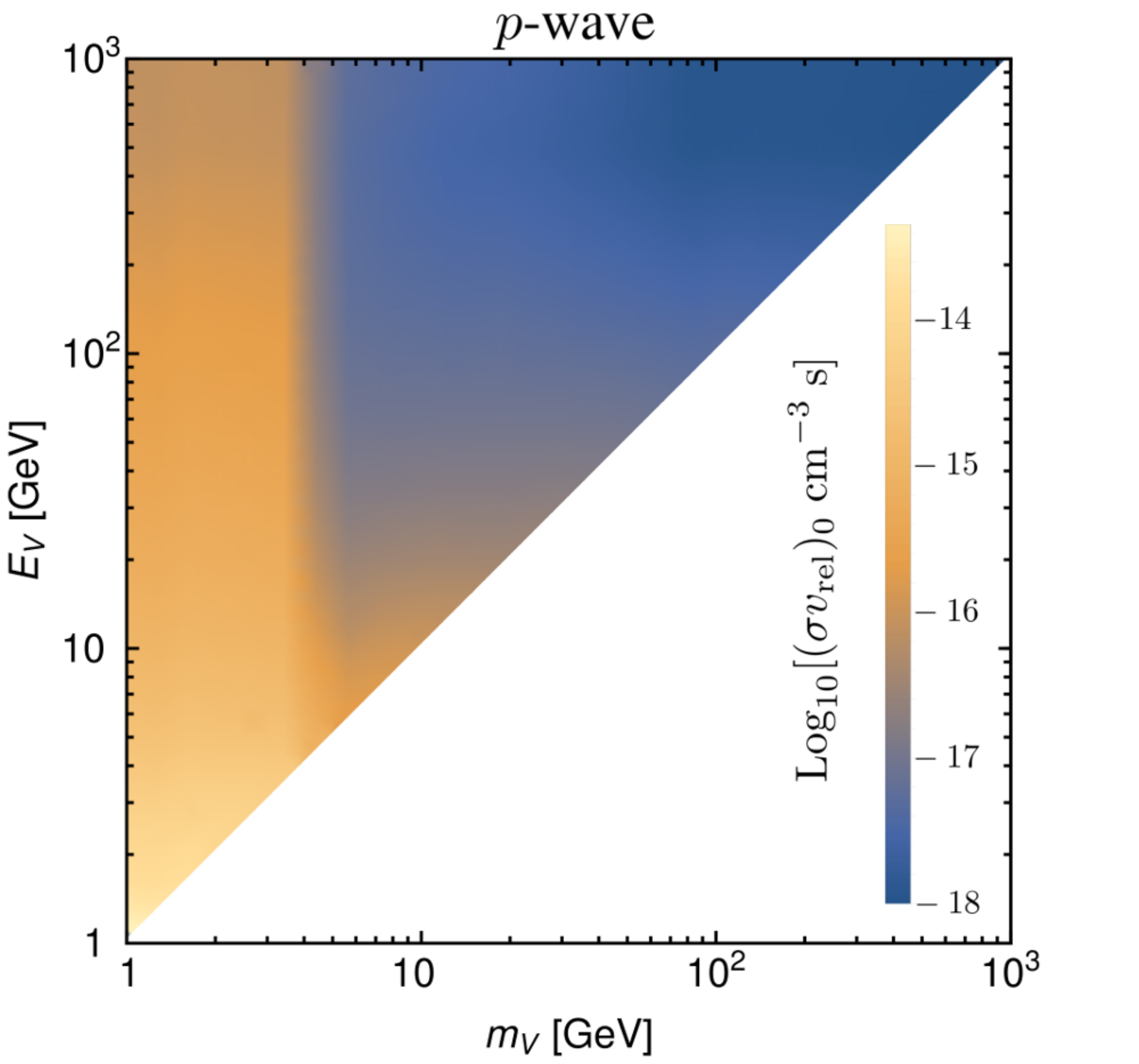}
\end{center}
\caption{\small Limits from Fermi-LAT  dSphs observations on the dark photon production cross section, \cref{eq:flux}, with prefactor normalisation $\fion^{2}/(\fion\mpd + \fion\me+[1-\fion]\mH)^{2} = 1/(100 \; \mathrm{GeV})^{2}$. Note the limit on the flux becomes much stronger around $\mV \sim 5$ GeV due to the more efficient production of ordinary photons. 
The constraint on the velocity-independent part of the cross section is $\sim 8$ orders of magnitude stronger when using the $s$-wave $J$-factors in comparison to the $p$-wave ones, as expected since $\vrel \sim 10^{-4}$ for dSphs. 
Note that the constraints obtained using the $s$-wave $J$-factors can be applied on the averaged $\<\sigma \vrel\>$, independently of the velocity dependence of the cross section, provided that the DM velocity dispersion is approximately constant within the regions of the halo that contribute significantly to the $J$-factors. \bigskip }
\label{fig:fermilimit}
\end{figure}

The constraints in terms of the dark photon mass and energy are shown in \cref{fig:fermilimit}. We provide the limits as a supplementary data file which can be used to constrain models with kinetically mixed dark photons. The constraints obtained using the $s$-wave $J$-factors apply on $\<\sigma \vrel\>_{\BSF}$, independently of the velocity scaling of the cross section, in the approximation where the DM velocity dispersion is nearly constant within the regions that contribute significantly to $J$. 
We also run the analysis for $p$-wave $J$-factors and show the resulting limits in \cref{fig:fermilimit}. In this case, $J$-factors values are taken from~\cite{Boddy:2019qak} and we model their distribution with a log-normal probability distribution function. For comparison with $s$-wave results, we use the same four dSphs for the stacked analysis. 

These constraints apply as long as the dark photons decay within the area encompassed in the $J$-factors, which corresponds to 0.5~deg circle centered on the dSph galaxy under consideration. The closest of the four dSphs used in the analysis is Ursa Minor, at a distance of about 60~kpc~\cite{Karachentsev_2004}. To be conservative, we shall require that the dark photons decay within 1/10 of the corresponding radius, \emph{i.e.}~$\gamma c \tau_{\mathsmaller{\V}} \lesssim 10^{18}$~m, taking into account their boost factor at production, $\gamma = \EV/\mV \simeq \ED/\mV$. This implies
\begin{equation}
\epsilon \gtrsim 
10^{-16}  
\(\frac{10}{g_{\dec}}\)^{1/2}  
\(\frac{10~{\rm GeV}}{\mV}\)^{1/2}
\(\frac{\ED}{\mV}\)^{1/2} ,
\label{eq:EpsilonMinForID}
\end{equation}
where $g_{\dec}$ stands for the accessible decay channels. Note that this rough estimate neglects resonant features in the dark photon decay.

\section{Comparison of constraints to model predictions}
\label{sec:comparison}
\subsection{DM annihilation in the symmetric limit}

\begin{figure}[p]
\begin{center}
\includegraphics[width=185pt]{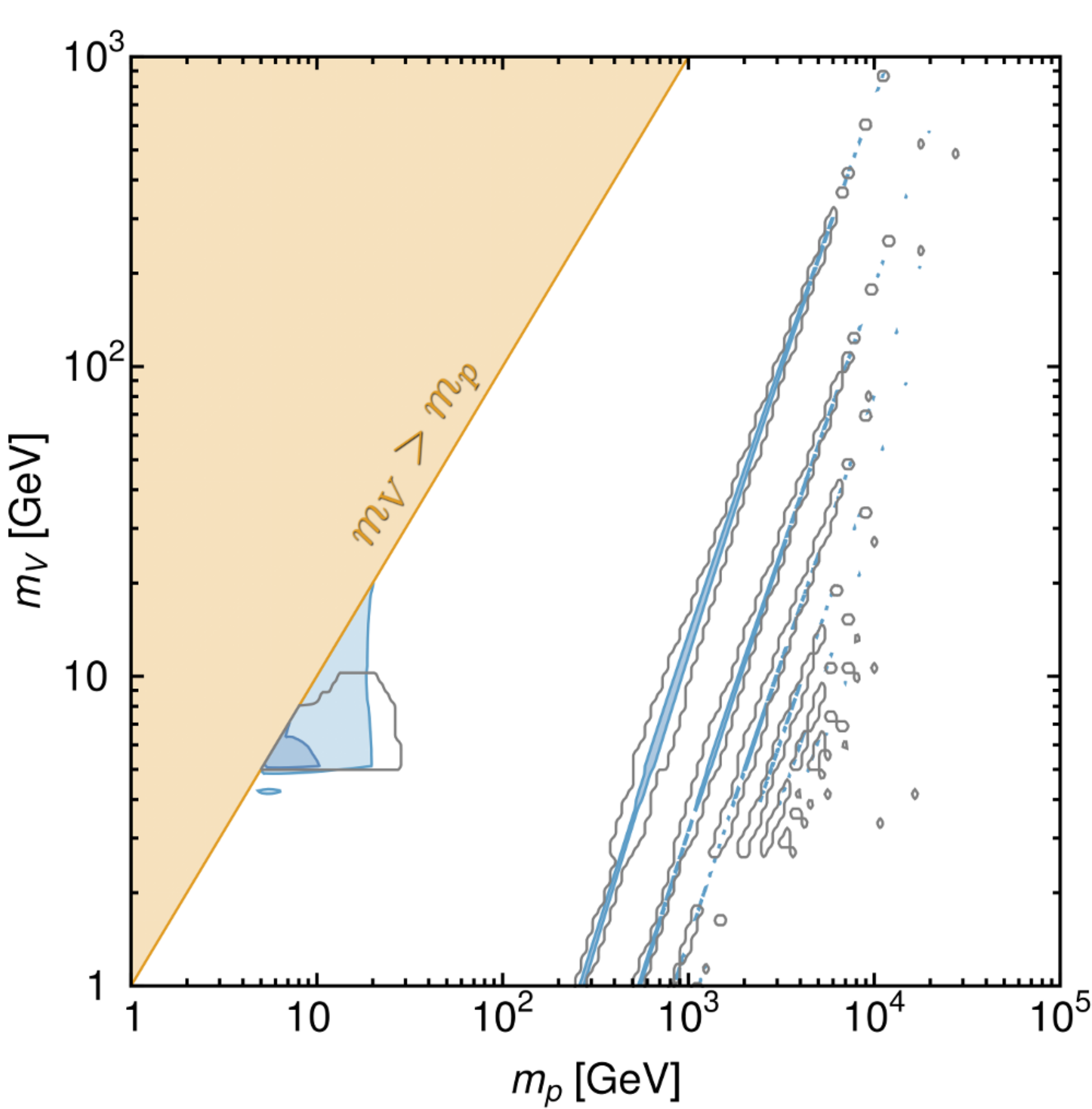}
\end{center}
\caption{\small Limits from Fermi-LAT dSphs on DM annihilation in the symmetric limit with equal numbers of $\p$ and $\bar{\p}$. Shaded regions are constrained. The dark (light) blue region includes (does not include) the profiling over the diffuse background. We have not averaged over the velocity distribution and simply set $\vrel = 20$ km/s. The limits approximately reproduce the constraints from~\cite[Fig. 1]{Baldes:2017gzu} shown outlined in gray from an analysis using fifteen dSphs and averaging over the velocity distribution.
The details of the analysis together with the number of dSphs used differ so it should not be surprising that the constrained regions do not overlap entirely.
\label{fig:symmetric}  \bigskip}

\begin{center}
\includegraphics[width=185pt]{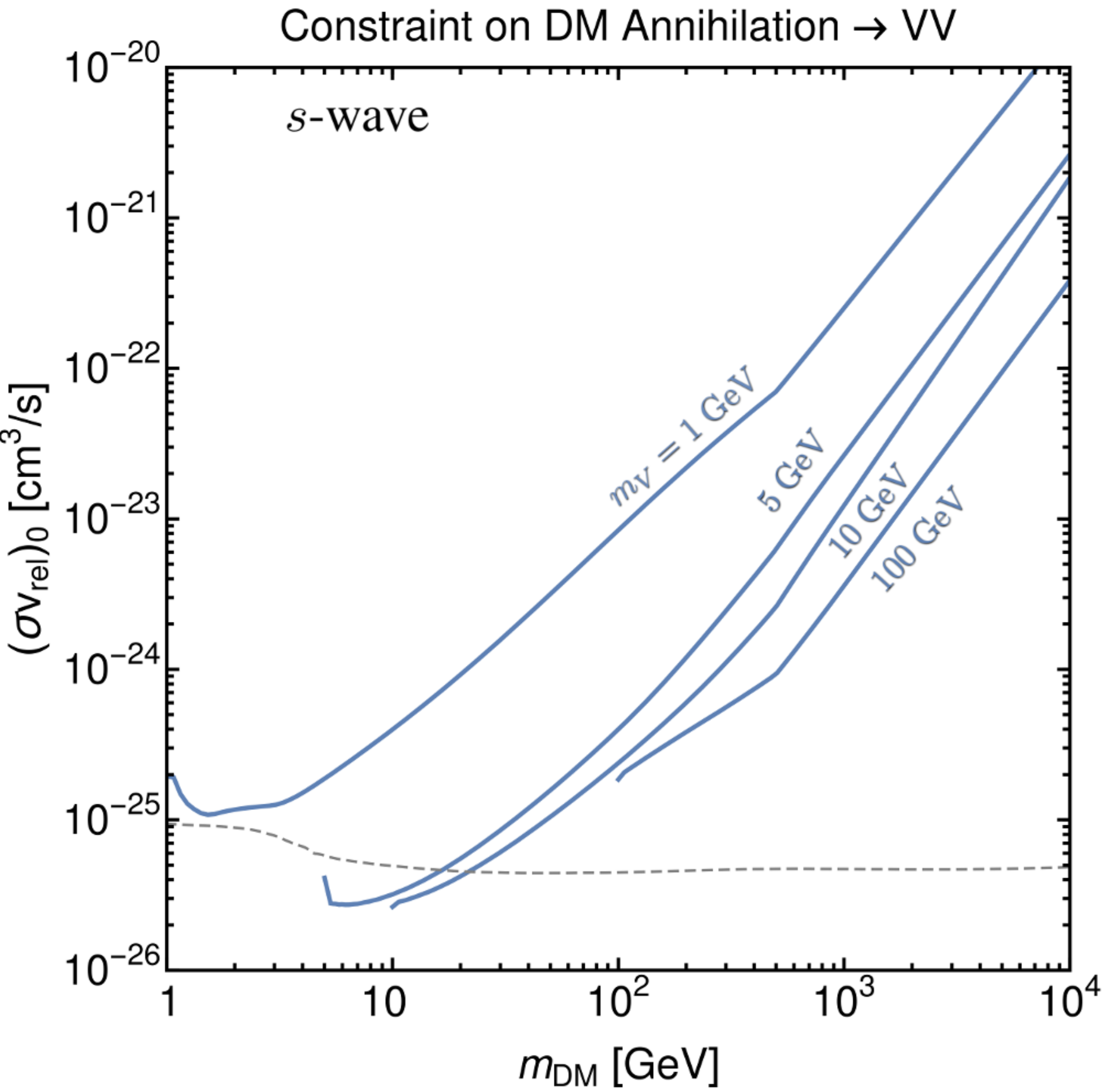}
\includegraphics[width=185pt]{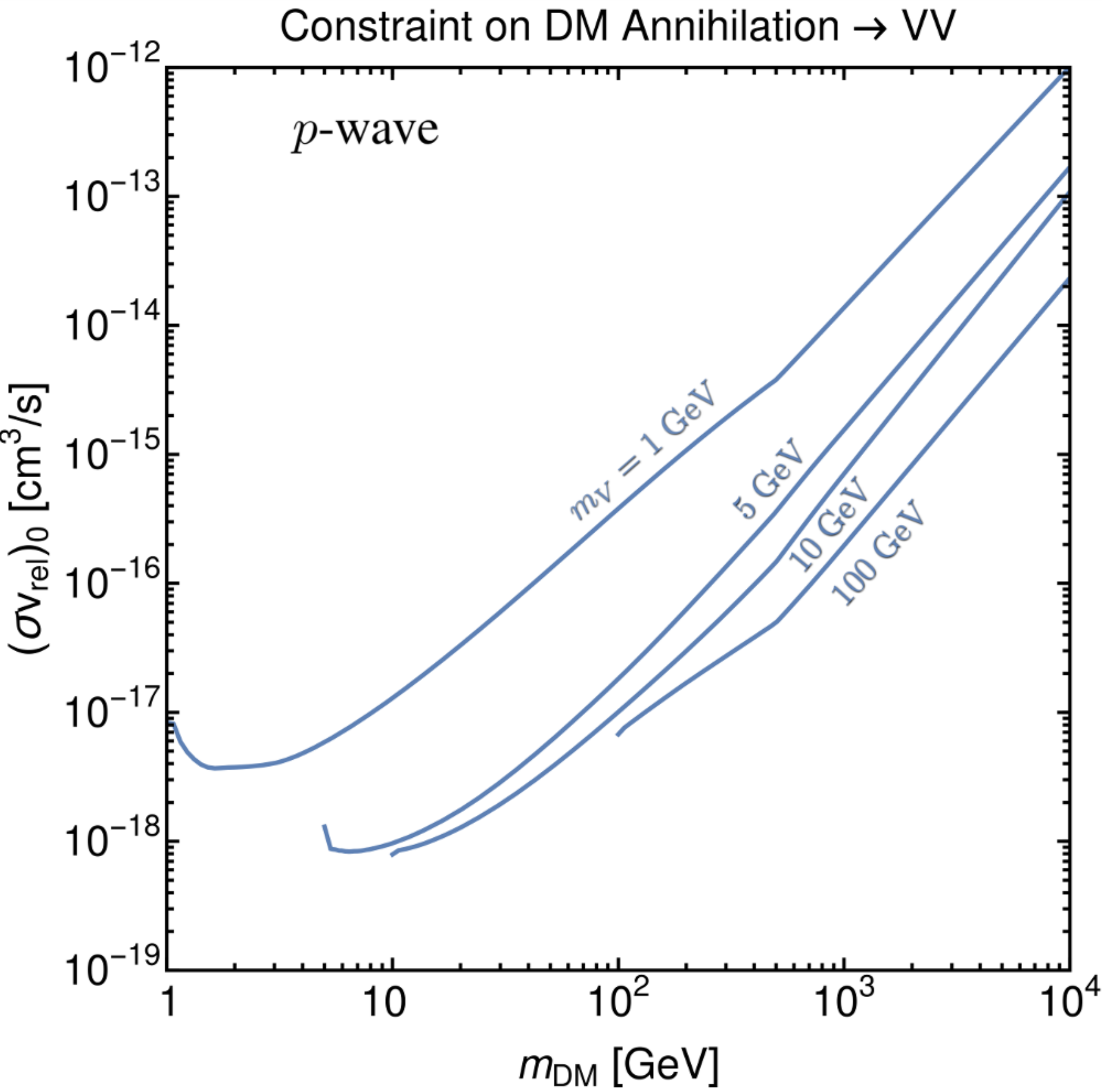}
\end{center}
\caption{\small Left: constraint on the generic $s$-wave DM annihilation cross section for different dark photon masses. The thermal relic line for non-self conjugate DM is also shown (dashed line). Right: same but for a generic $p$-wave cross section.
\label{fig:symmetric_generic}  \bigskip}

\end{figure}

We first use our results to constrain DM annihilation in the symmetric limit. For this we assume a standard secluded WIMP type scenario with equal number of $\p$ and $\bar{\p}$. The coupling $\aD$ is set to return the correct relic abundance~\cite{Baldes:2017gzw,vonHarling:2014kha}. We can then set $\EV = \mpd$ and include a multiplicative factor of two in the flux as each annihilation creates two dark photons and hence twice the number of visible photons as in our expressions for $dN_{\gamma}/dE$. The limits are shown in \cref{fig:symmetric}. Note that on the Sommerfeld resonances, which show up as the thin constrained regions on the right of the plot, the cross section can be much larger today than at freeze-out. Shown in \cref{fig:symmetric_generic} is the limit on the cross section itself for different choices of $\mV$. 

\begin{figure}[p]

\begin{center}
\includegraphics[width=185pt]{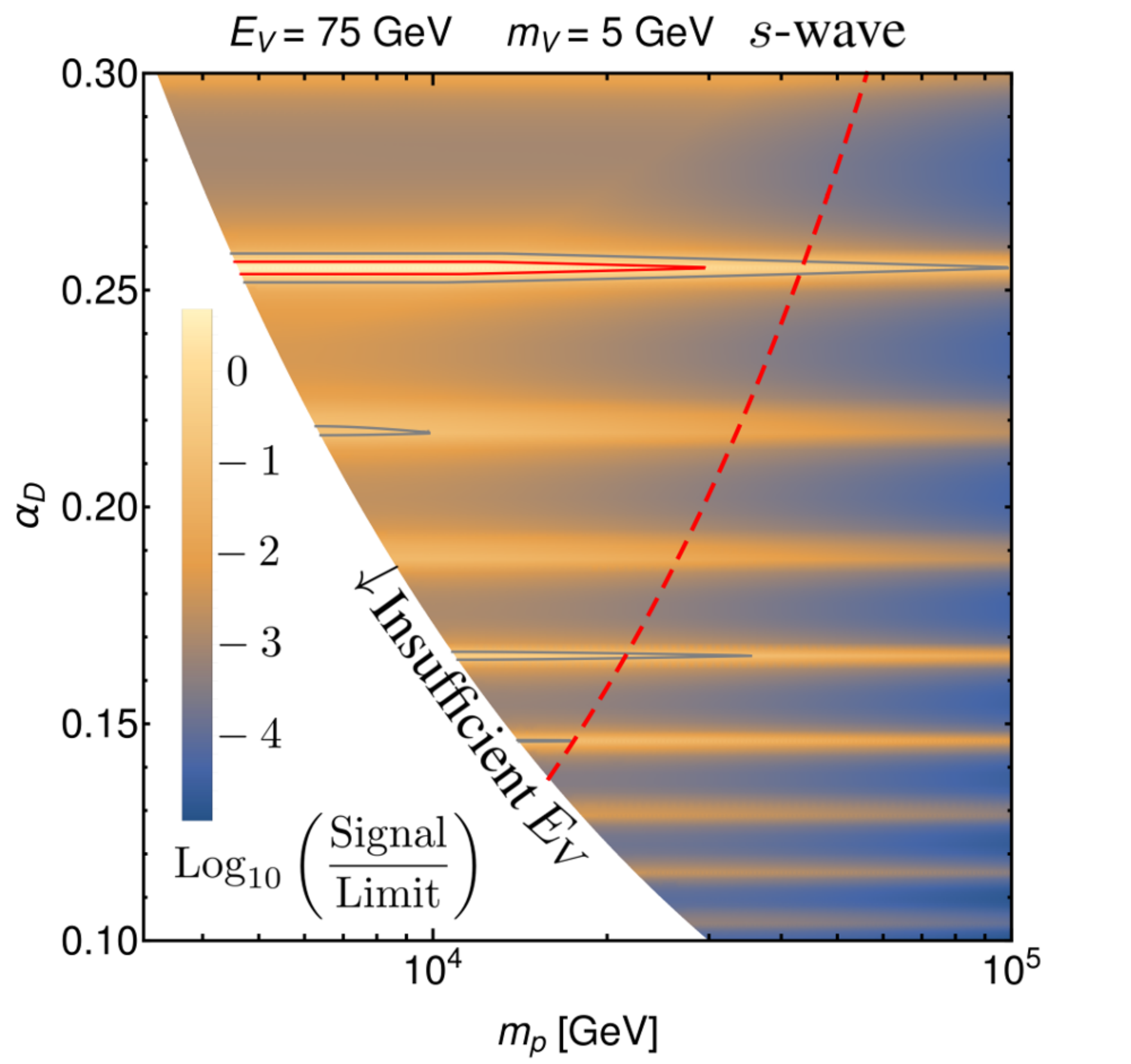} \quad
\includegraphics[width=185pt]{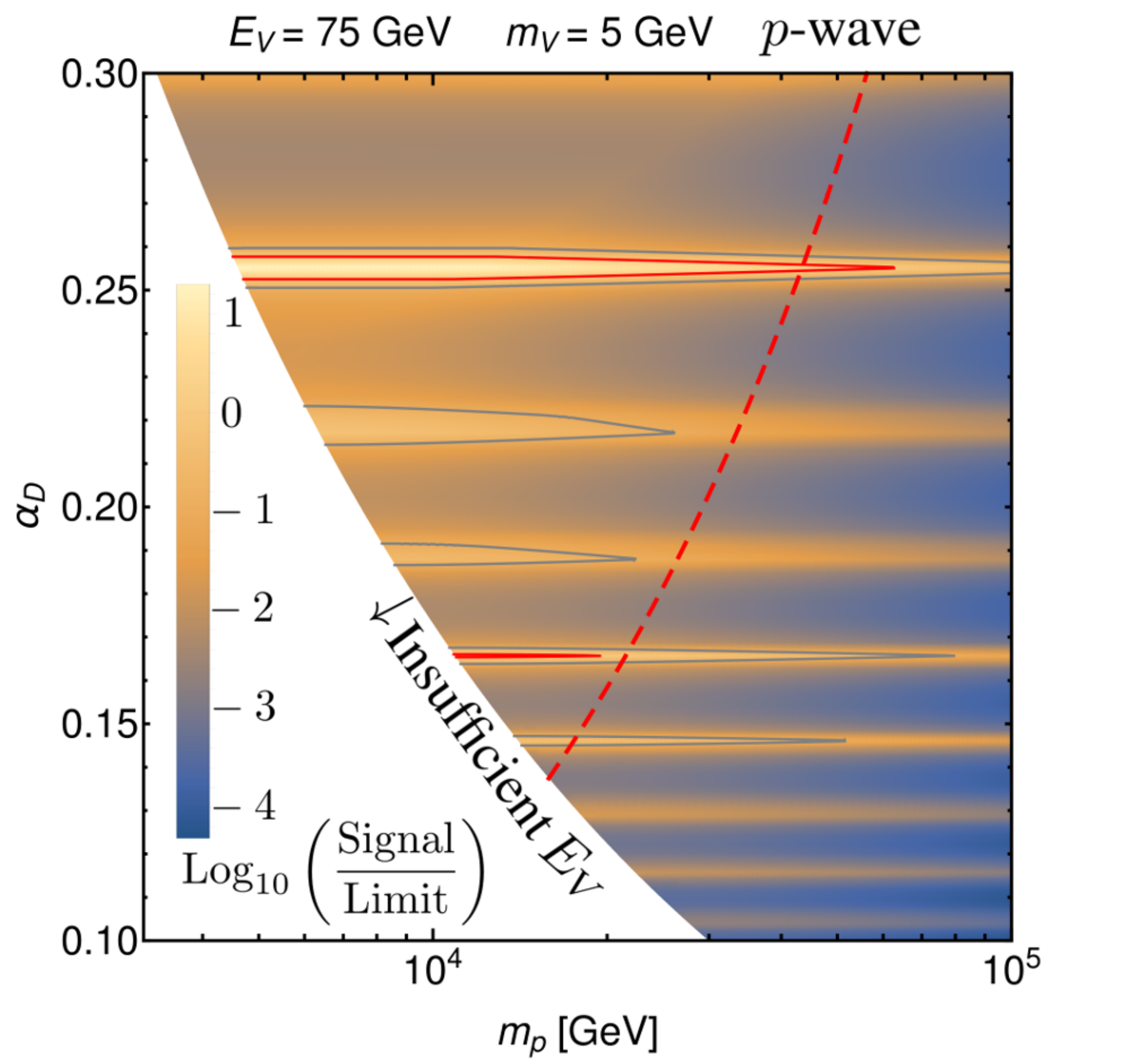} \\
\includegraphics[width=185pt]{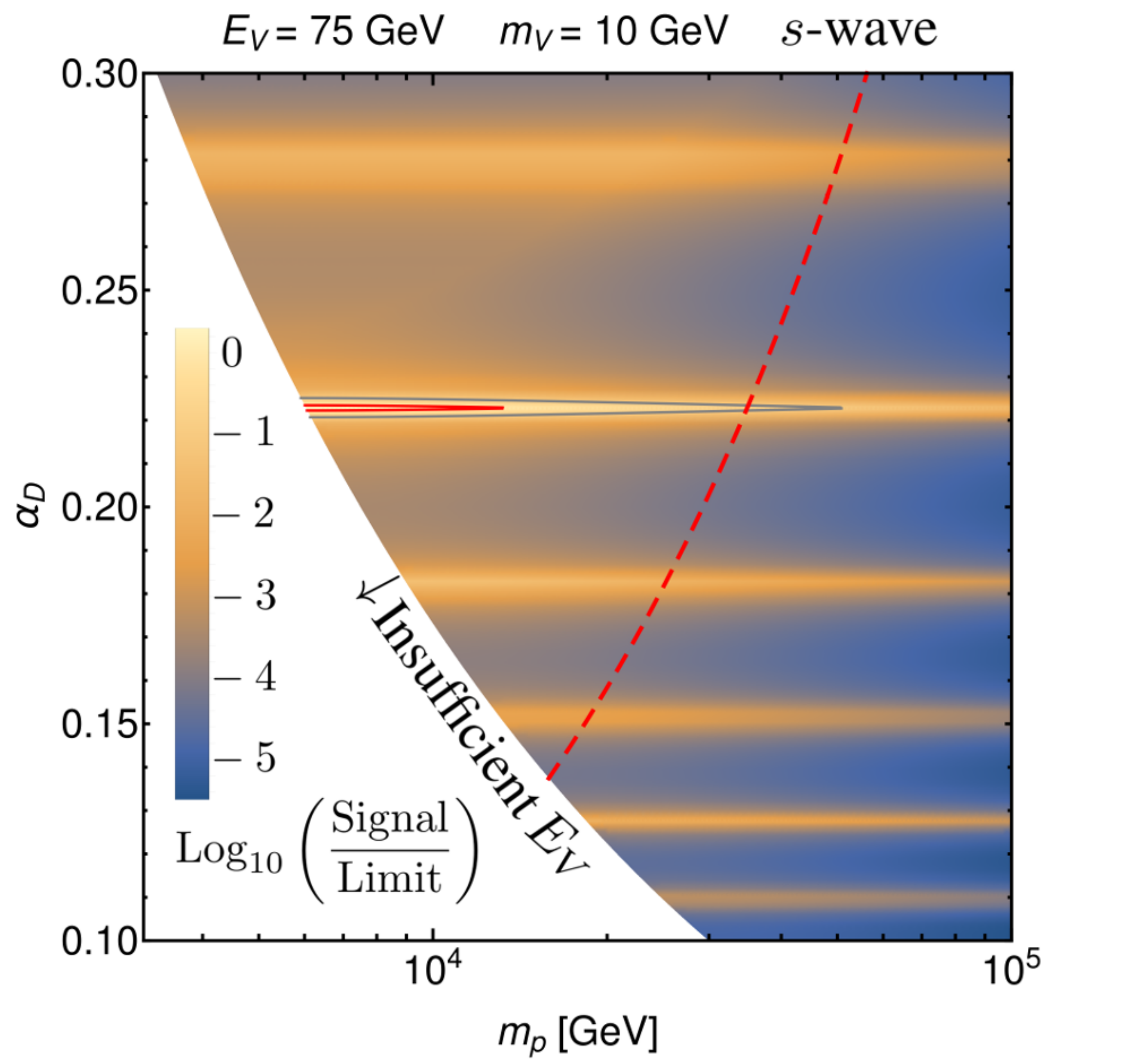} \quad
\includegraphics[width=185pt]{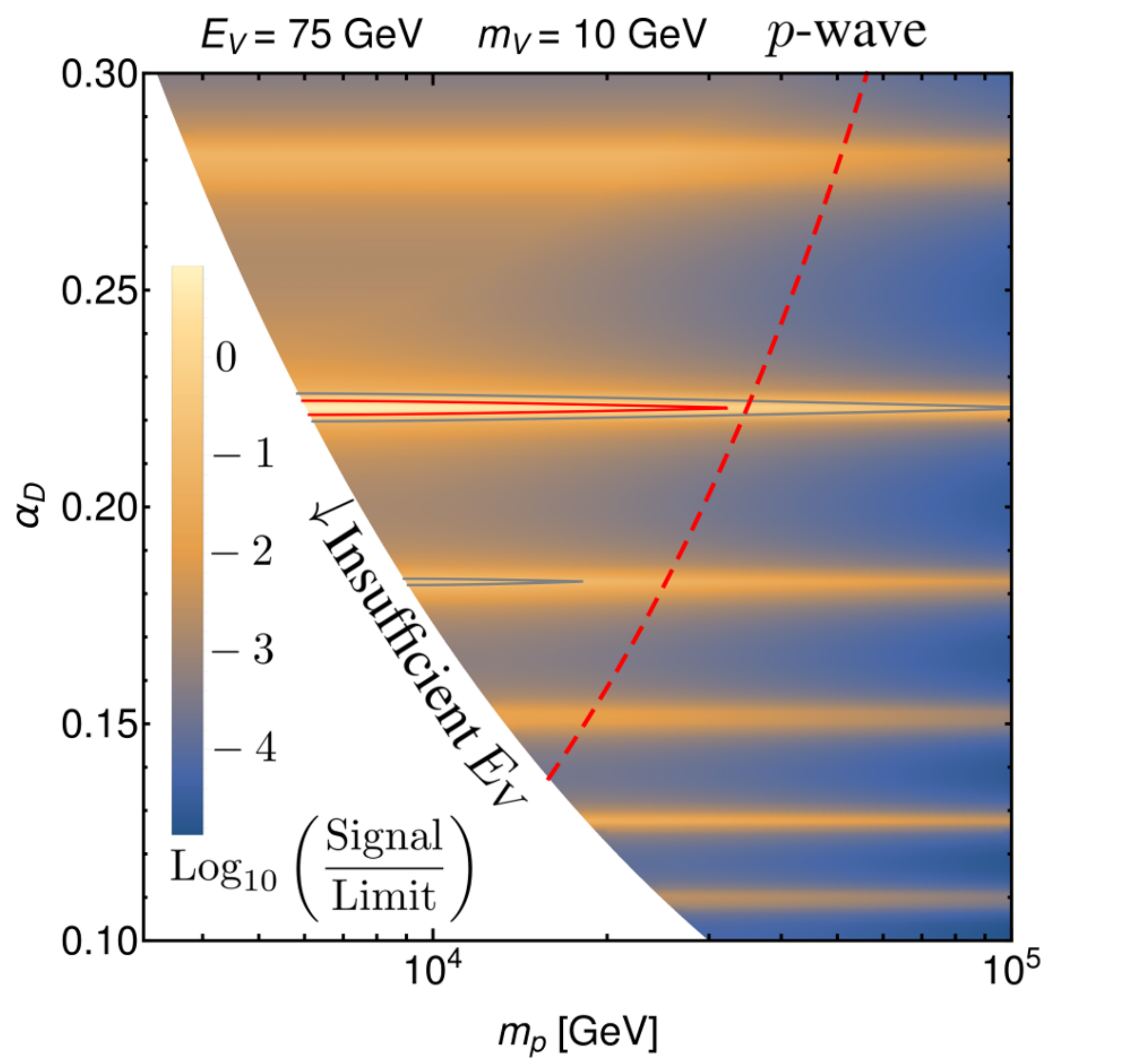} \\
\includegraphics[width=185pt]{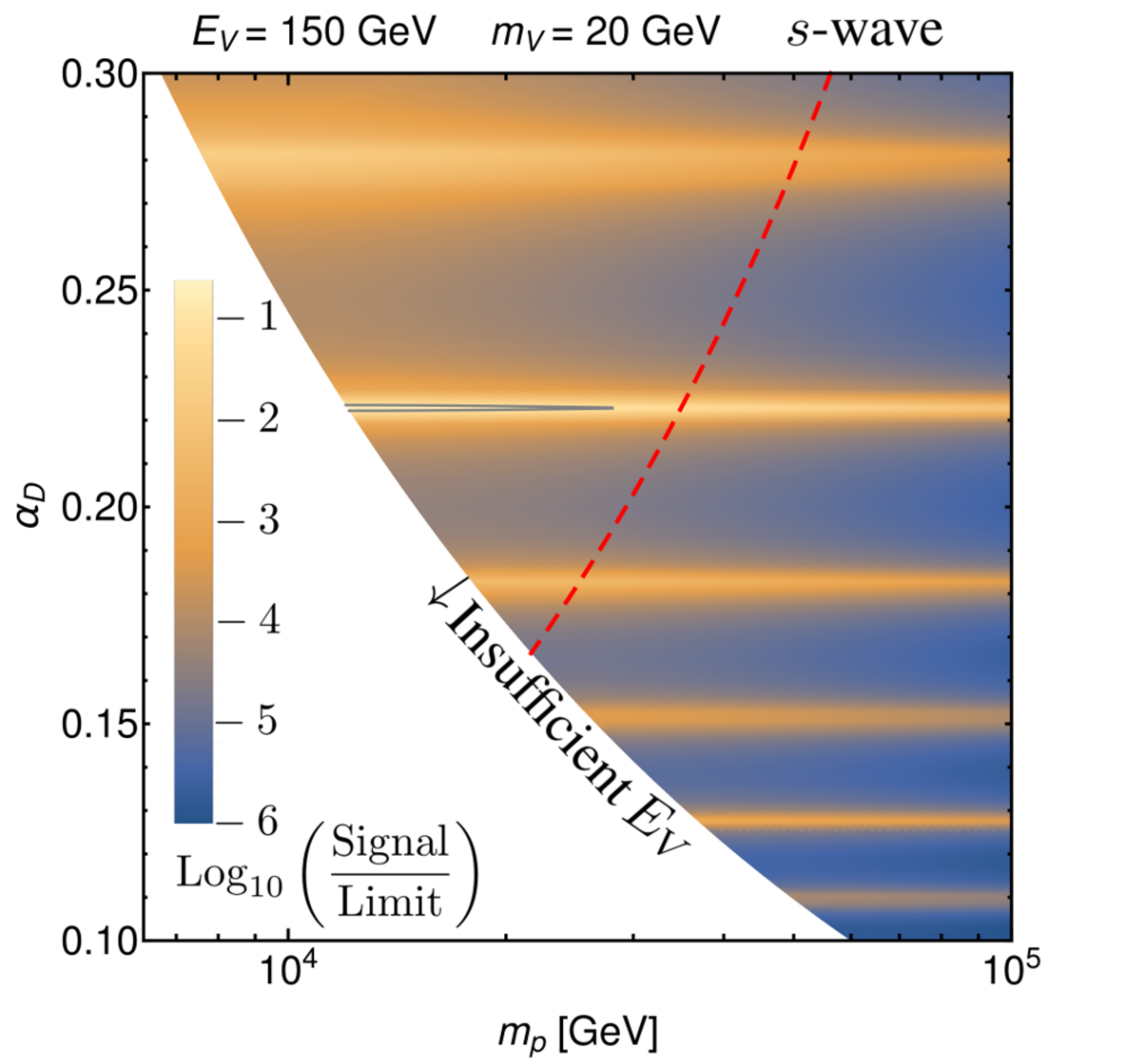} \quad
\includegraphics[width=185pt]{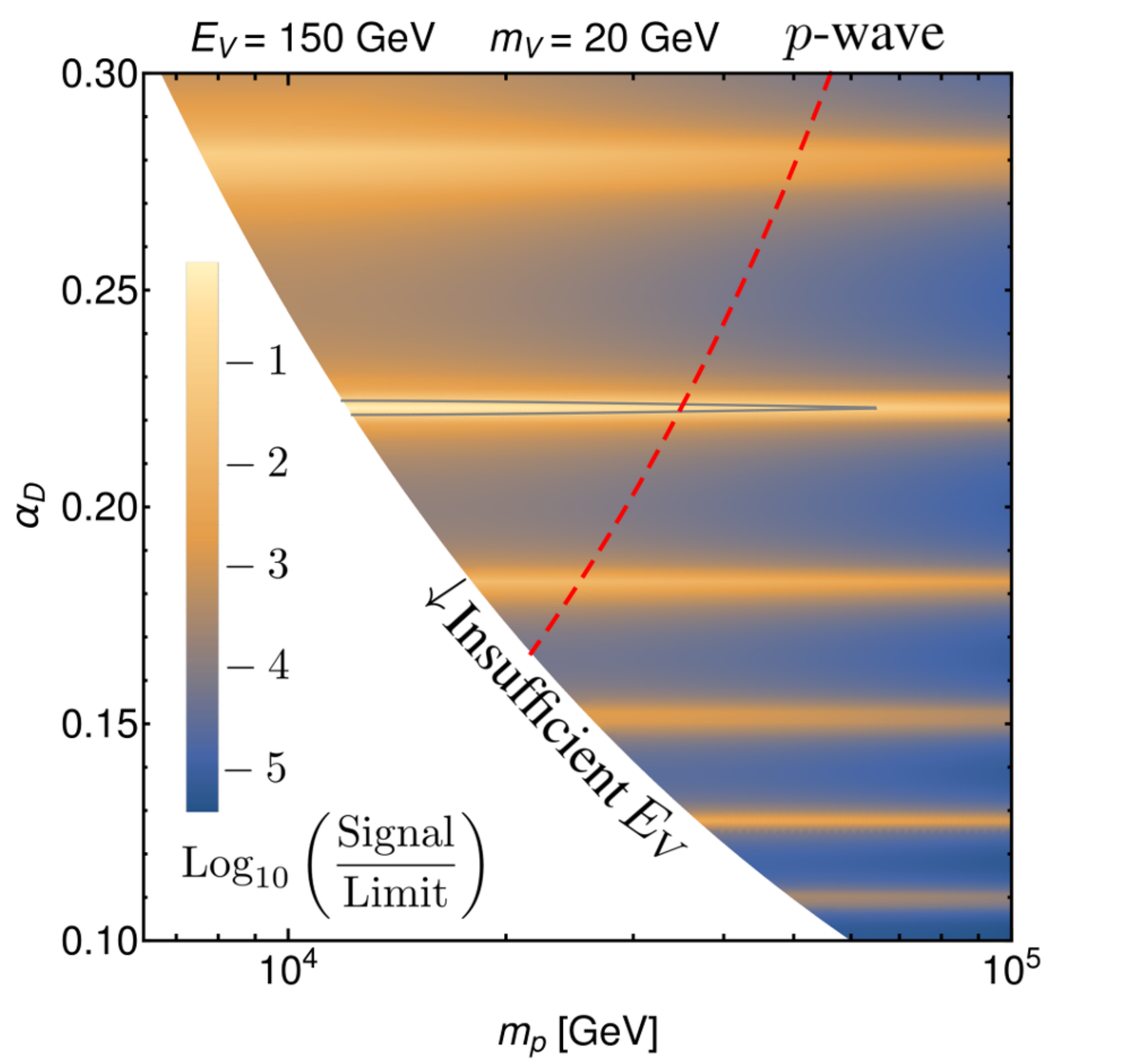}
\end{center}
\caption{\small Limits from Fermi-LAT dSphs on bound state formation in the dark QED asymmetric DM model for three sets of parameters. The constrained regions are shown by a red contour. The grey contours show the would-be constrained regions if the limit on the flux were improved by a factor of ten. The DM relative velocity is set to $\vrel = 20$ km/s. We have enforced $\me < \mpd$ which implies we cannot reach the required $\EV$ in the white regions of the plots. The red dashed line shows the minimum allowed coupling to avoid overclosure in a standard thermal history~\cite{Baldes:2017gzu,Baldes:2017gzw}. The constraint using the $s$-wave ($p$-wave) $J$-factor is shown on the left (right). The $p$-wave constraint is a factor of $\approx 4$ stronger.
\label{fig:ADMBSFconstraint} \bigskip}

\end{figure}

\subsection{Dark atom formation}

We now apply the constraints to the atomic bound state formation in our dark sector. The constraint is given in terms of $\mV$, and $\EV$. The underlying model parameters are $\mV$, $\aD$, $\mpd$, and $\me$. Here we visualise the parameter space by fixing  $\mV$ and $\EV$, varying $\aD$, $\mpd$, and choosing $\me$ in order to return the required $\EV$. Typical results, showing newly constrained regions of parameter space, are displayed in \cref{fig:ADMBSFconstraint}.

As can be seen, the novel constraints currently rule out only small areas of parameter space. For this reason, and considering the uncertainties on the DM velocity profile in the dSphs, we have not performed a velocity average over the DM distribution but simply set the velocity to an illustrative value from \cref{eq:vel1,eq:vel2}, namely $\vrel = 20$ km/s.

The analysis has been performed using the limits from both the $s$- and $p$-wave $J$-factors. With this choice of $\vrel$ the resulting constraints on the BSF cross section differ by a factor of $\approx 4$.  Note the condition $\vrel < \mV/\muD$ is satisfied over the entire range of the plots in \cref{fig:ADMBSFconstraint}. Nevertheless, the non-trivial $\vrel$ dependence of the cross section means the assumed $\vrel$ does not entirely factor out for the $p$-wave constraint, as would be the case for a pure $\vrel^{2}$ dependence. This shows the underlying error incurred through this approximate technique. To overcome this source of uncertainty it would be necessary to fully account for the non-trivial velocity dependence of $\sv_{\BSF}$ when determining the $J$-factor from the estimate of the underlying DM phase space distribution.

Limits could also be derived using observations of the Galactic Centre, which features a higher $\vrel$, and hence higher $\sv_{\BSF}$. Albeit, one must then deal with the complication of the well known excess in Fermi-LAT observations of the Galactic Centre over the standard background modelling, \emph{e.g.}~see~\cite{Daylan:2014rsa,Calore:2014xka,Linden:2016rcf,Macias:2016nev,Storm:2017arh,Bartels:2017vsx,Leane:2019xiy,Chang:2019ars,Leane:2020nmi,Buschmann:2020adf}.

\subsection{Variations}

\label{sec:variation}

\begin{figure}[t]
\begin{center}
\includegraphics[width=200pt]{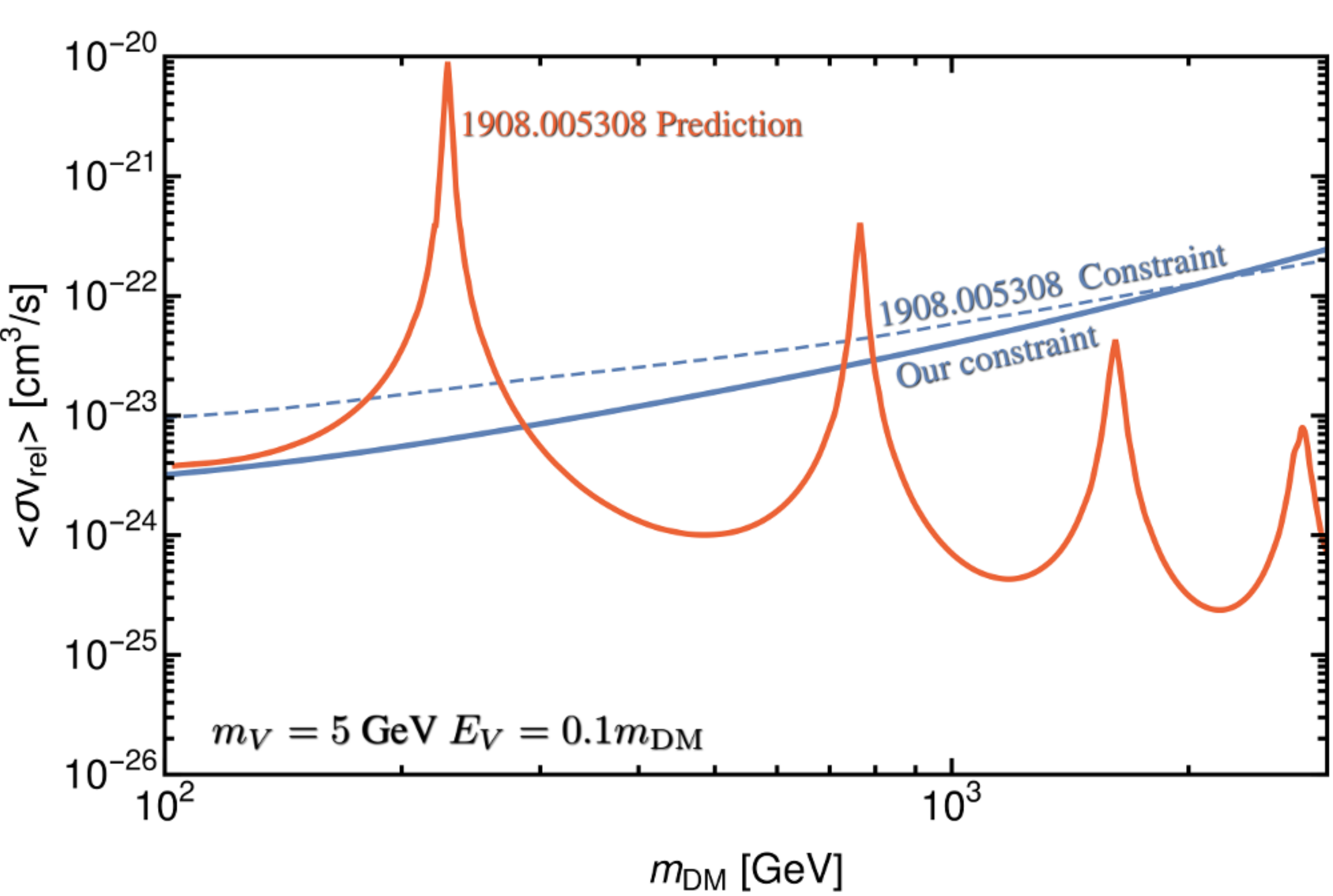}
\end{center}
\caption{\small Constraint on the bound state formation considered in Ref.~\cite{Mahbubani:2019pij} from our analysis and compared to the approximate constraint derived in~\cite{Mahbubani:2019pij} using the results of~\cite{Profumo:2017obk}. Although the constraints here are in rough agreement, our constraints can be applied to a wider range of dark photon masses. 
\label{fig:tesi_comparison} \bigskip}
\end{figure}

Finally we can consider variants of the above model. For example, if there is another dark sector force in addition to the $U(1)_D$, the binding energy of the composite state can be made larger, while keeping $\mpd$ small enough to not suppress the signal due to the falling number density, and keeping $\aD$ in the perturbative range. Such a setup has recently been considered in Ref.~\cite{Mahbubani:2019pij}.
Here, BSF occurs when the upper ($N^+$) and lower ($N^-$) components of a dark baryon isospin doublet, with opposite $U(1)_{D}$ charges, combine and emit a dark photon. The total binding energy is now no longer solely determined by the $U(1)_{D}$ but also involves an additional force, {\it e.g.} originating from a local dark $SU(3)_{D}$ symmetry. The cross section has been calculated in~\cite{Mahbubani:2019pij} and we extract it from their Fig.~3 for an example parameter point. We then confront it with our constraint from the dSphs in \cref{fig:tesi_comparison}.

We also compare to the approximate dSph constraint derived in~\cite{Mahbubani:2019pij}. This was found by using the scaling
\begin{equation}
\langle \sigma_{\BSF} \vrel  \rangle < 2\left( \frac{ \mDM }{ \EV } \right)^{2} \left[ \langle \sigma_{N^+N^- \to VV} \vrel \rangle \Big|_{\mDM \to \EV}\right],
\end{equation}
as in \cref{eq:Rescaling}.
For the constraint $\langle \sigma_{N^+N^- \to VV} \vrel \rangle$ the authors of~\cite{Mahbubani:2019pij} took the available limit for DM annihilating to $VV$ followed by the 100\% decay of $V \to \tau\tau$ from~\cite{Profumo:2017obk} and then weakened it by $1/0.1$. This last factor is included as a dark photon with $\mV \approx 5$ GeV decays into $\tau \tau$ with a branching fraction of around 0.1 (see \cref{fig:DPBR}). From \cref{fig:tesi_comparison} we see the constraint derived using these approximations is not too far off our constraint which takes into account the various decay channels of $V$ more precisely. The key point is that using our results such models can be constrained more widely, with fewer assumptions, and greater ease.

\section{Conclusions}

The radiative formation of DM bound states, as well as exothermic level transitions between bound levels, provide novel sources 
of signals that can be probed via indirect searches. The existence of bound levels -- a consequence of long-range interactions -- is an important feature of many self-interacting and/or asymmetric DM models. Unitarity arguments along with various model-dependent considerations suggest it is also a generic characteristic of (symmetric or asymmetric) thermal-relic DM in the multi-TeV mass regime and above. As our DM searches move beyond the paradigm of 100~GeV -- 1~TeV symmetric thermal-relic DM, identifying and exploring such novel signatures becomes essential. 

In this paper, we employed indirect searches to derive constraints on the formation of DM bound states that occurs with emission of a dark photon kinetically mixed with hypercharge. We used Fermi-LAT observations of dSphs, but our analysis can of course be extended to other experiments, such as H.E.S.S., or other celestial regions of interest, such as the Galactic Centre. 
Our results are cast in terms of the amount of energy dissipated and the DM mass, which determines the number density of the dark particles. While the radiated energy in DM annihilation is of the order of the DM mass, BSF occurs with dissipation of a smaller amount of energy that depends on the underlying dynamics. Our results are therefore applicable to a variety of DM models where BSF occurs via dark photon emission, and reproduce also constraints on DM annihilation into dark photons. In addition, we have discussed the recasting of existing constraints on DM annihilation into SM particles, to apply on BSF. 

In the course of this work we developed the treatment of a number of subtleties, namely the effects of the dark photon polarisation states, and the non-trivial velocity dependence of the bound state formation cross section. The latter means the conventionally given $J$-factors do not fully fit the requirements of the model. We estimated the error introduced by using an approximate technique. If limits eventually become more constraining on such models, a more careful treatment of the $J$-factors may become necessary. Further improvements can also be made by taking into account the effect of low lying QCD resonances on the photon spectrum produced in the cascades of the dark photon decay products~\cite{Plehn:2019jeo}.

We have considered and applied our constraints on a simple dark QED model of asymmetric DM that implies the existence of dark atoms forming via emission of dark photons. We determined the BSF cross section, the DM ionization fraction, and the $\gamma$-ray spectrum arising from the cascades of the dark photon decay products. The combination of these elements allowed us to predict the photon flux resulting from BSF as a function of the underlying model parameters. Thus allowing us to derive novel constraints on the parameter space. We found that the predicted flux typically lies below the derived limit, except for some resonance peaks at relatively large values of the dark coupling $\aD \gtrsim 0.1$. Furthermore, variations of the model can lead to somewhat larger signal predictions~\cite{Mahbubani:2019pij} which we also briefly explored.

We also showed that our constraints can be applied to the annihilation and the decay products of unstable bound states of symmetric DM. In agreement with previous studies~\cite{Cirelli:2016rnw}, we observed that in this case, the low-energy dark photon emitted in the formation of the DM bound states does not constrain the model any further due to the suppression of the DM number density by the large DM mass. However, this result does not preclude that the low-energy radiation emitted in the formation of (unstable) bound states can yield an observable signal. It is possible for example that in other DM models the spectral features of the low-energy radiation produced in BSF differ from those of the high-energy radiation emitted in DM annihilation or in the decay of unstable bound states, and render it competitive.

Crucial input in predicting the signals generated by BSF -- and in fact in predicting any manifestation of DM today -- is the preceding cosmological history. In the model of atomic DM considered here, the cosmological evolution determines the residual ionized component of DM that is available to form bound states today. A large BSF cross section may imply suppressed indirect signals today because the DM has already formed deeply bound atomic states in the early Universe. The details of the interplay between cosmology and phenomenology depend on the DM model and it is essential to compute these two self-consistently. For models that feature long-range interactions, the formation of stable or unstable bound states in the early Universe can critically affect all expected phenomenology of DM today~\cite{Petraki:2014uza,vonHarling:2014kha,Pearce:2015zca}.

\section*{Acknowledgements}
{
In memory of Mathieu Boudaud, our friend and colleague, who contributed to discussions that led to this work. 
We thank Andreas Goudelis, Julien Billard, Marco Cirelli, Julien Masbou, and Emmanuel Moulin for their coordinating efforts for the GPS working group and for useful discussions. 
N.L.R. thanks Bryan Webber for discussions on final state radiation.  
I.B. thanks Andrea Tesi for helpful correspondence.

\paragraph{Funding information} 
This work was initiated in the framework of the dark matter GPS (Groupement de Priorit\'es Scientifiques) working group of the IRN-Terascale.
I.B. is a postdoctoral researcher of the F.R.S.--FNRS with the project ``\emph{Exploring new facets of DM}."  
K.P. was supported by the ANR ACHN 2015 grant (``TheIntricateDark" project), and by the NWO Vidi grant ``Self-interacting asymmetric dark matter."
N.L.R. is supported by the Miller Institute for Basic Research in Science at the University of California, Berkeley.
This work made use of resources provided by the National Energy Research Scientific Computing Center, a U.S. Department of Energy Office of Science User Facility supported by Contract No. DE-AC02-05CH11231.

}

\begin{appendix}

\section{Further constraints on the dark photon}
\label{sec:darkphotonconstraints}

\begin{figure}[t]
\begin{center}
\includegraphics[width=195pt]{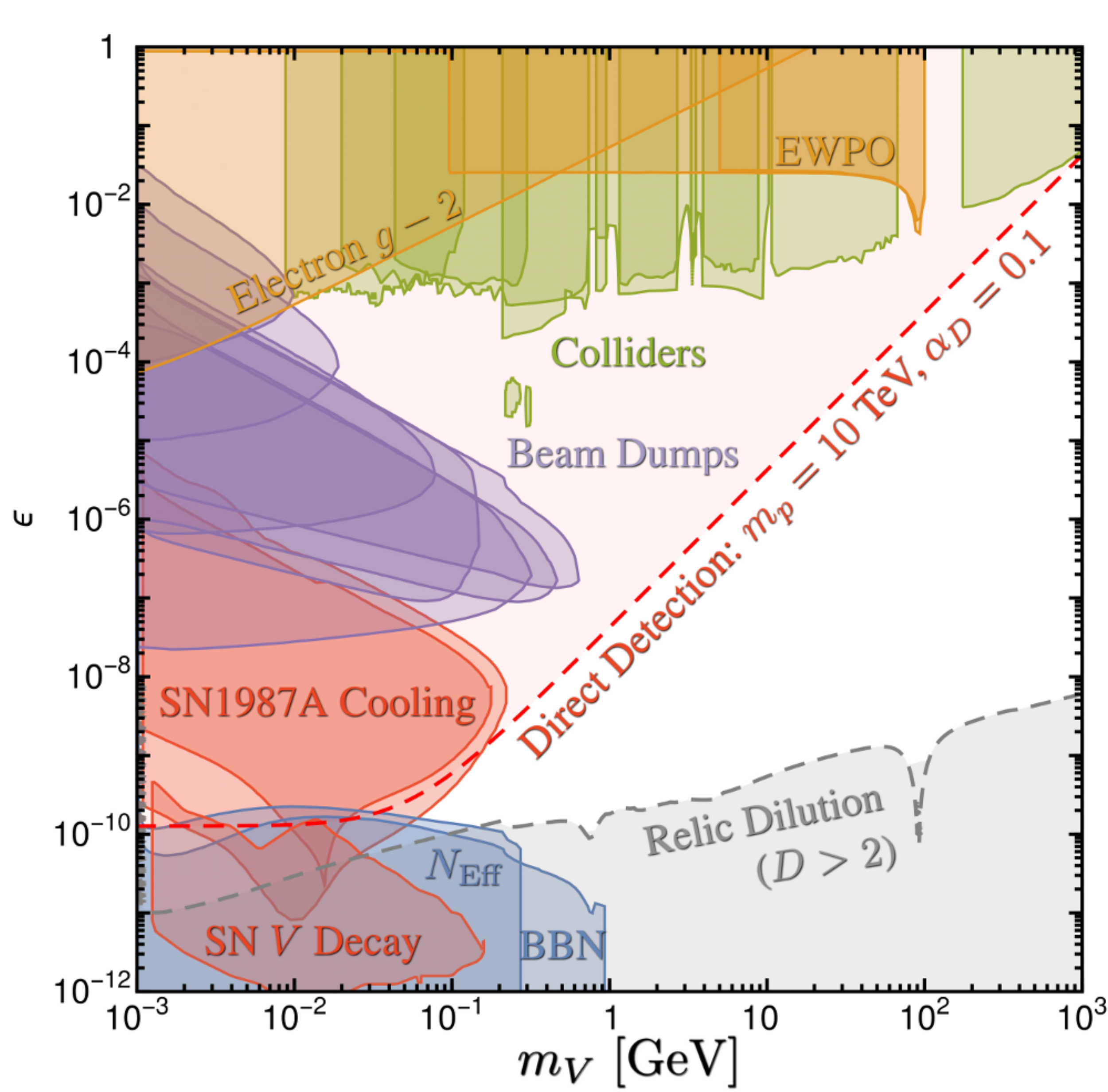}
\end{center}
\caption{\small The constraints on the dark photon parameter space. Details of the individual constraints can be found in the text. The cut-off in the EWPO, BBN, and $N_{\rm eff}$ constraints is artificial and originates from a limited plot range in~\cite{Hook:2010tw,Curtin:2014cca,Hufnagel:2018bjp}.}
\label{fig:darkphoton} \bigskip
\end{figure}

The leading constraints come from a number of sources. In \cref{fig:darkphoton} we have chosen to show the more stringent constraints, also including the latest supernova and BBN (Big Bang nucleosynthesis) limits, important for smaller $\epsilon$.

\begin{itemize}

\item \textbf{Electron $g-2$}. The strongest constraint in the top left corner of the plot comes from the anomalous magnetic moment of the electron~\cite{Endo:2012hp}.

\item \textbf{Electroweak Precision Observables (EWPO)}. The constraint from precision observables has been derived in~\cite{Hook:2010tw} and \cite{Curtin:2014cca}, which give consistent results.

\item \textbf{Colliders/accelerators (prompt or short decay lengths)}. The leading constraints come from NA48/2~\cite{Batley:2015lha}, Mainz Microtron A1~\cite{Merkel:2014avp} (which uses fixed target electron scattering), BABAR~\cite{Lees:2014xha}, LHCb~\cite{Aaij:2019bvg,Ilten:2016tkc,Ilten:2018crw}, and ATLAS~\cite{Cline:2014dwa,ATLAS:2013jma}. The fine detail of the BABAR and LHCb constraints, due to the excellent energy resolution of the detectors, has been smoothed over to give the approximate constraint.

\item \textbf{Beam dumps (long decay lengths)}. Limits come from electron and proton beams. The limits shown were found in~\cite{Blumlein:2011mv,Gninenko:2012eq,Andreas:2012mt,Blumlein:2013cua}.

\item \textbf{Supernovae}. The traditional constraint comes from limiting excess cooling in SN1987A~\cite{Chang:2016ntp}. Recently a stronger constraint has been set by considering energy transfer by dark photons from the centre of the supernova to the outer layers, which can affect the explosion~\cite{Sung:2019xie}. At lower values of $\epsilon$ a constraint has been set by considering dark photons escaping Galactic supernovae~\cite{DeRocco:2019njg}.

\item \textbf{BBN/CMB}. The constraints have recently been updated for particles decaying electromagnetically using a BBN code~\cite{Hufnagel:2018bjp,Forestell:2018txr} (a comparable limit is derived from CMB $N_{\rm Eff}$ measurements --- late decaying dark photons do not heat the SM neutrinos, lowering $N_{\rm Eff}$~\cite{Hufnagel:2018bjp}). Taking the limit on the lifetime from~\cite{Hufnagel:2018bjp}, we can convert it into a limit on $\epsilon$ for a given dark photon mass. Note the limit shown is actually somewhat conservative, as our dark photons will have a $\sim 40$\% higher temperature at BBN than what is assumed in~\cite{Hufnagel:2018bjp} (see above). The results qualitatively match those derived analytically in~\cite{Cirelli:2016rnw}. Quantitatively, the results from the BBN code are somewhat more stringent. The sharp cut-off at high masses is due to the limited range of the plot in~\cite{Hufnagel:2018bjp}, although the limit is known to become progressively weaker as $\mV$ increases~\cite{Cirelli:2016rnw}.

\item \textbf{Relic Dilution}. Although not a constraint, we show on the plot the area in which relics are diluted by the entropy injection, due to the long lived dark photon~\cite{Cirelli:2018iax}. Somewhat arbitrarily, we show a contour for which $Y_{\DM} \to Y_{\DM}/2$ following dark photon decay. 

\item \textbf{Direct detection}. The DM will induce nuclear recoils in direct detection experiments via t-channel exchange of the dark mediator $V$ and the SM $Z$ boson. We derive a constraint from XENON1T data as described in \cref{sec:directdetection},  
considering only the ionised DM component. 
Unlike the other constraints, the direct detection limit also depends on $\aD$ and $\mpd$, 
which determine the scattering of dark protons on the target, as well as on $\me$ which affects the residual DM ionisation fraction. 
The parameters chosen for the example shown in \cref{fig:darkphoton} correspond to a rather stringent exclusion contour.
Analogously to the indirect detection signals, the direct detection rate does not increase monotonically with the coupling $\aD$; large $\aD$ may imply the efficient formation of deeply bound dark atoms in the early Universe, whose interaction with the target nuclei is partially screened due to their zero net charge.

\end{itemize}

Some overall remarks are now in order regarding \cref{fig:darkphoton}. We conclude that even with a choice of $\aD$ and $\mpd$ which results in a stringent direct detection constraint there is still unexcluded parameter space for which the dark photon does not lead to additional dilution of relic densities. Note also that since the publication of~\cite{Baldes:2017gzu} the area relevant for self-interacting DM, $\mV \lesssim \mathcal{O}(10)$~MeV, has largely been ruled out from the updated BBN and SN constraints (at that time a window around $\epsilon \sim 10^{-10}$ was still open and also not excluded by direct detection). We therefore do not consider large DM self-interactions in this work. For large couplings there is also the issue of apparent unitarity violation due to the breakdown of the perturbative expansion together with the possibility of low lying Landau poles.

\begin{itemize}
\item \textbf{Unitarity and Landau poles} \begin{figure}[t]
\begin{center}
\includegraphics[width=200pt]{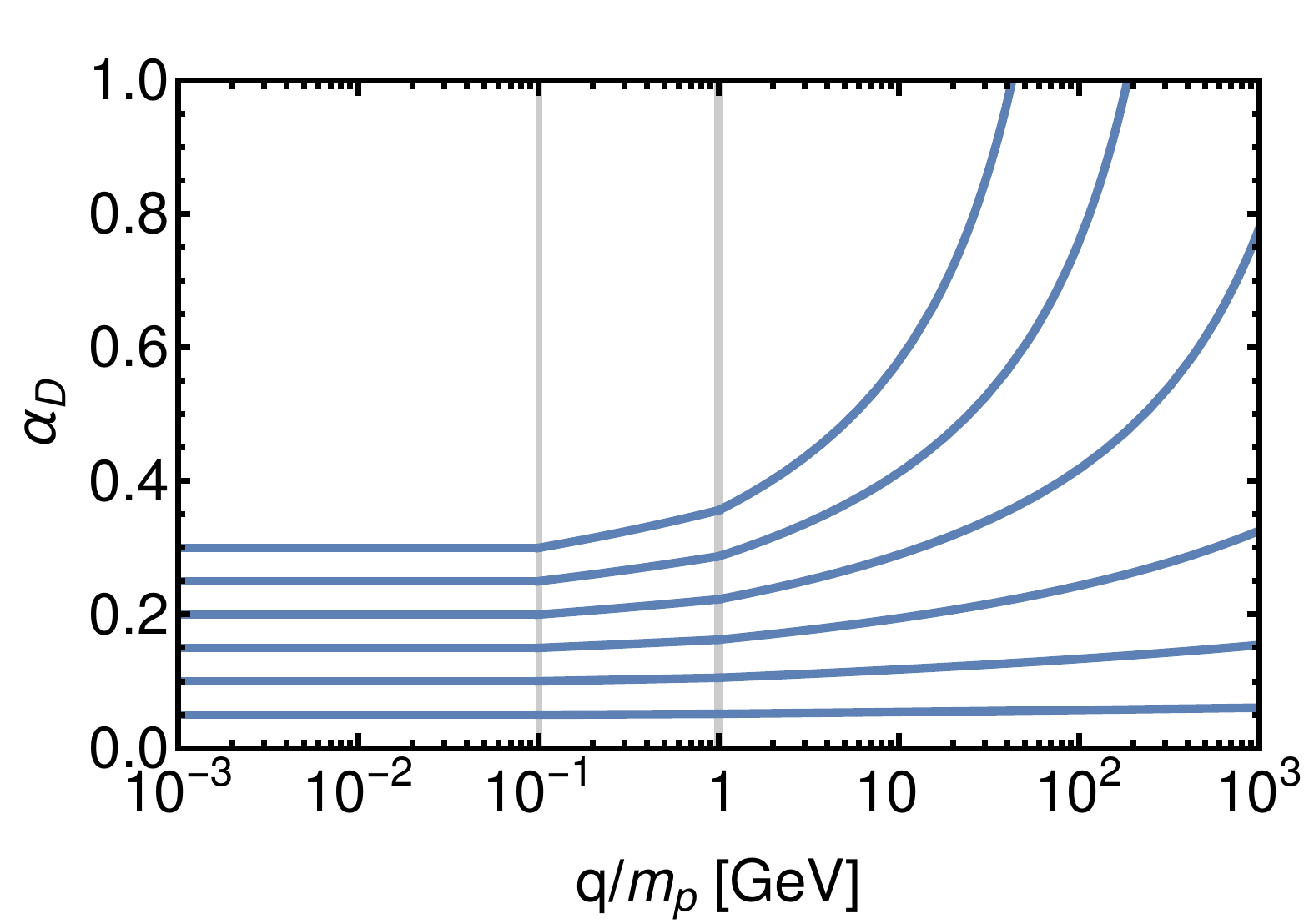}
\end{center}
\caption{\small Running of the dark gauge coupling with $\me = \mpd/10$. For $\aD(\mpd) \lesssim 0.2$ we are safe from a Landau pole for a few orders of magnitude above $\mpd$. The non-relativistic DM annihilation cross section naively violates unitarity at $\aD=0.68$, indicating our perturbative expansion has broken down for such large values of $\aD$, and that presumably our perturbative results are no longer to be trusted already somewhat below this value.  \bigskip}
\label{fig:running}
\end{figure}

The running of the dark gauge coupling is described by~\cite{Davoudiasl:2015hxa}
\begin{equation}
\frac{d \, \aD }{ d \, \mathrm{ln} \, q } = \beta_{D}(\aD,n_{F}),
\end{equation}
where $q$ is the renormalisation scale, and the $\beta$ function is analogous to QED and given at two-loop level by~\cite{DeRafael:1974iv,Broadhurst:1992za,Dunne:2002ta,Baikov:2012rr}
\begin{equation}
\beta_{D}(\aD,n_{F}) = \frac{ \aD^{2} }{ 2\pi } \left( \frac{4}{3}n_{F} + \frac{\aD}{\pi}n_{F}\right),
\end{equation}
and $n_{F}$ are the number of Dirac fermions. To be concrete, for $\mpd < q$ we have $n_{F} = 2$, for intermediate values $\me < q < \mpd$ we have $n_{F} =1$, and for $q < \me$ we have $n_{F}=0$. The result of an evaluation of the running is shown in \cref{fig:running}. As can be seen, we are free from low lying Landau poles provided $\aD(\mpd) \lesssim 0.2$, although slightly higher values are also possible depending on the demand placed on the range for a valid EFT above $\mpd$. Note our perturbative expansion leads to apparent unitarity violation in the DM annihilation cross section for $\aD \gtrsim 0.68$~\cite{Baldes:2017gzw,Baldes:2017gzu}.
The two constraints, no low lying Landau pole and no unitarity violation, therefore lead to roughly the same ballpark constraint on $\aD$.
\end{itemize}

\section{Direct detection}
\label{sec:directdetection}

\begin{figure}[t]
\begin{center}
\includegraphics[width=390pt]{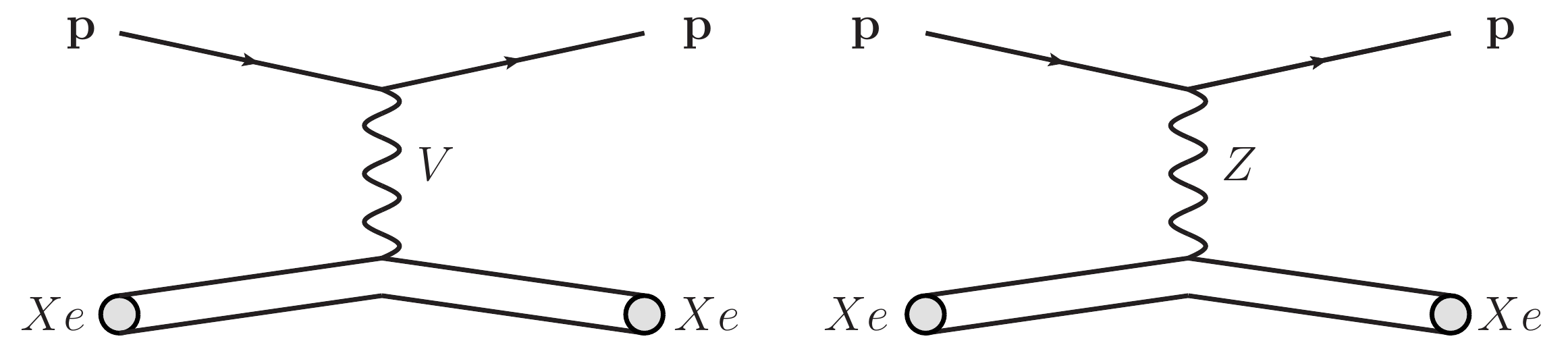}
\end{center}
\caption{\small Feynman diagrams contributing to the process of dark matter induced nuclear recoils of Xenon.}
\label{fig:DD_Xe} \bigskip
\end{figure}

Nuclear recoils are induced through $t$-channel exchange of the dark mediator $\V$ and the SM $Z$ boson, as depicted in \cref{fig:DD_Xe} (the photon has no tree level coupling to the DM). 
Here we do not attempt a thorough analysis of the DM direct detection, and will consider only the ionised component of DM. Due to their neutrality, the interaction of dark atoms with target nuclei is partially screened, although it may still be significant; we refer to~\cite{Cline:2012is,Kahlhoefer:2020rmg} for details. 

The spin-independent dark ion - target nucleus cross-section is given by
\begin{align}
\frac{ d \sigma }{ dE_{R} } = \frac{ M_{T} F_{\rm Helm}^{2} }{ 2\pi \vrel^{2} } &
\Bigg\{   \frac{ g_{V \p } [(A_{T}-Z_T)c_{V n}+Z_{T}c_{V p}] }{ 2M_{T}E_{R} + \mV^2 } \nonumber \\
& \qquad \qquad \qquad + \frac{ g_{Z \p } [(A_{T}-Z_T)c_{Z n}+Z_{T}c_{Z p}] }{ 2M_{T}E_{R} + M_Z^2 } \Bigg\}^{2},
\label{eq:DDfull}
\end{align}
where $E_{R}$ is the recoil energy, $M_T$ is the mass of the target nucleus, $g_{V \p }$ ($g_{Z \p }$) is the effective coupling of the $V$ ($Z$) to the dark matter, $c_{V p,n }$ ($c_{Z p,n }$) is the effective coupling of the $V$ ($Z$) to the SM proton and neutron, $A_T$ ($Z_{T}$) is the atomic mass (electric charge) of the target nucleus, and $F_{\rm Helm}$ is the Helm form factor~\cite{Helm:1956zz,Lewin:1995rx}. 
The effective couplings to the DM are given by
\begin{align}
g_{V \p }  =  \frac{ g_{\mathsmaller{D}} }{ \sqrt{1-\frac{ \epsilon^{2} }{ c_{w}^{2} }}} s_{\alpha}, \qquad \qquad
g_{Z \p }  =  \frac{ g_{\mathsmaller{D}} }{ \sqrt{1-\frac{ \epsilon^{2} }{ c_{w}^{2} }}} c_{\alpha} .
\end{align}
where $\alpha$ is a mixing angle which brings the massive neutral gauge bosons into diagonal form whose approximate expression can be found in \cref{eq:appalpha} (the full expression~\cite{Curtin:2014cca} is used in our code).
The couplings of the vector bosons to the nucleons can be derived from
\begin{align}
c_{Vn} = (c_{Vd_L}+c_{Vd_R})+(c_{Vu_L}+c_{Vu_R})/2, \\
c_{Vp} = (c_{Vu_L}+c_{Vu_R})+(c_{Vd_L}+c_{Vd_R})/2, \\
c_{Zn} = (c_{Zd_L}+c_{Zd_R})+(c_{Zu_L}+c_{Zu_R})/2, \\
c_{Zp} = (c_{Zu_L}+c_{Zu_R})+(c_{Zd_L}+c_{Zd_R})/2.
\end{align}
Here the couplings to the chiral components of the fields are given by~\cite{Curtin:2014cca}
\begin{align}
c_{Vf} &=  \frac{g}{c_{w}} \left(-s_{\alpha} [ c_{\theta}^{2}T_{3f} - s_{\theta}^{2}Y_f ] + \eta c_{\alpha} s_{\theta} Y_f \right), \label{eq:Vcoupling} \\
c_{Zf} &= \frac{g}{c_{w}} \left( c_{\alpha} [ c_{\theta}^{2}T_{3f} - s_{\theta}^{2}Y_f ] + \eta s_{\alpha} s_{\theta} Y_f \right),  \label{eq:Zcoupling}
\end{align}
where we remind the reader that $T_{3f}$ ($Y_f$) is the eigenvalue of the weak isospin (weak hypercharge) of the chiral field $f$, and $\eta$ is given in \cref{eq:eta}.
It is well known to dark photon aficionados that in the limit $\mV \ll M_{Z}$, the $c_{Vf}$ couplings to the fermions become proportional to $\epsilon Q_{f}$. Furthermore, the $Z$ exchange becomes suppressed compared to the $V$ exchange, due to the far more massive propagator. The above cross section then reduces to an electromagnetic one suppressed by an $\epsilon^{2}$ factor and modulo the finite $\mV$ mass. In this limit we may write
\begin{equation}
\frac{ d \sigma }{ dE_{R} } \to \frac{ M_{T} F_{\rm Helm}^{2} }{ 2\pi \vrel^{2} } \left( \frac{ \epsilon g_{\rm EM} g_{D} Z_{T} }{ 2M_{T}E_{R} + \mV^{2} } \right)^{2},
\label{eq:EM1}
\end{equation} 
where $g_{\rm EM}$ is the electromagnetic coupling strength. This cross section has been used in a number of previous studies, {\it e.g.}~\cite{Cirelli:2016rnw,Baldes:2017gzu}. Amusingly, the full cross section, \cref{eq:DDfull}, reduces to the same limiting behaviour also for heavier dark mediator masses. To see this, note that for the non-electromagnetic type coupling of $V$ to be in effect, $\mV \gtrsim 10$ GeV $\gg 10 $ MeV $\gtrsim 2E_{R}M_T$, as the recoil energy is limited by the non-relativistic velocities of the DM in the halo. We can therefore ignore the momentum exchange in the propagators. In the limit of a small mixing, the couplings can be approximated by
	\begin{align}
	g_{V \p } & \simeq g_{\mathsmaller{D}} , \qquad  \qquad
	& g_{Z \p }  \simeq - \frac{ \epsilon g_{\mathsmaller{D}} t_{w}}{ 1- \delta^{2} }   \\
	c_{Vf} & \simeq \epsilon t_{w} \left( \frac{c^{\rm SM}_{Z f}}{1-\delta^{2}} + \frac{gY_{f}}{c_{w}} \right) \qquad \qquad
	& c_{Zf}  \simeq c^{\rm SM}_{Z f} - \frac{ \epsilon^{2}t_{w}^{2}gY_{f} }{ 1-\delta^{2} }
	\end{align}
where $\delta \equiv \mV/M_Z$, and $c^{\rm SM}_{Z f}$ is the SM coupling of the $Z$ boson to chiral fermion $f$, which can be found by using \cref{eq:Zcoupling} and taking the appropriate limit. Substituting the above approximate forms into \cref{eq:DDfull} and ignoring the momentum exchange, one finds the different couplings and masses associated with the two propagators simplify down to
	\begin{equation}
	\frac{ d \sigma }{ dE_{R} } \to \frac{ M_{T} F_{\rm Helm}^{2} }{ 2\pi \vrel^{2} } \left( \frac{ \epsilon g_{\rm EM} g_{D} Z_{T} }{ \mV^{2} } \right)^{2}+\mathcal{O}(\epsilon^{6}),
	\end{equation} 
independent of $\delta$, which is just the same as \cref{eq:EM1} albeit with no momentum exchange. 

The rate of nuclear recoils per unit of fiducial target mass is given by
	\begin{equation}
	\frac{{\rm d}R_{\rm T}}{{\rm d}E_{\rm R}}= \frac{\xi_{\rm T}}{{m_{\rm T}}}\frac{\rho_\odot}{\mpd+\me}\int_{v_{\rm min}}^{v_{\rm esc}} \hspace{-.4cm}{\rm d}^3 v \, v f_{\rm E}(\vec v)\frac{{\rm d}\sigma}{{\rm d}E_{\rm R}}(v,E_{\rm R})  \ ,
	\label{eq:DDNR}
	\end{equation}
where $\xi_{\rm T}$ is the mass fraction of the target nucleus. Here a number of astrophysical parameters enter for which we assume the standard halo model: $\rho_\odot =0.3$ GeV/cm$^3$ is the local DM energy density, $f_{\rm E}(\vec v)$ is the DM speed distribution in the Earth's frame, given a Maxwellian DM velocity distribution in the halo frame with peak DM speed $v_{0} = 220$ km/s and $v_{\rm Earth}$ = 232 km/s,  $v_{\rm esc}=544$ km/s is the Milky Way's escape speed, and $v_{\rm min}$ is the minimum speed for which DM particles can provide a given recoil energy $E_{\rm R}$~\cite{Green:2011bv}. 

We find the limit on the model by confronting it with the latest XENON1T results~\cite{Aprile:2018dbl}. Constraints from LUX~\cite{Akerib:2016vxi} and PANDAX~\cite{Cui:2017nnn,Ren:2018gyx} are expected to give similar results. At low DM masses, CRESST-III~\cite{Abdelhameed:2019hmk}, CDMS~\cite{Agnese:2017jvy}, CDEX~\cite{Jiang:2018pic}, and DarkSide~\cite{Agnes:2018ves} provide more stringent constraints, see {\it e.g.} the analysis in~\cite{Cirelli:2016rnw,Baldes:2017gzu,Li:2019drx}. Alternatively the Migdal effect can be exploited~\cite{Kouvaris:2016afs,Ibe:2017yqa,Dolan:2017xbu,Aprile:2019jmx}. Here we shall focus on the limits for $\mpd \gtrsim 50$ GeV using a simple analysis. More sophisticated analyses taking into account the shape of the spectrum are of course possible~\cite{Hambye:2018dpi}.

Note we have multi-component DM in our model. For sufficiently heavy masses for the DM components, $\mpd, \me \gtrsim 50$ GeV, away from threshold effects, this increases the expected number of scattering events by a factor of two, as $\ne = \np$. This holds provided $\me \ll \mpd$, which we assume here, so the former component is negligible for the total DM energy density, otherwise there is a suppression as can be seen in \eqref{eq:DDNR}. This factor of two is included in our limit. For a more detailed study of direct detection of multi-component DM see~\cite{Herrero-Garcia:2018qnz}.

To set a limit we use \cref{eq:DDfull,eq:DDNR} convoluted with the best fit total efficiency of the detector, shown in Fig.~1 of ~\cite{Aprile:2018dbl}, to find the expected number of events in XENON1T for our model given $\epsilon$, $\aD$, $\mpd$, and $\mV$. The XENON1T collaboration has reported 14 events in their nuclear recoil signal reference region in 278.8 days of exposure time of their 1.3 tonnes of fiducial mass, see the second column, table I of~\cite{Aprile:2018dbl}. The estimated background is $7.36 \pm 0.61$ events.  We take the $90 \%$ C.L. limit which corresponds to DM contributing 12.8 events~\cite{Kavanagh:2016pyr}. We find an exclusion by demanding the expected number of events at a given parameter point in our model not exceed 12.8. The result of such a procedure is shown in \cref{fig:darkphoton}. Although this is a simplified procedure, for DM masses $\mpd \gtrsim 30$ GeV, it returns a limit on the generic spin-independent cross section matching that of the XENON1T analysis within a factor of two. Thus it is sufficiently accurate for our purposes here.

\section{Dark sector temperature}

\label{sec:TD}

\begin{figure}[t]
\begin{center}
\includegraphics[width=190pt]{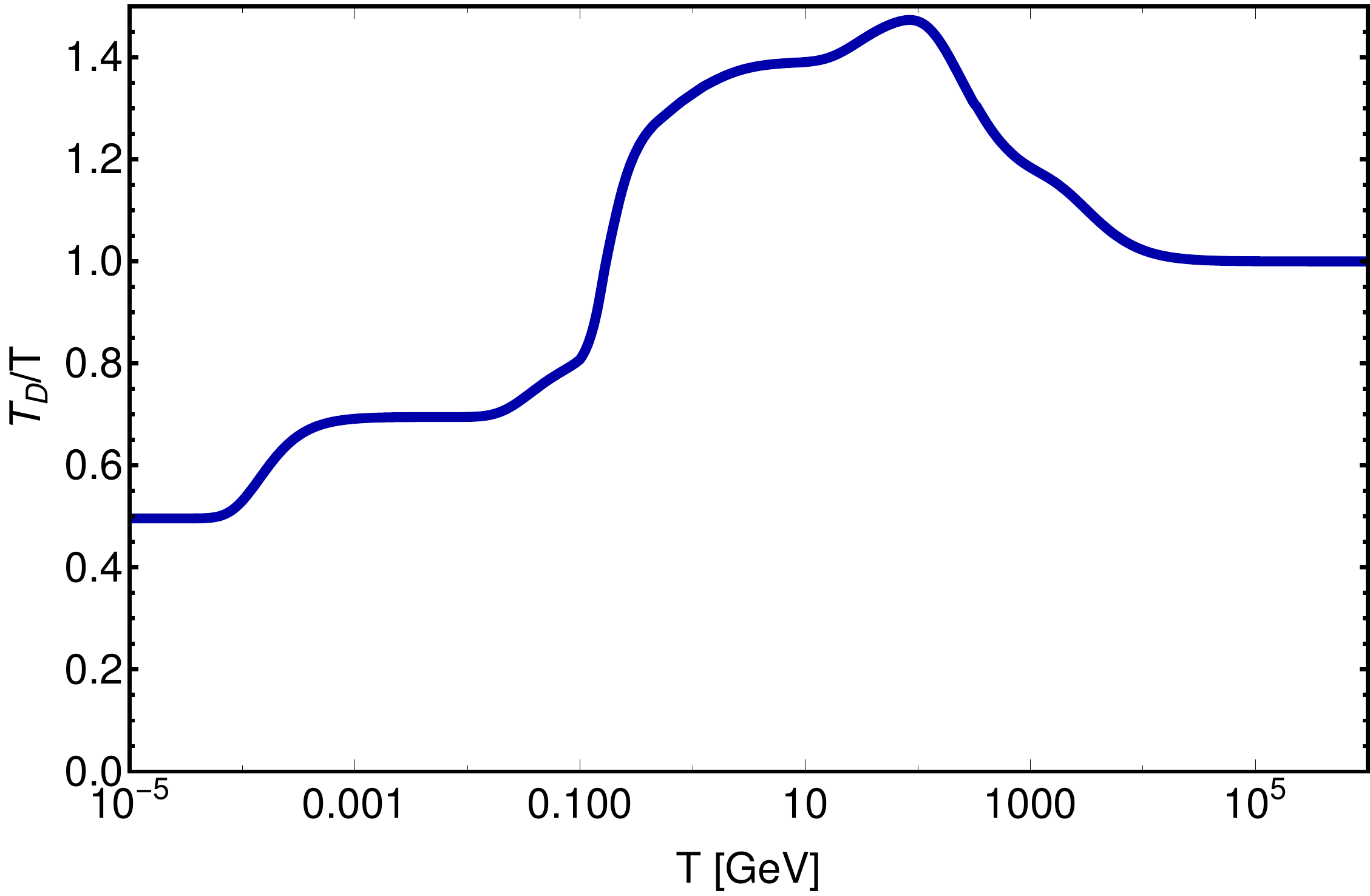}
\includegraphics[width=190pt]{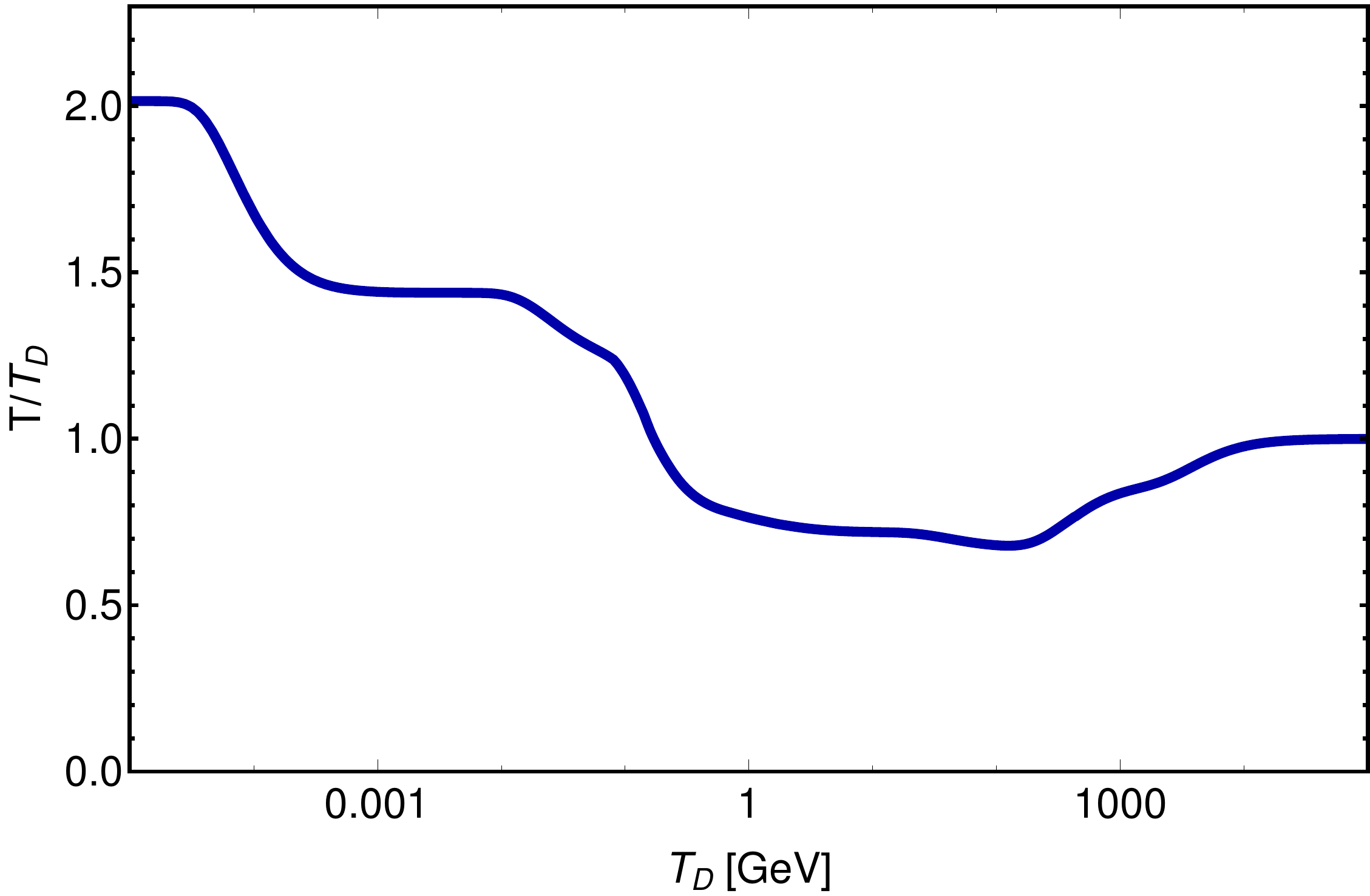}
\end{center}
\caption{\small Left: the evolution of the dark-to-visible temperature ratio $\tau\equiv \TD/T$, as a function of $T$, for $\tau_{i}=1$ at $T=10^{4}$ TeV, $\mpd = 10$ TeV and $\me=1$ TeV, calculated using an iterative approach.
Right: the evolution of $1/\tau$ vs $T_{D}$ for the same choice of parameters. \bigskip}
\label{fig:phioverT}
\end{figure}

Let us denote the visible sector temperature with $T$ and the dark sector temperature $\TD$. Following from independent conservation of entropy in each sector, the temperature ratio is given by
\begin{equation}
\tau \equiv \frac{\TD}{T} = \left( \frac{h_{\rm SM}(T)}{h_{\rm SM}(T_{i})} \frac{h_{\rm D}(T_{i})}{h_{\rm D}(T)} \right)^{1/3} \tau_{i}
\label{eq:entconsv}
\end{equation}
where $\tau_{i}$ is the initial temperature ratio at temperature $T_{i}$ and $h_{\rm SM}$ ($h_{\rm D}$) count the effective entropic degrees-of-freedom in the SM (dark) sector. Prior to the decay of the dark photons, the effective entropic degrees-of-freedom in the dark sector may be modelled as
\begin{equation}
h_{D} = 3 + \frac{7}{8} \times 4 \times \left[\tilde{n} \left( \frac{ \me }{ \TD } \right) \right] + \frac{7}{8} \times 4 \times \left[\tilde{n}\left( \frac{ \mDM }{ \TD } \right) \right],
\label{eq:darkdof}
\end{equation}
where we model the disappearance of a massive species from the thermal bath with the ratio of the number density to the massless number density, $\tilde{n}(x)=(x)^{2}K_{2}(x)/2$, where $K_{2}(x)$ is the modified Bessel function of the second kind of order two.

Now we wish to find $\tau(T)$. Due to $\TD$ entering on both sides of \cref{eq:entconsv}, one can not trivially evaluate $\tau(T)$ analytically. 
Nevertheless, as long as $\tau$ does not depart too far from unity, one can easily estimate it by taking into account the various mass thresholds in the dark sector, together with the SM degrees-of-freedom. To obtain a more accurate evaluation of $\tau(T)$, an iterative approach can be used. The result of such an evaluation is shown in \cref{fig:phioverT}. Very similarly one can of course also find $\tau$ as a function of $\TD$.

\section{Boosting spectra between frames}
\label{sec:boostdetails}
Here we provide a derivation of the boosted spectrum result provided in Eq.~\ref{eq:boostedspec}. Before converting to dimensionless parameters, the boosted spectrum can be written as
\be
\frac{dN}{dE} = \int_0^{\mV/2} dE_0\, \int_{-1}^1 dz\,p(z)\, \frac{dN}{dE_0}(E_0)\, \delta \left[ E - E_0 \frac{\EV}{\mV} \left( 1 + z \sqrt{1 - \frac{\mV^2}{\EV^2}} \right) \right]\,,
\ee
where $E$ and $E_0$ are the photon energy in the observer and $\V$ rest frames, respectively. In detail, we know that the energy of the photon in the observer frame is given by Eq.~\ref{eq:boostedphotonenergy}. This energy depends on both the energy and angle of the photon in the $\V$ rest frame, each of which are drawn from the distributions $dN/dE_0$ and $p(z)$ respectively. We obtain the full spectrum by simply marginalising over both of these distributions. Converting to dimensionless quantities, we have
\bea
\frac{dN}{dx} 
= &\int_0^1 dx_0\, \int_{-1}^1 dz\,p(z)\, \frac{dN}{dx_0}(x_0)\, \delta \left[ x - \frac{1}{2} x_0 \left( 1 + z \sqrt{1 - \epsilon_B} \right) \right] \\
= &\frac{2}{\sqrt{1 - \epsilon_B}} \int_0^1 \frac{dx_0}{x_0}\, \int_{-1}^1 dz\,p(z)\, \frac{dN}{dx_0}(x_0)\, \delta \left[ z - \frac{2 x/x_0 - 1}{\sqrt{1 - \epsilon_B}} \right]\,.
\eea
Recall $\epsilon_B = (\mV/\EV)^2$, $x_0 = 2 E_0/\mV$, and $x = E/\EV$.

Now we will use the $\delta$-function to perform the angular $z$ integral. To do so, we must consider where the $\delta$-function has support. To begin with, as $x_0 \in [0,1]$ generically, we have
\be
0 \leq x \leq  \frac{1}{2} \left( 1 + \sqrt{1 - \epsilon_B} \right)\,.
\ee
For the $\delta$ function to have support, we require
\be
\frac{2x}{\epsilon_B} (1-\sqrt{1 - \epsilon_B}) \leq x_0 \leq \frac{2x}{\epsilon_B} (1+\sqrt{1 - \epsilon_B})\,.
\ee
Accordingly, we conclude
\bea
\frac{dN}{dx} 
= &\frac{2}{\sqrt{1 - \epsilon_B}} \int_{x_0^{\rm min}}^{x_0^{\rm max}} \frac{dx_0}{x_0}\, p\left(\frac{2 x/x_0 - 1}{\sqrt{1 - \epsilon_B}} \right)\, \frac{dN}{dx_0}(x_0)\,, \\
x_0^{\rm min} = & \frac{2x}{\epsilon_B} (1-\sqrt{1 - \epsilon_B})\,, \\
x_0^{\rm max} = & {\rm min} \left[ 1,\, \frac{2x}{\epsilon_B} (1+\sqrt{1 - \epsilon_B}) \right]\,,
\eea
which is the result quoted in the main text. Note if we are not considering polarised $\V$ decays, but just averaging over all polarisations, then we take $p(z) =1/2$, and the result is equivalent to (B3)/(B4) of~\cite{Elor:2015tva}. Similarly, in the large hierarchies limit ($\epsilon_B \to 0$), this reduces to (14) of the same work.

\section{Analytic results for Final State Radiation}
\label{sec:analyticfsr}

We want to determine the spectrum of photons resulting from final state radiation of the form $V \to \ell^+ \ell^- \gamma$, where $\ell = e, \mu$. Conventionally in the literature, the form used is
\be
\frac{dN}{dx} = \frac{\alpha_{\rm EM}}{\pi} \frac{1+(1-x)^2}{x} \left[ \ln \left( \frac{1-x}{\epsilon_{l}} \right) - 1 \right]\,.
\label{eq:FSRexpanded}
\ee
See, for example, (A2) of~\cite{Elor:2015tva}. Recall here $x = 2E_{\gamma}/\mV$ and $\epsilon_{l} = m_{\ell}^2/\mV^2$. The above is an expansion in $\epsilon_{l}$, so it assumes $\epsilon_{l} \ll 1$. Nevertheless, we are considering small vector masses, all the way to $\epsilon_{l} \sim 1$, and thus this approximation will not be valid. Thus we need a more general result. 

We can obtain this from the calculation in~\cite{Ioffe:1978dc} for $e^+ e^- \to Q \bar{Q} g$, where $Q$ is a heavy quark. From that result, we determine
\bea
\frac{dN}{dx} 
=\frac{\alpha_{\rm EM}}{\pi} &\left[ \frac{1+(1-x)^2-4\epsilon_{l} (x+2\epsilon_{l})}{x \left( 1 + 2 \epsilon_{l} \right) \sqrt{1-4\epsilon_{l}}} \ln \left[ \frac{1+\sqrt{1-4\epsilon_{l}/(1-x)}}{1-\sqrt{1-4\epsilon_{l}/(1-x)}} \right] \right. \\
&\left.
- \frac{1+(1-x)^2+4\epsilon_{l}(1-x)}{x \left( 1 + 2 \epsilon_{l} \right) \sqrt{1-4\epsilon_{l}}} \sqrt{1-\frac{4\epsilon_{l}}{1-x}} \right]\,.
\eea
From this form we can see straightforwardly, that in the limit $\epsilon_{l} \to 0$, this reduces to Eq.~\eqref{eq:FSRexpanded} up to $\mathcal{O}(\epsilon_l)$ corrections. The full result shown here also agrees with the calculation in~\cite{Coogan:2019qpu}.

\begin{figure}[t]
\begin{center}
\includegraphics[scale=0.25]{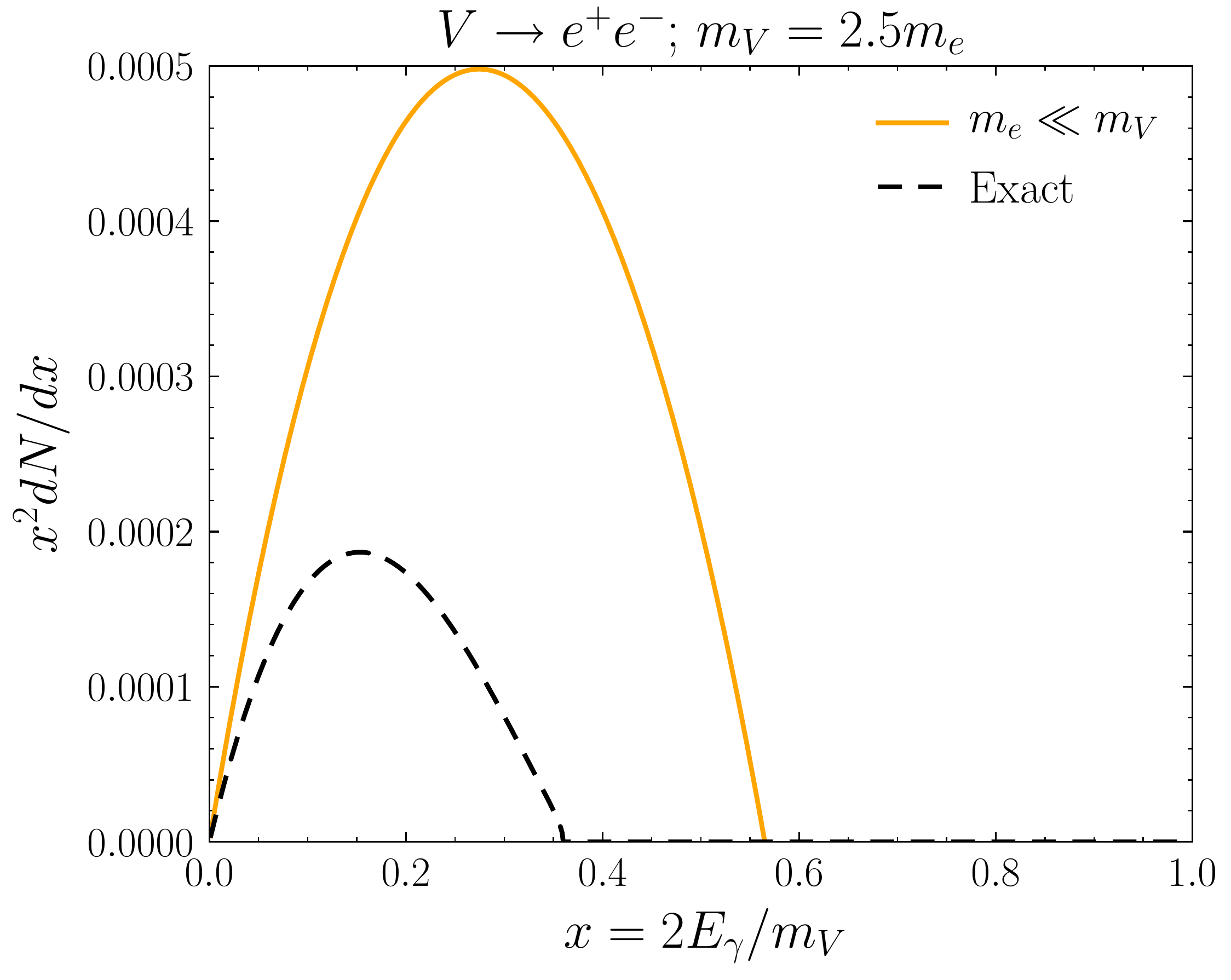}\hspace{0.2cm}
\includegraphics[scale=0.25]{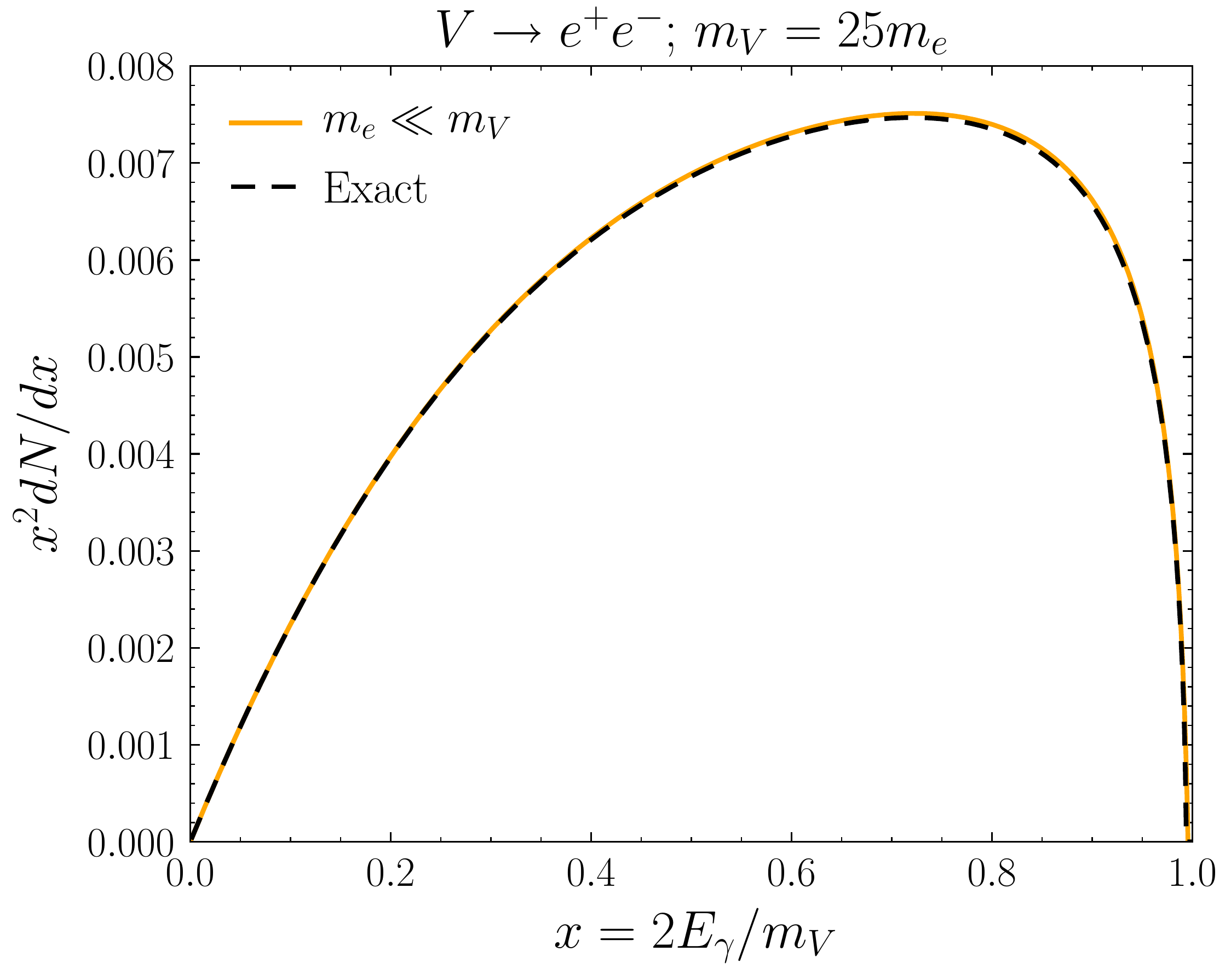}\hspace{0.2cm}
\end{center}
\caption{\small Comparison of the exact and approximate expressions in the $V$ rest frame for two different choices of $\mV$. \bigskip}
\label{fig:restcomparison}
\end{figure}

From the full result, we can determine the kinematic limits on the photon energy. The minimum photon energy is 0, whereas the maximum is when the photon is emitted in the opposite direction of the $\ell^+ \ell^-$, which are collinear and of equal energy. Then we have
\be
2 \sqrt{E_{\ell}^2-m_{\ell}^2} = E_{\gamma}\,,
\ee
or
\be
2 \sqrt{x_{\ell}^2-4\epsilon_{l}} = x\,.
\ee
Energy conservation gives $x_{\ell} = 1-x/2$, so that
\bea
4 ((1-x/2)^2-4\epsilon_{l}) = x^2\,,
\eea
which rearranges to give a maximum of $x=1-4\epsilon_{l}$, and hence $x \in [0,1-4\epsilon_{l}]$. We can see from the above that if $x > 1-4\epsilon_{l}$, then $\sqrt{1-4\epsilon_{l}/(1-x)}$ becomes imaginary. In the $\epsilon_{l} \to 0$ limit, we have $x \in [0,1]$.
Numerically, we can compare the exact and approximate expressions. This is done for two different values of $\mV$ in \cref{fig:restcomparison} (all spectra in the $V$ rest frame). We see that for $\mV \sim 2 m_e$ there is a significant difference.

\end{appendix}

\bibliography{BSF_refs.bib}

\nolinenumbers

\end{document}